\documentclass[a4paper,12pt]{book}

\usepackage{mathtools}
\usepackage{amsfonts}
\usepackage{fullpage}
\usepackage{braket}
\usepackage[english,greek]{babel}
\usepackage{verbatim}
\usepackage[utf8]{inputenc}
\usepackage{textalpha}
\usepackage{caption}
\usepackage{subcaption}
\DeclareMathOperator{\Tr}{Tr} 
\usepackage{hyperref}
\usepackage{enumitem}
\usepackage{setspace}

\begin{document}
\selectlanguage{english}

\bibliographystyle{bibliography/aip.bst}

\begin{titlepage}
\begin{center}
\vspace*{1cm}
 
\Huge
\textbf{Quantum chaos in many-body systems without a classical analogue}
 
\vspace{0.5cm}
\LARGE
 
\vspace{1.5cm}

Fotis I. Giasemis
 
\vfill

\includegraphics[width=0.3\textwidth]{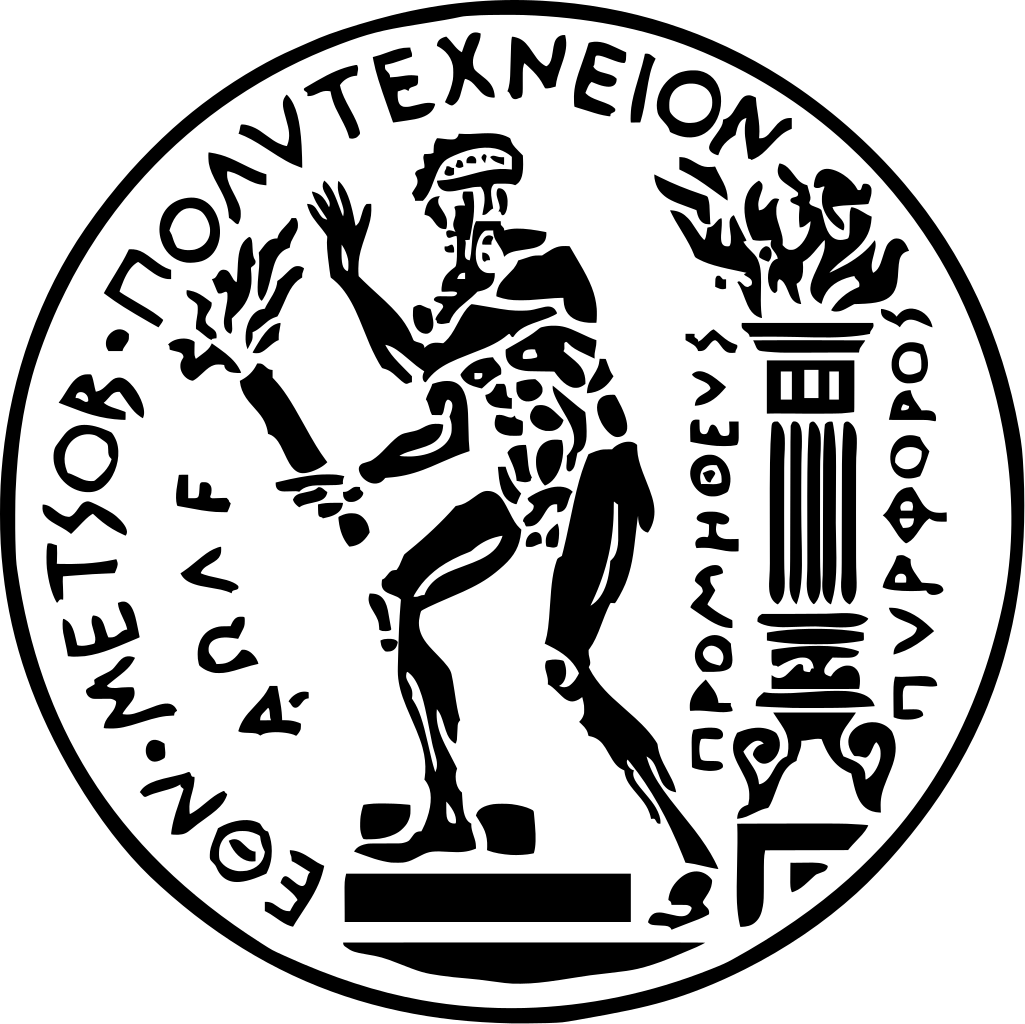}
\vfill

{\Large
\begin{tabular}[t]{@{}l@{\hspace{3pt}}p{.32\textwidth}@{}}
Examiners: & Alex Kehagias  \\
& Yannis Kominis \\
& Astero Provata \\
\end{tabular}%
\begin{tabular}[t]{@{}l@{\hspace{3pt}}p{.3\textwidth}@{}}
Supervisors: & Alex Kehagias \\
& Achilleas Lazarides \\
& Spyros Sotiriadis
\end{tabular}%
}

\vspace{.8cm}

Dissertation Submitted in Partial Fulfilment of the\\ Requirements for the Degree of\\ Master of Applied Mechanics
 
\vspace{.8cm}

Department of Mechanics\\
School of Applied Mathematical and Physical Sciences\\
National Technical University of Athens\\
\vspace{.4cm}
Athens, June 17, 2022
 
\end{center}
\end{titlepage}

\onehalfspacing

\pagebreak

\chapter*{Abstract}
In classical systems, chaos is clearly defined via the behavior of trajectories. In quantum systems with a classical analogue one finds that the transition from regular to chaotic dynamics is signified by a change in the spectral statistics. This has been found to remain true for quantum systems with no classical analogue, including many-body systems. Furthermore, quantum chaotic systems explore all the allowed configurations in the Hilbert space, i.e. they are ergodic, while integrable systems, and systems in the many-body localized phase, are restricted to a certain subspace of the available phase space, and hence strongly break ergodicity. In this dissertation, we study the intermediate behavior between ergodicity and localization, i.e. the weak breaking of ergodicity. The model examined is the PXP spin chain model, where spins are allowed to flip only under certain kinetic constraints. We start by reproducing some already established results. First, we explore the eigenstate thermalization hypothesis (ETH) for this model and demonstrate the existence of a small number of states, throughout the PXP spectrum, that violate the ETH. Then we study the level-spacing statistics of the model, a well-known quantum chaos diagnostic, which turns out to be close to semi-Poisson and approach Wigner--Dyson statistics for large system sizes. Moreover, we examine various aspects of the model that have not been studied before. For example, the eigenvector component statistics, another quantum chaos diagnostic, for the PXP model turn out to be non-Gaussian. Finally, we perform a quench, in order to study how the energy spreads throughout the system, and observe ballistic fronts. 

\vspace{1cm}
\noindent
\textbf{Keywords: quantum chaos, quantum many-body scars, weak ergodicity breaking, PXP, spin chain model, eigenstate thermalization hypothesis, quantum thermalization, level-spacing statistics, eigenvector component statistics, quantum quench, energy density front}

\chapter*{Abstract in Greek}

\selectlanguage{greek}

Σε κλασικά συστήματα, το χαός ορίζεται από την συμπεριφορά των τροχιών. Σε κβαντικά δυναμικά συστήματα με κλασικό ανάλογο, η μετάβαση από κανονική σε χαοτική δυναμική, σηματοδοτείται από αλλαγή στην στατιστική των ιδιοενεργειών. Αυτό ισχύει ακόμη και σε κβαντικά συστήματα χωρίς κλασικό ανάλογο, συμπεριλαμβανομένων των συστημάτων πολλών σωμάτων. Επιπλέον, τα κβαντικά χαοτικά συστήμα εξερευνούν όλο τον φασικό χώρο διαθέσιμο, δηλαδή είναι εργοδικά, ενώ τα ολοκληρώσιμα συστήματα, και συστήματα σε {\selectlanguage{english}MBL} φάση, είναι περιορισμένα σε ένα συγκεκριμένο μέρος του συνολικού διαθέσιμου φασικού χώρου, και επομένως σπάνε ισχυρά την εργοδικότητα. Σε αυτή την εργασία μελετάμε την ενδιάμεση συμπεριφορά μεταξύ της εργοδικότητας και της μη εργοδικότητας, δηλαδή το ασθενές σπάσιμο της εργοδικότητας. Το μοντελό που μελετάται είναι η {\selectlanguage{english}PXP} σπιν αλυσίδα, όπου τα σπιν μπορούν να αλλάξουν μόνο υπό κάποιες συνθήκες. Ξεκινάμε αναπαράγοντας κάποια ήδη γνωστά αποτελέσματα. Πρώτα ελέγχουμε την εγκυρότητα μίας βασικής υπόθεσης στην θεωρία της κβαντικής δυναμικής πολλών σωμάτων, την λεγόμενη {\selectlanguage{english}eigenstate thermalization hypothesis (ETH)}, επιβεβαιώνοντας την ύπαρξη κάποιων καταστάσεων που σπάνε την {\selectlanguage{english}ETH}. Έπειτα, μελετάμε την στατιστική των αποστάσεων μεταξύ των ιδιοενεργειών του μοντέλου, ένα γνωστό τεστ για τον αν ένα κβαντικό σύστημα είναι χαοτικό ή όχι. Επίσης, ερευνούμε διάφορες ιδιότητες του μοντέλου που δεν έχουν εξετασθεί. Για παράδειγμα, η στατιστική των συνιστωσών των ιδιοδιανυσμάτων για το {\selectlanguage{english}PXP} μοντέλο είναι μη γκαουσιανή. Τέλος, μελετάμε την διάδοση της ενέργειας στο σύστημα και παρατηρούμε γραμμική συμπεριφορά.

\vspace{1cm}
\noindent
\textbf{Λέξεις κλειδιά: κβαντικό χάος, εργοδικότητα, κβαντικές ουλές, {\selectlanguage{english}PXP}, σπιν αλυσίδα, κβαντική θερμοποίηση}

\selectlanguage{english}

\listoffigures

{\renewcommand{\baselinestretch}{1}\tableofcontents}

\vfill
\pagebreak
\subsection*{Acknowledgements}
Throughout the writing of this dissertation I have received a great deal of support and assistance. I would like to thank my supervisors, Prof Alex Kehagias, Dr Achilleas Lazarides and Dr Spyros Sotiriadis, whose expertise was invaluable in formulating the research questions and methodology. Your insightful feedback and interesting discussions pushed me to sharpen my thinking and motivated me to gain a deeper understanding of the field of quantum many-body dynamics.

\chapter{Introduction}

Chaos theory became an established science in the second half of the 20th century, even though it was proposed as early as 1880 by Henri Poincare \cite{Poincare}. This theory describes a certain class of dynamical systems that may exhibit high sensitivity to initial conditions. Even a minor perturbation can lead to an exponential growth of perturbations in the initial conditions, resulting in a completely different trajectory. As a result of this sensitivity, the behavior of these so-called chaotic systems appears random, which seems in contradiction to the fact that these systems are deterministic. 

At the same time, quantum mechanics, already developed in the first half of the 20th century, challenged the classifications based on the behavior of the trajectories. Quantum chaos studies this incompatibility: how chaotic classical systems can be described in terms of quantum theory. It is now generally accepted that in quantum systems the transition from regular to chaotic dynamics is signified by a change in the spectral statistics \cite{haake}. 

More recently, significant efforts have been focused on understanding \textit{quantum thermalization}, i.e. how isolated quantum systems approach equilibrium. This interest has come hand in hand with the experimental advances in controllable, quantum-coherent systems of ultracold atoms \cite{Kinoshita:2006ts,Michael:2015ub}, trapped ions \cite{Smith:2016uo}, and nitrogen-vacancy spins in diamond \cite{Kucsko:2018tn}. These systems allow one to realize highly tunable lattice models of interacting spins, bosons, or fermions, and to characterize their quantum thermalization \cite{M.:2016wg}.

The process of thermalization is believed to be governed by the properties of the system's many-body eigenstates via the eigenstate thermalization hypothesis (ETH). Although there is no formal proof for the ETH, it has been demonstrated numerically for various system of spins, fermions and bosons, in one (1D) and two dimensions (2D) \cite{Rigol:2007vf,Rigol:2008vq}. More specifically, it is suggested that in many cases when the system thermalizes, \textit{all} of its highly excited states obey the ETH \cite{Kim:2014uk}, i.e., they are typical thermal states and akin to random vectors.

However, not all quantum systems obey the ETH. In integrable systems, and systems in the many-body localized phase \cite{DAlessio:2016tw,Serbyn:2013ug}, the ETH is strongly violated, because of the presence of an extensive number of operators $K_i$, which commute with the system's Hamiltoninian $ [H,K_i] = 0 $. The presence of these operators prevents the system from exploring all allowed configurations in the Hilbert space, leading to strong ergodicity breaking. 

Despite the progress in theoretical understanding of ergodic systems, and systems that strongly violate the ETH, much less in known about the existence of intermediate behaviors. Can ergodicity be broken \textit{weakly}? Indeed, recently, an example of this was discovered experimentally \cite{Bernien:2017to}. A Rydberg atom platform \cite{Bernien:2017to,Schauss:2012tk,Labuhn:2016wc} was used to realize a quantum model with kinetic constraints induced by strong nearest-neighbor repulsion between atoms in excited states. In this thesis we numerically study the effective Hamiltonian for this system, according to \cite{Turner:2018to}, the so-called PXP.

In Section \ref{section:pxp}, we describe the PXP model. Then, in Chapters \ref{ch:eth}-\ref{ch:level}, we reproduce some already established results. In the former, we explore ETH for this model and demonstrate the existence of a small number of states, throughout the PXP spectrum, that violate the ETH. In the latter, we study the level-spacing statistics of the model, a well-known diagnostic for quantum chaos. Moreover, in Chapters \ref{ch:eigvec}-\ref{ch:transport}, we study various aspects of the model that have not been studied before. In Chapter \ref{ch:eigvec}, we study the eigenvector component statistics, which turn out to be non-Gaussian. This non-Gaussianity is unexpected since chaotic systems have eigenvectors akin to Gaussian random vectors. In Chapter \ref{ch:transport}, we perform a quench on the system and study its response. By studying the spreading of energy, a conserved quantity, throughout our system, we can make conclusions about whether the system is diffusive or non-diffusive. The ballistic (linear) fronts observed also come as a surprise since the system, prepared in a generic state, is diffusive.

\section{Kinetically Constrained PXP Model} \label{section:pxp}
We study the following spin-$1/2$ Hamiltonian which models a Rydberg atom chain in the limit where the nearest-neighbor interaction is much larger than the detuning and the Rabi frequency \cite{Lesanovsky:2012tq,Turner:2018ts}.

\begin{equation} \label{pxp}
H_{\text{PXP}} = \sum_{i=1}^L P_{i-1}X_{i}P_{i+1},
\end{equation}
where $X_i$, $Y_i$, $Z_i$ are the Pauli operators, $L$ denotes the length of the chain and we work in units $\hbar = 1$. We assume that each atom can be either in the ground state, $\ket{\circ}$, or the excited (Rydberg) state, $\ket{\bullet}$, of a single atom. The operator $X_i =\ket{\circ} \bra{\bullet} + \ket{\bullet} \bra{\circ} $ creates or removes an excitation at site $i$, and projectors $P_i = \ket{\circ} \bra{\circ} = (1-Z_i)/2$, written in terms of $Z_i = \ket{\bullet} \bra{\bullet} - \ket{\circ} \bra{\circ}$, ensure that two adjacent atoms are not in the excited state simultaneously. For example, $P_1X_2P_3$ acting on $\ket{\circ \circ \circ }$ gives $\ket{\circ \bullet \circ}$, while it annihilates any of the configurations $\ket{\bullet \circ \circ}$, $\ket{\circ \circ \bullet}$, $\ket{\bullet \circ \bullet}$. For the derivation of the PXP Hamiltonian see \cite{Turner:2018to}.

The Hamiltonian in Eq. (\ref{pxp}) does not allow for relaxation of several adjacent Rydberg atoms. In other words, states of the form $\ket{ \cdots \bullet \bullet \cdots}$ break our chain into two parts that are completely independent. The Hilbert space is broken into disconnected parts, and often instead of studying the whole Hilbert space we study only the reduced Hilbert space without any adjacent Rydberg atoms. With that constraint, and for periodic boundary conditions (PBC), the Hilbert space dimension is equal to $\mathcal{D} = F_{L-1} + F_{L+1}$, where $F_n$ is the $n$th Fibonacci number. On the other hand, for open boundary conditions (OBC), $\mathcal{D} = F_{L+2}$. Thus, the Hilbert space is very different from, for example, the spin-$1/2$ chain where the number of states grows as $2^L$. Also, the model of Eq. (\ref{pxp}) has spatial inversion symmetry $I$ which maps $i \rightarrow L-i+1$, as well as translation symmetry, with PBC. For some calculations, we will explicitly resolve these symmetries in order to fully diagonalize systems with up to $L = 32$ sites.

Experiment \cite{Bernien:2017to} and simulations on small systems \cite{Sun2008} have shown that the relaxation under time evolution with the Hamiltonian in Eq. (\ref{pxp}) strongly depends on the initial state of the system. In particular, starting from period-2 charge density wave states $$ \ket{ \mathbb{Z}_2 } = \ket{ \bullet \circ \bullet \circ \cdots},\ \ket{ \mathbb{Z}_2 ' } = \ket{ \circ \bullet \circ \bullet \cdots}, $$ the system shows surprising long-time oscillations, while starting from the state $ \ket{\circ} $ the system quickly relaxes and shows no revivals. The latter behavior is characteristic of a thermalizing system, while the former might suggest that the system is non-ergodic. Given that the model in Eq. (\ref{pxp}), with PBC imposed, is translation invariant and has no disorder, many-body localization cannot be the cause of this weak breaking of ergodicity.

\chapter{Thermalization and Special Eigenstates} \label{ch:eth}
In this chapter, reproducing results from \cite{Turner:2018to}, we investigate properties of eigenstates of the Hamiltonian in Eq. (\ref{pxp}) using exact diagonalization methods \cite{Sandvik}. First, we directly test the ETH. We find that the majority of the eigenstates appear thermal, apart from a small subset of special eigenstates that strongly violate ETH. We will call these states, following \cite{Turner:2018to}, quantum many-body scars (QMBSs). Then, we calculate the overlap of the eigenstates of the PXP Hamiltonian with the $ \ket{ \mathbb{Z}_2 } $ product state. We observe that there exists a set of eigenstates with anomalously large overlap. The states at the top of the tower-like structures coincide with the aforementioned special eigenstates.

\section{Exact diagonalization methods} \label{section:ed}
Exactly diagonalizing a spin chain model is a challenging task because of the dimensionality of the matrix representing the Hamiltonian of the model. For a spin-$1/2$ model, like PXP, where each site has 2 possible configurations, up or down, the Hilbert space dimension grows as $\mathcal{D} = 2^L$, which means that for $L=30$ we have $ \mathcal{D} = 1,073,741,824 $.

\par However, for the PXP model, because of the model being local and due to the symmetries that it has, the matrix representing the Hamiltonian of the system is sparse and block-diagonal. The latter means that the Hamiltonian is broken up into pieces that are disconnected, with each block having fixed quantum numbers relating to the symmetries of the model. For example, for inversion symmetry, there exists an eigenbasis where the Hamiltonian $H$ of the model splits into a symmetric ($+$) block and an antisymmetric ($-$) block 
$$
H=\left[
\begin{array}{c|c}
+ &  \\ \hline
 & -
\end{array}\right].
$$
Now, instead of diagonalizing the full $H$ we can focus on the $+$ block, for example, and diagonalize this instead. This reduces the dimension by half. We can do this for the other symmetries as well. The PXP model, Eq. (\ref{pxp}), with PBC, has translational symmetry and we can resolve this symmetry by further restricting the Hilbert space to the zero-momentum ($0$) sector. And finally, we can insist that there are no adjacent Rydberg excitations in our system. All of the above, reduce the Hilbert space dimension to, as we saw, $\mathcal{D} = F_{L-1} + F_{L+1}$. Therefore, for $L=30$, the dimension is reduced down to $1,860,498$.

\par
\begin{figure}
\centering
\includegraphics[width=\textwidth]{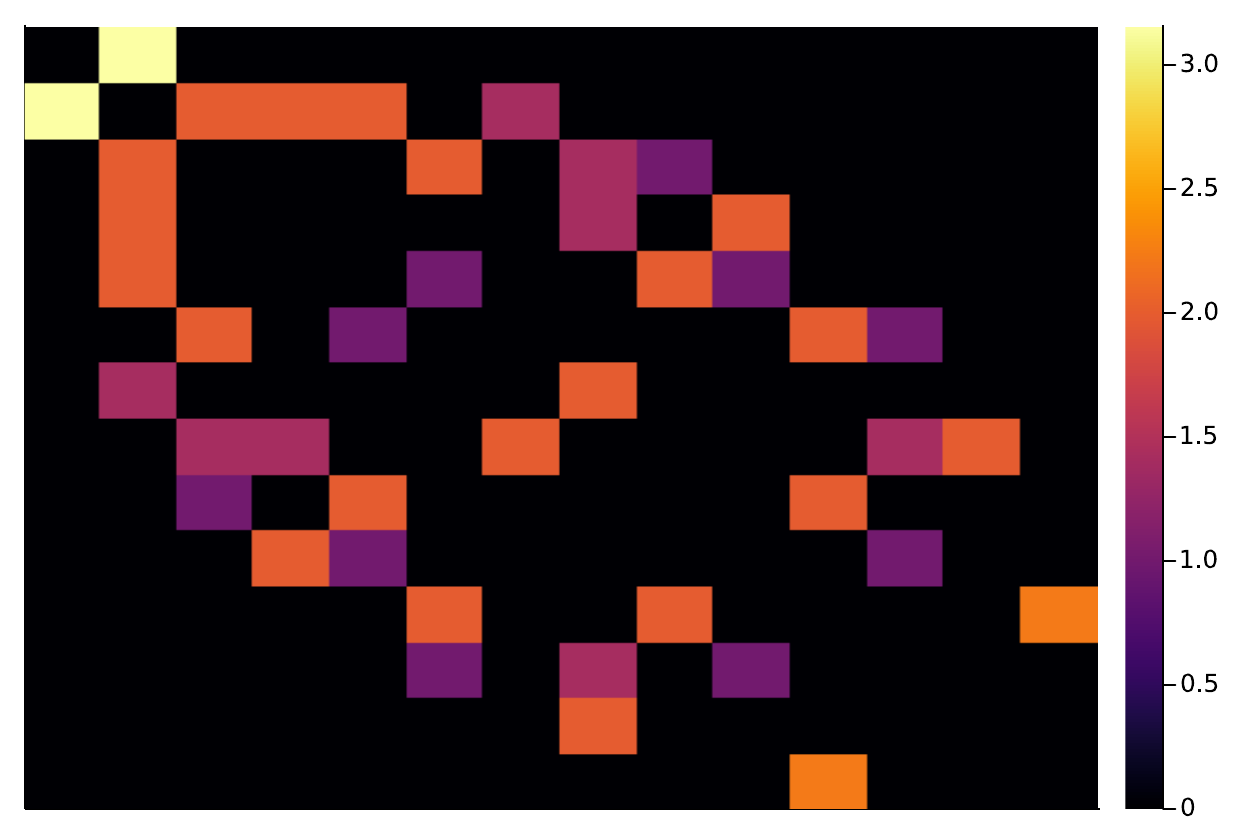}
\caption[PXP Hamiltonian for $L=10$]{Heatmap of the values of the matrix that represents the PXP Hamiltonian for $L=10$, in the zero-momentum, inversion-symmetric, no adjacent excitations sector. The dimension of the reduced Hilbert space in this case is $ \mathcal{D} = 14$, and therefore the Hamiltonian is a matrix of dimension $14 \times 14$.}
\label{fig:pxp}
\end{figure}

 We implement \cite{Sandvik} to do the above restrictions to the zero-momentum ($0$), inversion symmetric ($+$), and with no adjacent excitations sector. The code, written in \textsf{Julia}, can be found in the Appendix. We verified that the Hamiltonian is correct by starting with the full PXP Hamiltonian, projecting to the lower subspaces, and then comparing the two sets of eigenvalues. In Fig. \ref{fig:pxp}, we see a heatmap of the PXP Hamiltonian for $L=10$. The dimension of the reduced Hilbert space, i.e. zero-momentum, inversion symmetric, no adjacent excitations sector, in this case is $ \mathcal{D} = 14$, which is significantly smaller than the full Hamiltonian which would have dimension $\mathcal{D} = 2^{10} = 1,024.$

\section{Breakdown of ETH in special eigenstates}

Thermalization in ergodic systems is explained by the powerful conjecture regarding the nature of eigenstates: the eigenstate thermalization hypothesis (ETH) \cite{Srednicki:1994uz,Deutsch:1991uc,Rigol:2008vq}. The ETH states that in ergodic systems, the individual excited eigenstates have thermal expectation values of physical observables, which are identical to those obtained using the microcanonical and Gibbs ensembles. The expectation value of a physical observable associated with an operator $O$ is given by the diagonal matrix element $ O_{\alpha \alpha} = \bra{\alpha} O \ket{\alpha} $, where $\ket{\alpha}$ is an eigenstate of $H$, $ H \ket{\alpha} = E_{\alpha} \ket{\alpha}$. Furthermore, ETH can be formulated as an ansatz for the matrix elements of observables in the basis of the eigenstates of the Hamiltonian \cite{Srednicki:1999vu,Srednicki:1996vb,DAlessio:2016tw}: 
\begin{equation} \label{eth}
O_{mn} = O(\bar{E}) \delta_{mn} + e^{-S(\bar{E})/2} f_{O}(\bar{E},\omega) R_{mn},
\end{equation}
where $\bar{E} \equiv (E_m + E_n)/2$, $\omega \equiv E_n - E_m$, and $S(E)$ is the thermodynamic entropy at energy $E$. Crucially, $O(\bar{E})$ and $f_{O}(\bar{E},\omega)$ are smooth functions of their arguments, the value $O(\bar{E})$ is identical to the expectation value of the microcanonical ensemble at energy $E$ and $R_{nm}$ is a random real or complex variable with zero mean and unit variance ($\overline{R^2_{mn}} = 1$ or $\overline{|R_{mn}|^2} = 1$, respectively). While there is no rigorous understanding of which observables satisfy the ETH and which do not, it is generally expected that Eq. (\ref{eth}) holds for all physical observables. We note that it has been verified in several low-dimensional models \cite{Steinigeweg:2013ti,Khatami:2013tb,Beugeling:2015vx}, while it was found to break down in many-body localized systems \cite{Serbyn:2015wu,Serbyn:2017wn}.

\begin{figure} 

\includegraphics[width=\textwidth]{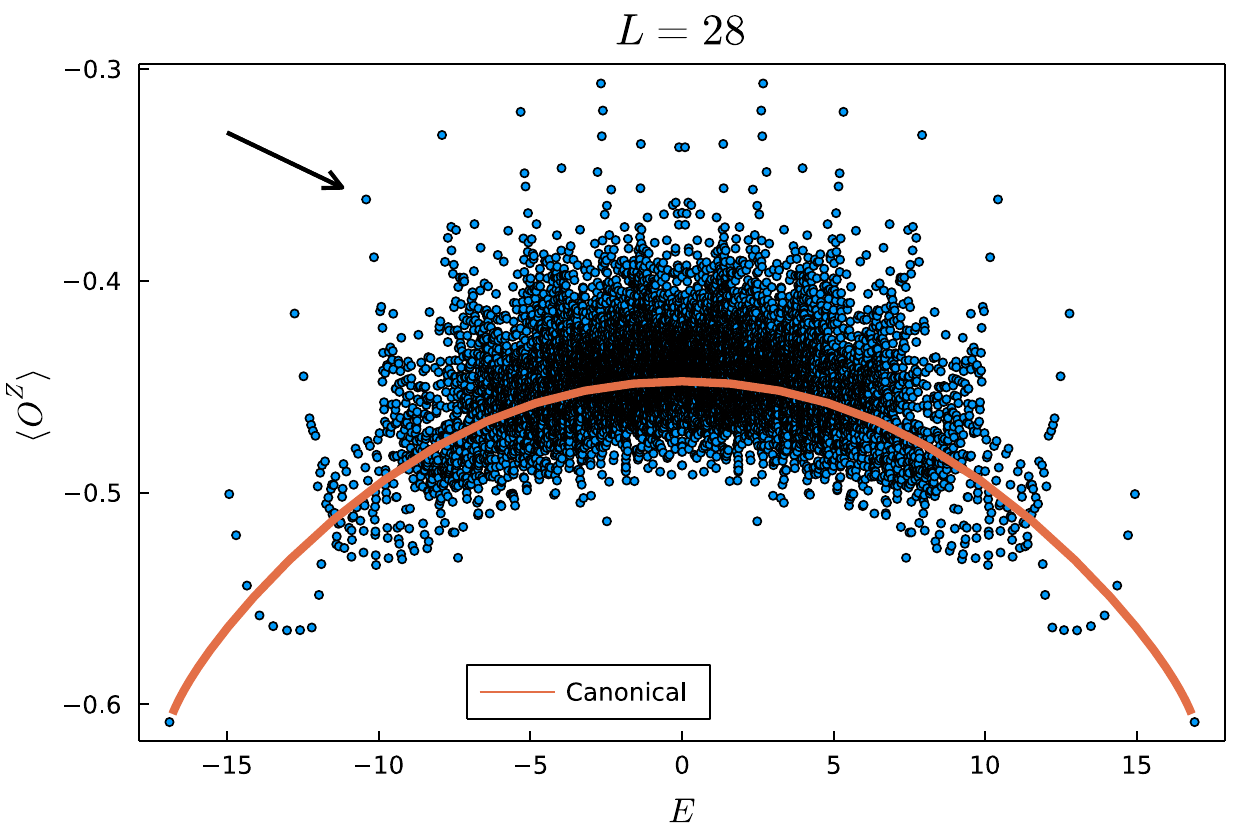}
\caption[Strong violation of the ETH]{Strong violation of the ETH revealed by the eigenstate expectation values $ \langle O^Z \rangle$, with $ O^Z = (1/L) \sum_{j=1}^L Z_j $, plotted as a function of energy. While the majority of the points are concentrated around the canonical ensemble prediction, the band of special eigenstates is also clearly visible. For these eigenstates, $ \langle O^Z \rangle $ strongly deviates from the canonical prediction at the corresponding energy, see Section \ref{sec:canonical} for the calculation of the canonical curve. The system contains $L=28$ atoms in the zero-momentum, inversion-symmetric, no adjacent excitations sector. We see a total of 15, $L/2 + 1$ for $L=28$, special eigenstates (one of them shown with an arrow), at the top of the tower-like structures. }
\label{fig:eev}

\end{figure}

In Fig. \ref{fig:eev} we test the ansatz (\ref{eth}) for the diagonal matrix elements, also known as eigenstate expectation values (EEVs), of the local operator $O^Z = (1/L) \sum_{j=1}^L Z_j $ in the PXP model in Eq. (\ref{pxp}). Fig. \ref{fig:eev} shows that most of the expectation values are close to the canonical prediction $ O(\bar{E}) $, which is calculated from the Gibbs states defined by the density matrix $\rho \propto \exp(-\beta H)$. The value of $\beta \in (-\infty, +\infty)$ is extracted by relating the observable expectation value to the mean energy in the Gibbs ensemble. However, in Fig. \ref{fig:eev}, we can also see that are some states that clearly violate the ETH. For OBC, the number of these states is $L+1$, while for PBC, the number is $L/2 + 1$ in the zero-momentum sector, and $L/2$ in the $\pi$-momentum sector, resulting in the same total count \cite{Turner:2018to}. These special states, can be viewed as parent states that define the ETH-breaking ``towers'' visible in the figure.

\subsection{Canonical curve} \label{sec:canonical}

In order to calculate the thermal curve, we start with the construction of the reduced PXP Hamiltonian (Section \ref{section:ed}) and with a similar construction for the $O^Z$ operator in the eigenstate basis of the Hamiltonian. Then, for various values of $\beta$ we calculate the density matrix $\rho = \frac{1}{\mathcal{Z}} e^{-\beta H} $ for the system. In the Hamiltonian eigenstate basis, this operator simplifies to a diagonal operator with diagonal elements $$ \rho_{ii} = \frac{1}{\mathcal{Z}} e^{-\beta E_i},$$ 
with $\mathcal{Z} = \sum_i e^{-\beta E_i}$, where the sum is over the eigenenergies of the Hamiltonian $\{E_j\}_j$. The expectation value of an operator $A$, in a mixed state described by the density operator $\rho$ can be calculated using the trace $$ \langle A \rangle = \Tr (\rho A) .$$ Therfore, in order to calculate the expectation values of the energy, $\langle E \rangle$, and of the observable, $ \langle O^Z \rangle $, we use
$$ \langle E \rangle = \Tr (\rho\ V^T H V ), $$
$$ \langle O^Z \rangle = \Tr (\rho\ V^T O^Z V ), $$
where we have used $ V^T \bullet V $, with $V$ the matrix of the eigenvectors of $H$, in order to express the operator in the eigenstate basis of $H$.

\section{Scaling of the standard deviation of \texorpdfstring{$O^Z$}{O Z} EEVs}
Then, we looked at how the standard deviation of the EEVs of the $O^Z = (1/L) \sum_{j=1}^L Z_j $ operator scales as a function of the dimension of the Hilbert space. We do this in the following way. First, we take the points in Fig. \ref{fig:eev}, and we split the energy $E$ in intervals or ``windows''. For each window, we calculate the mean, and subtract it from all the points in the window, such that the distribution is centered around zero for all values of $E$. Finally, we calculate the standard deviation $\sigma$ of this distribution, which is essentially the ``width'' of the EEV curve, in Fig. \ref{fig:eev}. For a chaotic system, due to the ETH, we expect this width to scale as $$ \sigma  \sim \mathcal{D}^{-1/2}, $$
with $\sigma \rightarrow 0$ in the thermodynamic limit $\mathcal{D} \rightarrow \infty$.

\begin{figure}
\centering
\includegraphics[width=\textwidth]{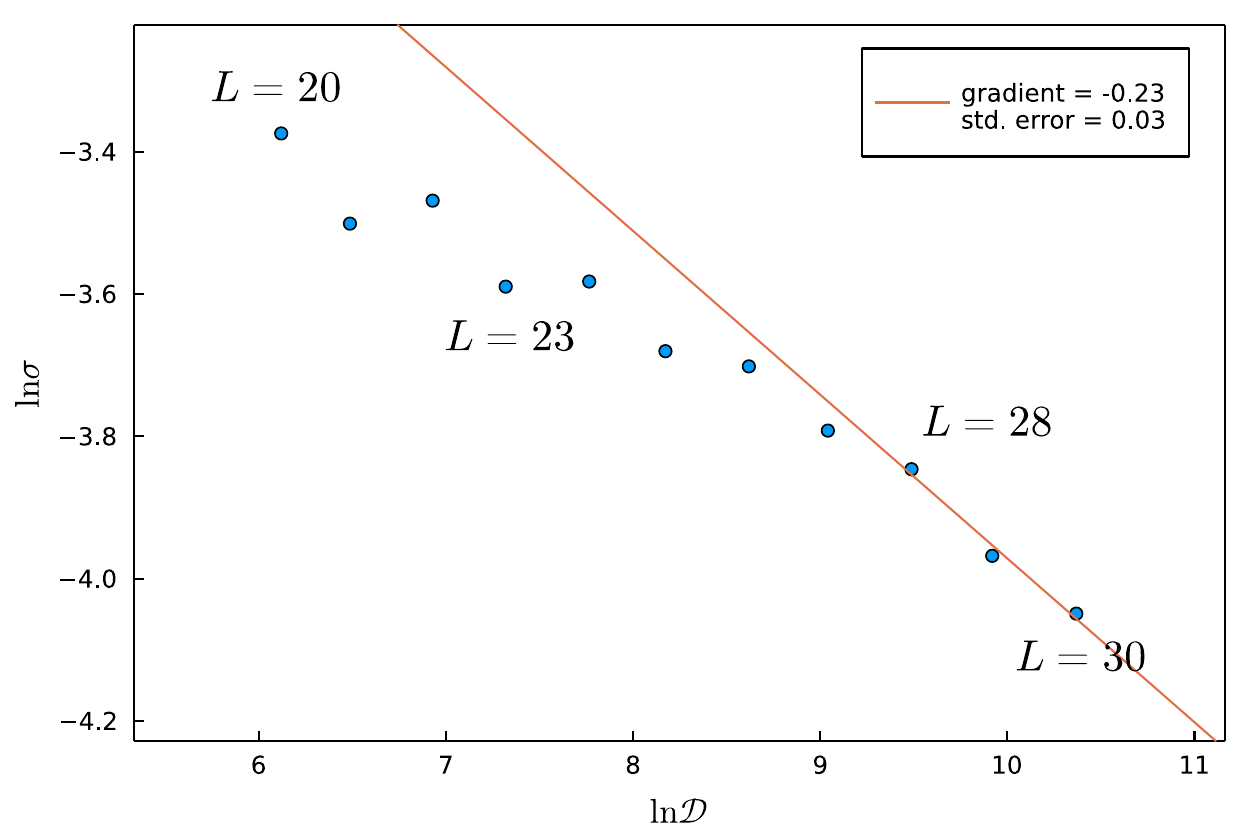}
\caption[Standard deviation $\sigma$ of the EEVs of the $O^Z$ operator]{Standard deviation $\sigma$ of the EEVs of the $O^Z = (1/L) \sum_{j=1}^L Z_j $ operator for various system sizes $L$, as a function of the dimension of the Hilbert space $\mathcal{D}$ in the zero-momentum, inversion symmetric, no adjacent excitations sector. The regression is performed only on the last 3 points.}
\label{reg:20-30}
\end{figure}

For the PXP model, in Fig. \ref{reg:20-30}, we plot the standard deviation $\sigma$ for various system sizes $L$ as a function of the Hilbert space dimension $\mathcal{D}$. We perform a regression only on the last 3 points, because the previous ones do not seem to fall on a straight line. We have $$ \sigma \sim \mathcal{D}^{-0.23},$$ for the system sizes we have been able to reach. In \cite{Turner:2018to}, the corresponding exponent is $-0.34$. The discrepancy is probably due to the fact that we were limited up to sizes $L=30$. However, the important observation that we verified is that this scaling is indeed slower than expected for a chaotic system.

\section{Overlap of special eigenstates with product states} \label{sec:z2}

The PXP model breaks the ETH because of the existence of a relatively small number of highly atypical, nonthermal eigenstates. These states are distinguished by anomalous matrix elements of local observables (Fig. \ref{fig:eev}) as well as subthermal entanglement entropy (see \cite{Turner:2018to}, section III. B.). However, since we have $L+1$ such states among the exponentially many thermalizing eigenstates, one might expect that these states do not have direct physical relevance, as they might be hidden by the large number of typical eigenstates. Below we show that this is not the case because these special eigenstates have anomalously high overlaps with certain product states. This implies that superpositions of special eigenstates can be experimentally prepared and probed using a global quench. For example, a class of product states which was studied in recent experiments \cite{Bernien:2017to} are the charge-density-wave (CDW) states $$ \ket{\mathbb{Z}_k} = \ket{\cdots \underbrace{ \bullet \circ \cdots \circ}_\text{k} \bullet \cdots } $$ where the atoms in the excited state are separated by $k-1$ atoms in the ground state. In this section we focus on the period-2 CDW ($\mathbb{Z}_2$ or N\'eel) state.

\begin{figure} 

\includegraphics[width=\textwidth]{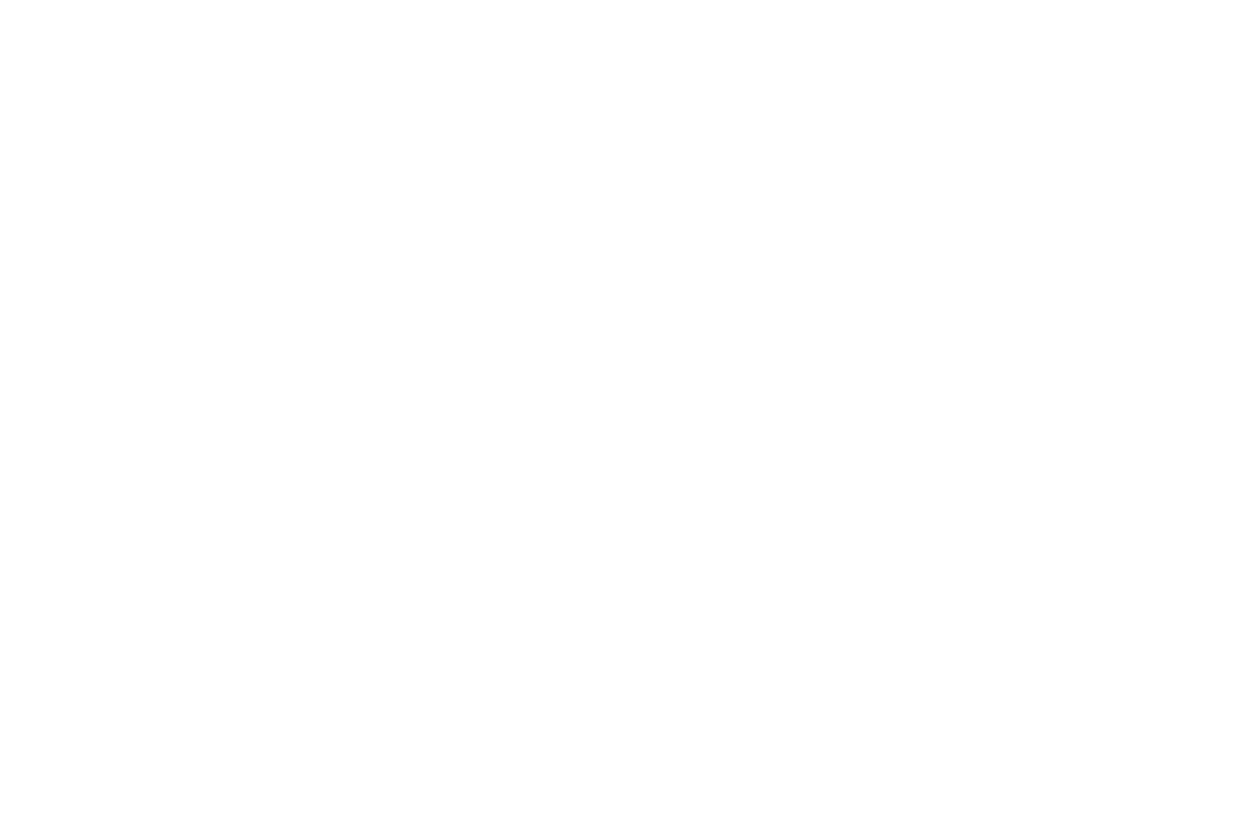}
\caption[Overlap of eigenstates with $\ket{\mathbb{Z}_2}$]{Overlap of all the different energy eigenstates $\ket{\psi}$ of the PXP Hamiltonian, with the $\ket{\mathbb{Z}_2}$ product state. Data are shown for $L=28$ sites in the zero-momentum, inversion symmetric, no adjacent excitations sector. The states with high overlap with the $\ket{\mathbb{Z}_2}$ state, lying on top of the ``towers'' observed, are identified with the states at the top of the ETH-breaking ``towers'' in Fig. \ref{fig:eev}, one of which is shown with an arrow.}
\label{fig:overlap}

\end{figure}

Fig. \ref{fig:overlap} shows the squared overlap between eigenstates in the PXP model and the $\ket{ \mathbb{Z}_2 }$ product state. From this plot, we see a subset of eigenstates that have a high overlap with the $\ket{\mathbb{Z}_2}$ product state. Looking at the $x$-axis, we can identify these special states with the ones that break ETH in Fig. \ref{fig:eev}, as well as the ones with subthermal entanglement entropy \cite{Turner:2018to}.

\chapter{Level statistics} \label{ch:level}
In this chapter, we reproduce results about the level-spacing statistics of the PXP model found in \cite{Turner:2018to}. The level-spacing statistics of a Hamiltonian is a diagnostic for whether the system is ergodic or integrable. Essentialy, we are studying the distribution of the differences between adjacent energy levels (eigenvalues of the Hamiltonian). Given the Hamiltonian of a finite-dimensional system, we sort the energies of the full spectrum in ascending order
$$ E_1 \leq E_2 \leq \cdots \ E_{\mathcal{D}-1} \leq E_{\mathcal{D}} ,$$
where $\mathcal{D}$ is the dimension of the Hilbert space, as well as the number of eigenvalues $ \{ E_i \}_{i=1}^{\mathcal{D}} $. Then, we calculate the $\mathcal{D}-1$ spacings 
$$ \delta_n = E_n - E_{n-1} ,$$
and divide them by the local mean level spacing, which is an average of sufficiently many spacings around the energy where the spacing is located at. This procedure is called the unfolding of the spectrum and yields normalized spacings 
$$ s_n = \frac{\delta_n}{\Delta E_n} ,$$
where the average is one ($\frac{1}{N} \sum_n s_n = 1 $). Then, one can study the histrograms of the spacings, i.e. $p(s)$. By normalizing using the local mean we have removed system size dependence, since otherwise $s_n \propto \frac{1}{N}$. For integrable systems, the spacing is highly degenerate, the eigenvalues are uncorrelated with each other and the normalized spacing follows Poisson statistics. On the other hand, for non-integrable systems, the spacing is not degenerate because of level repulsion, and the level statistics are described by random matrix theory (RMT) \cite{bgs}, and more specifically they follow the Wigner--Dyson distribution. The latter was conjectured in 1984 by Bohigas, Giannoni and Schmit and since then has been tested and confirmed in many different setups. To date, only non-generic counterexamples are known to violate this conjecture \cite{DAlessio:2016tw}. Therefore, the emergence of Wigner--Dyson statistics for the level-spacing is often considered as a defining property of quantum chaotic systems, whether these systems have a classical counterpart or not.

\begin{figure} 

\includegraphics[width=\textwidth]{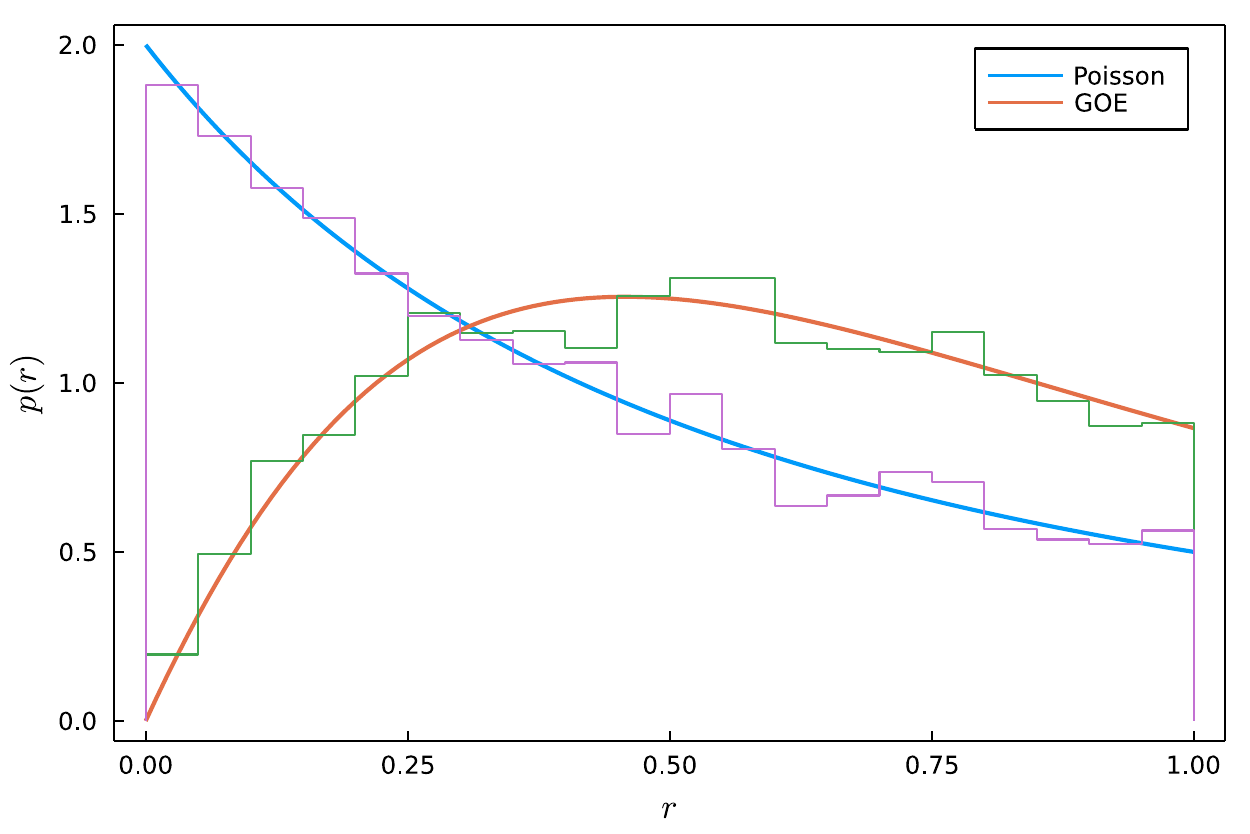}
\caption[Level spacing statistics, Poisson vs GOE]{Level spacing statistics, using the quantity $ r_n = \min\{\delta_n, \delta_{n+1}\} / {\max\{\delta_n, \delta_{n+1}\}}$ defined in \cite{OH2007}. We compare the theoretical prediction (solid curve), surmised in \cite{Atas:2013wy},  with the numerical calculation (histogram) for random matrices of size $\mathcal{D} = 6,000$. For the Poisson case, we generate a diagonal matrix with real random entries, while for the GOE case, we generate a random matrix $M$ of size $6,000 \times 6,000$ with real entries, and then symmetrize it using $(M + M^T)/2$.}
\label{fig:poisson+goe}

\end{figure}

Because the unfolding procedure can reduce the precision of the method \cite{Gomez:2002vh}, another method has proven more widely used. In order to define a better observable, which appears to be numerically the most accessible quantity that shows a clear, well-understood difference between the diffusive (chaotic) and non-diffusive phase (integrable), Oganesyan and Huse \cite{OH2007} proposed the quantity
$$ r_n \equiv \frac{\min\{\delta_n, \delta_{n+1}\}}{\max\{\delta_n, \delta_{n+1}\}} ,$$
where $ 0 \leq r_n \leq 1 .$
The average of the distribution of this quantity was shown \cite{Atas:2013wy} to be a stable and good predictor of quantum chaos. In Table \ref{tab:rstat}, we see the different values for the two cases. For an integrable system, we have Poisson statistics and an average of $ \langle r \rangle \approx 0.38629 $, while for a non-integrable system with time-reversal symmetry, the statistics are that of the Gaussian orthogonal ensemble (GOE) and $ \langle r \rangle \approx 0.53590 $. For the GOE, we define an ensemble of random matrices drawn from the Gaussian distribution \cite{DAlessio:2016tw}
$$ P(\hat{H}) \propto \exp{ \left[ - \frac{1}{2a^2} \Tr{\hat{H}^2} \right] }   ,$$
where $a$ sets the overall energy scale, and all entries in the Hamiltonian are real and satisfy $H_{ij} = H_{ji}$. For a description of the unitary and symplectic ensembles see \cite{DAlessio:2016tw}. The Poisson and the GOE curves are surmised, in \cite{Atas:2013wy}, to be
\begin{itemize}
\item[] $$ \text{Poisson: } p(r) = \frac{2}{(1+r)^2},$$
\item[] $$ \text{GOE: } p(r) = \frac{27}{4} \frac{r+r^2}{(1+r+r^2)^{5/2}}.$$
\end{itemize}
In Fig. \ref{fig:poisson+goe}, for the two cases, we compare the theoretical prediction (solid curve) with the numerical calculation (histogram) for random matrices of size $\mathcal{D} = 6,000$. For the Poisson case, we generate a diagonal matrix with real random entries, while for the GOE case, we generate a random matrix $M$ of size $6,000 \times 6,000$ with real entries, and then symmetrize it using $(M + M^T)/2$.

\begin{table}[]
\centering
\begin{tabular}{p{2.5cm} p{2.5cm} p{2.5cm} p{2.5cm} }
\hline
\hline
& ensemble & Poisson  & GOE \\
 \hline
observable &  &   &   \\
 \hline 
 $ \langle r \rangle $ & & $2 \ln 2 - 1$ & $4 - 2 \sqrt{3}$  \\
 & & $ \approx 0.38629 $  & $ \approx 0.53590$  \\
 \hline
 \hline
\end{tabular}
\caption[Values of the average $\langle r \rangle$ for Poisson and GOE]{Values of the average $\langle r \rangle$ for Poisson and GOE statistics, according to \cite{Atas:2013wy}.}
\label{tab:rstat}
\end{table}

\section{PXP}
\begin{figure} 

\includegraphics[width=\textwidth]{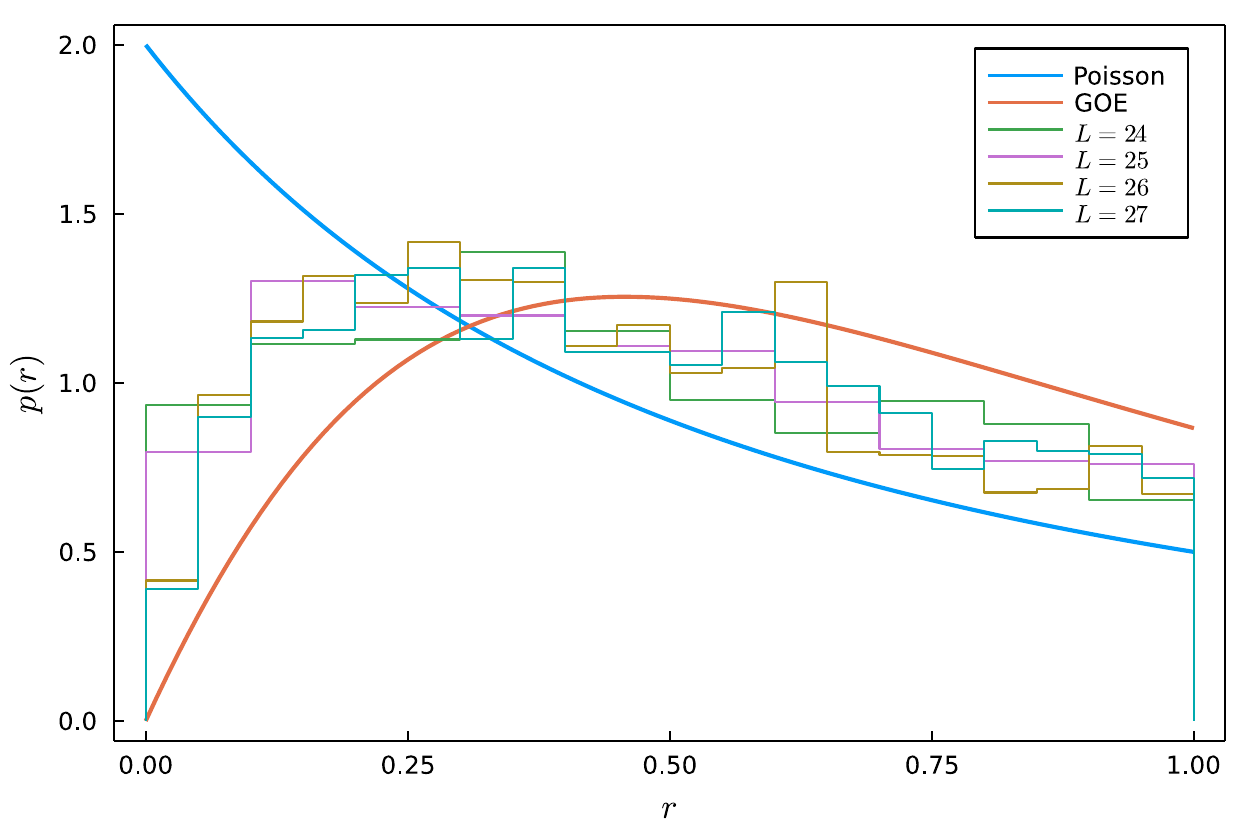}
\caption[Level spacing statistics, PXP]{Level spacing statistics, using the quantity $ r_n = \min\{\delta_n, \delta_{n+1}\} / {\max\{\delta_n, \delta_{n+1}\}}$ defined in \cite{OH2007}, for the PXP model, for various system sizes $L$. We see that even for relatively small system size, there is pronounced level repulsion and that the distribution is neither Poisson nor GOE. In fact, it was demonstrated in \cite{Turner:2018ts} that this distribution is close to the semi-Poisson distribution \cite{Bogomolny:1999vp}, characterized by level repulsion and an exponential tail. This is in sharp contrast with integrable systems, which always have Poisson level statistics. Moreover, when we increase the system size, we see that the statistics approach more and more the Wigner--Dyson distribution. The Wigner–Dyson (GOE) level statistics, along with ballistic growth of entanglement \cite{Turner:2018to}, rule out the integrability-based explanation of the non-ergodic dynamics in the model in \ref{pxp}.}
\label{fig:levels-PXP}

\end{figure}

A possible explanation for the observed ergodicity breaking and the special states of the PXP model could be some approximate integrability. To investigate this possibility we can study the level statistics of the model. Fig. \ref{fig:levels-PXP} reveals that even for relatively small system size, there is pronounced level repulsion and that the distribution is neither Poisson nor GOE. In fact, it was demonstrated in \cite{Turner:2018ts} that, for system sizes $L \leq 28,$ this distribution is close to the semi-Poisson distribution \cite{Bogomolny:1999vp}, characterized by level repulsion and an exponential tail. This is in sharp contrast with integrable systems, which always have Poisson level statistics. Moreover, when we increase the system size, we see that the statistics approach more and more the Wigner--Dyson distribution. The Wigner–Dyson (GOE) level statistics, along with ballistic growth of entanglement \cite{Turner:2018to}, rule out the integrability-based explanation of the non-ergodic dynamics in the model in \ref{pxp}.

\chapter{Eigenvector component statistics} \label{ch:eigvec}
Another diagnostic, which has not yet been studied for the PXP model, for whether a quantum system is integrable or not is the statistics of the eigenvector components of its Hamiltonian. Berry, in 1977 \cite{Berry:1977vx}, conjectured that the coefficients of high energy eigenstates of a quantum system in a generic basis corresponding to a chaotic classical system are independent Gaussian random variables, similar to the distribution of eigenstates in the corresponding random matrix ensemble \cite{rmt}. The connection between random matrix theory and realistic systems was made later, in \cite{Deutsch:1991uc}, showing that perturbing a Hamiltonian with a random matrix leads to thermalization. Later, Srednicki showed that if Berry's conjecture is satisfied, a gas of hard core particles has a distribution of velocities that approaches the Maxwell-Boltzmann distribution for large systems. Therefore, it was concluded that Berry's conjecture is a requirement for thermalization in quantum systems \cite{Srednicki:1994uz}. It was this intuition that motivated Srednicki to propose the ETH ansatz \cite{Srednicki:1996vb}. The $R_{\alpha \beta}$ term in the ansatz in Eq. (\ref{eth}) is justified through Berry's conjecture.
\section{PXP}
\begin{figure} 
     \centering
     \begin{subfigure}[t]{.8\textwidth}
         \centering
         \captionsetup{justification=raggedright, singlelinecheck=false, font=large, labelfont=bf}
         \caption{} \label{fig:eigvecPXP}
         \includegraphics[width=\textwidth]{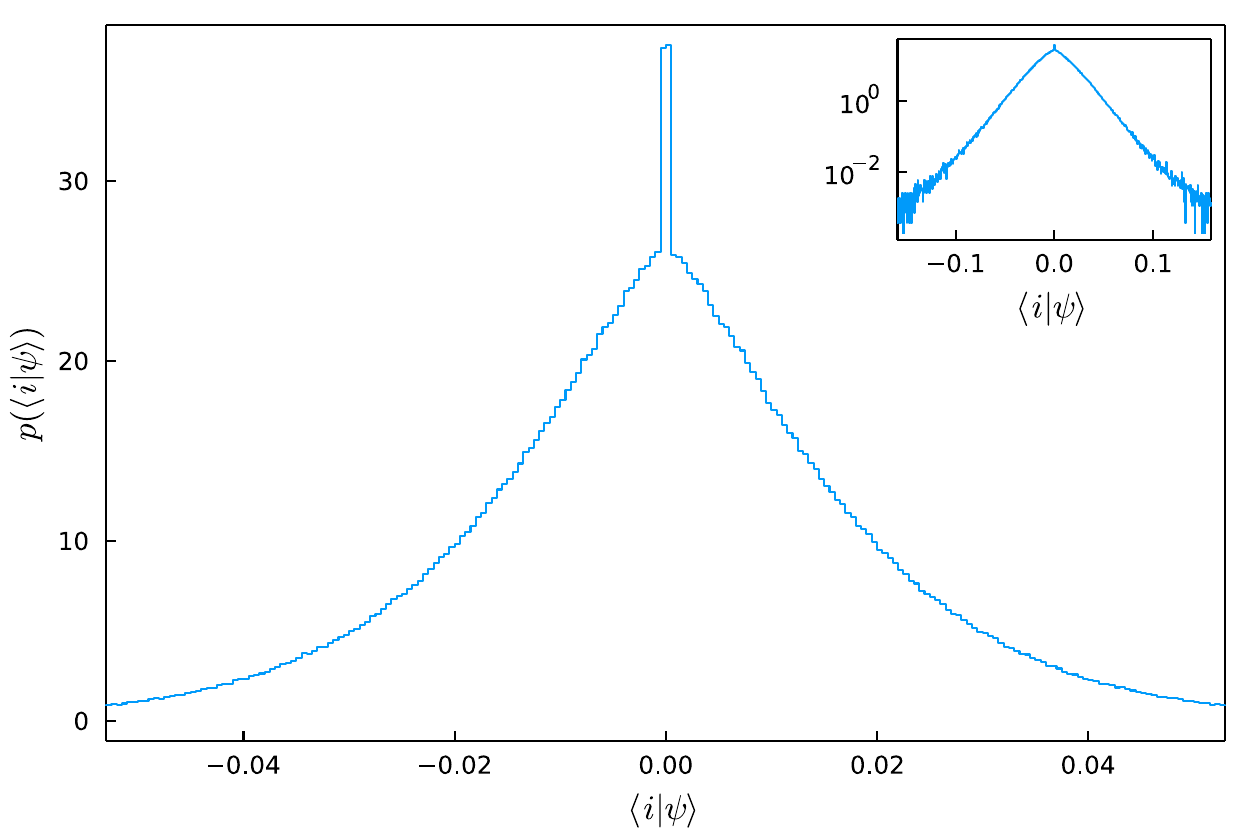}
     \end{subfigure}     
     \vfill
     \begin{subfigure}[t]{.8\textwidth}
         \centering
         \captionsetup{justification=raggedright, singlelinecheck=false, font=large, labelfont=bf}
         \caption{} \label{fig:eigvecGOE}
         \includegraphics[width=\textwidth]{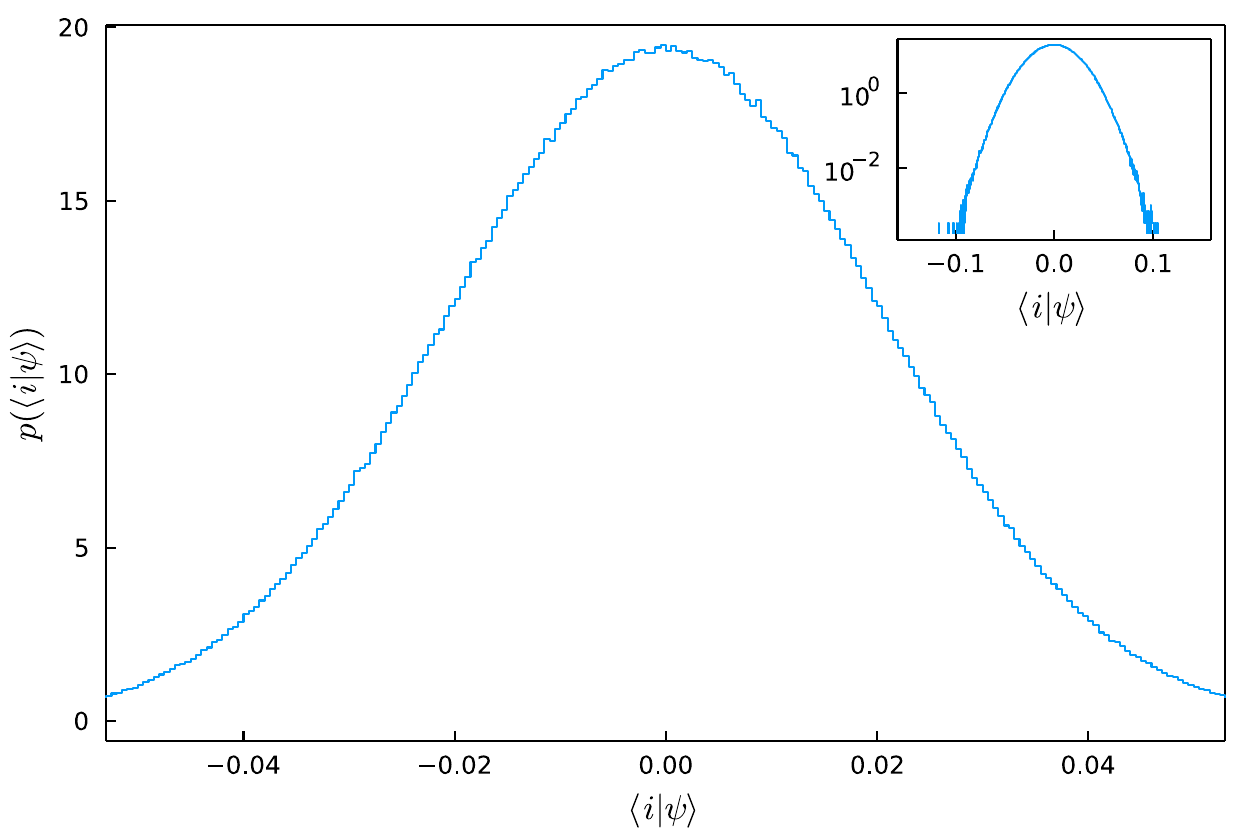}
     \end{subfigure}
     \caption[Eigenvector component statistics, PXP vs GOE]{(a) Eigenvector component statistics for the PXP Hamiltonian, for $L=24$. The dimension of the matrix in the zero-momentum, inversion symmetric, no adjacent excitations sector is $ \mathcal{D} = 2,359$. The inset is the same plot but with a logarithmic $y$-axis. The distribution is clearly not Gaussian, and we can also see a spike around $ \langle i | \psi \rangle = 0 $ present due to zero modes. (b) Eigenvector component statistics for a random (GOE) matrix of dimension $ \mathcal{D} = 2,359 $. The inset is the same plot but with a logarithmic $y$-axis. The distribution is clearly Gaussian.}
\label{eig_comp}
\end{figure}

To directly test Berry's conjecture for the PXP model, we calculate the distribution of the coefficients (components) $ \langle i | \psi \rangle $ of the eigenstates $ \ket{ \psi } $ in the spin basis $ \ket{ i } $ in Fig. \ref{eig_comp}. Surprisingly, the conjecture is clearly violated. The fact that Berry's conjecture is not satisfied does not necessarily imply that the ETH cannot be satisfied, even with non-Guassian fluctuations \cite{Luitz:2016tk}. However, the absence of a Thouless plateau in the off-diagonal matrix elements, along with the slow decay of fluctuations in diagonal matrix elements $ \overline{\Delta O^Z} $ and deviations from purely Wigner--Dyson level statistics, suggests that thermalization of the bulk of eigenstates in the PXP model may not follow the ETH \cite{Turner:2018to}.

\begin{figure}
\centering
\includegraphics[width=\textwidth]{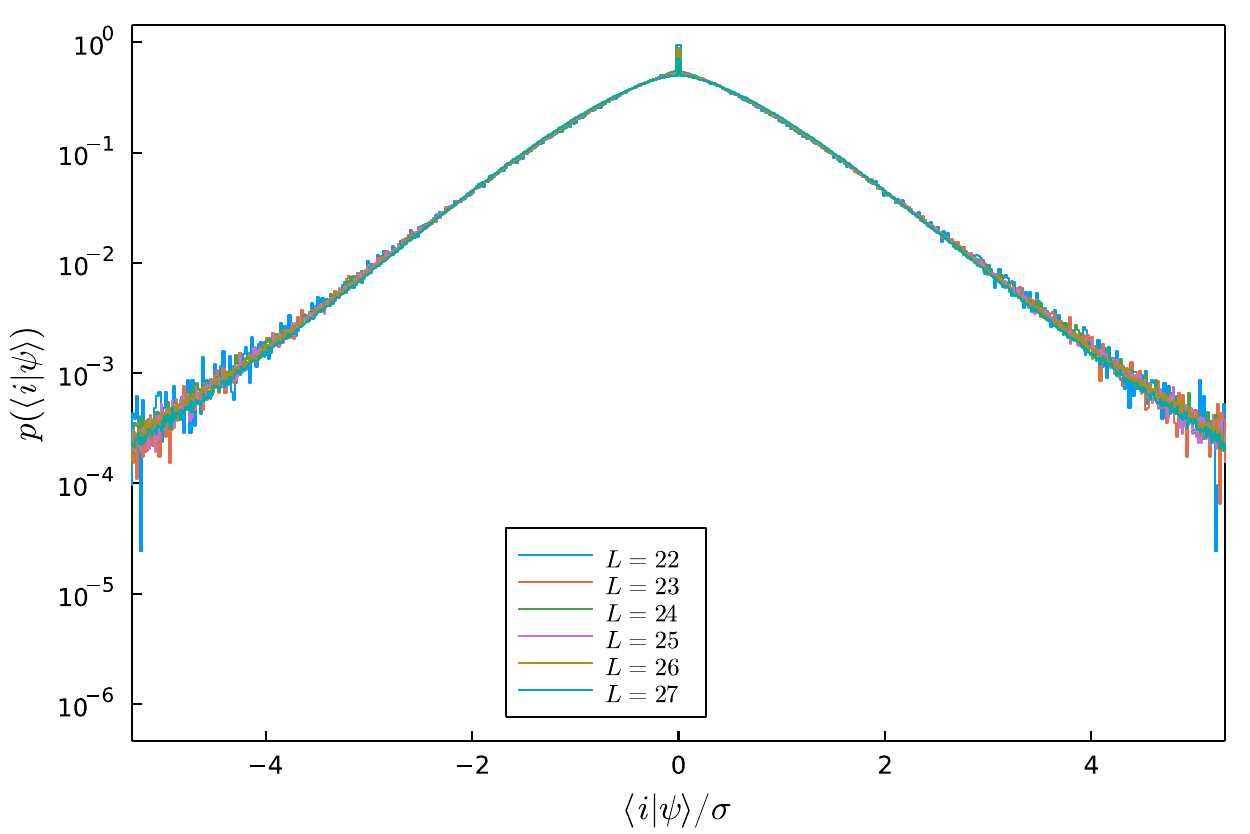}
\caption[Eigenvector component statistics, PXP]{Eigenvector component statistics for the PXP Hamiltonian, for various system sizes $L$, in the zero-momentum, inversion symmetric, no adjacent excitations sector.}
\label{fig:eigvec-Ls}
\end{figure}

Moreover, in Fig. \ref{fig:eigvec-Ls} we check if the distribution of the eigenvector components changes as we increase the system size $L$. We plot different values of $L$ superimposed, after normalizing by diving the eigenvector components with the standard deviation of the distribution. For $L \leq 28$ the distribution looks like it does not change shape. However, it might be that for $L>28$ that is no longer true. One possible scenario is that the distribution tends to a Gaussian for values $L>30$, but it remains to be checked.

\section{Zero-energy states}
A prominent feature in Fig. \ref{fig:eigvecPXP} is the peak on top of the non-Gaussian distribution of eigenvector components. This peak is caused by a large number of states annihilated by the PXP Hamiltonian, $ H \ket{\psi} = 0 $, which form a degenerate subspace of zero modes. The number of zero modes is a Fibonacci number, which means that this number grows exponentially with system size. For OBC and even system size $L$ this degeneracy is $ Z = F_{\frac{L}{2}+1} $. When instead $L$ is odd, we have $ Z = F_{\frac{L-1}{2}} $ \cite{Turner:2018ts,Turner:2018to}.

\begin{figure} 
     \centering
     \begin{subfigure}[t]{.75\textwidth}
         \centering
         \captionsetup{justification=raggedright, singlelinecheck=false, font=large, labelfont=bf}
         \caption{} \label{fig:bulk+scars}
         \includegraphics[width=\textwidth]{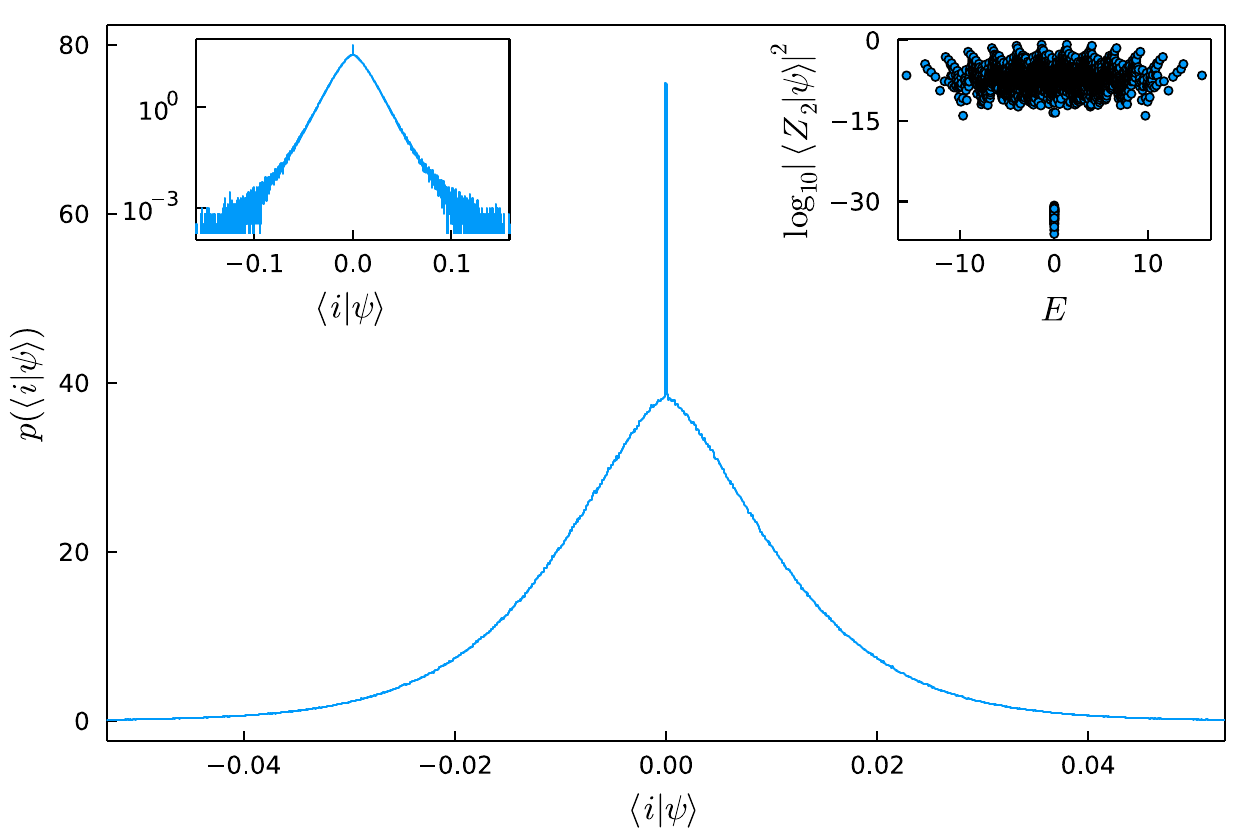}
     \end{subfigure}     
     \vfill
     \begin{subfigure}[t]{.75\textwidth}
         \centering
         \captionsetup{justification=raggedright, singlelinecheck=false, font=large, labelfont=bf}
         \caption{} \label{fig:bulk}
         \includegraphics[width=\textwidth]{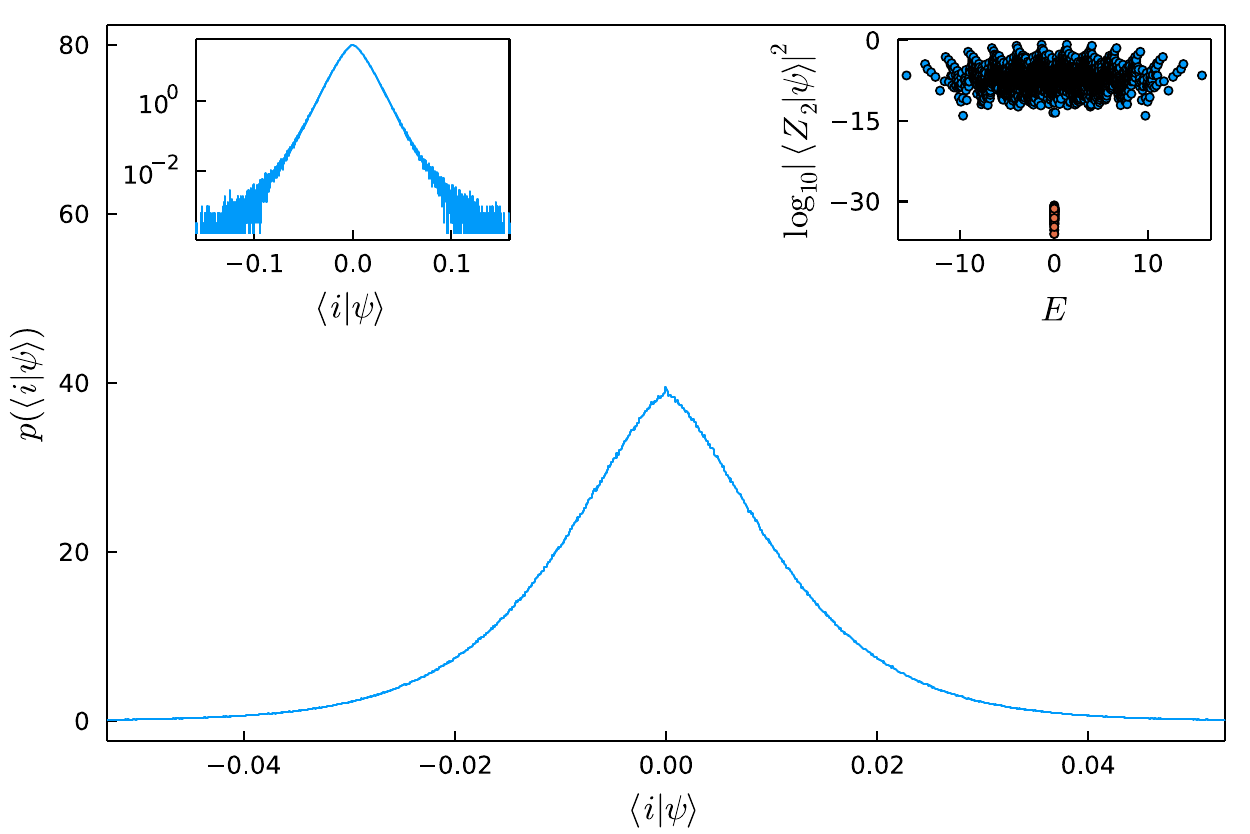}
     \end{subfigure}
     \caption[Eigenvector component statistics, excluding zero-energy states]{(a) Eigenvector component statistics for the PXP Hamiltonian, including the whole spectrum. The system contains $L=26$ atoms in the zero-momentum, inversion-symmetric, no adjacent excitations sector. We can clearly see a spike around $ \langle i | \psi \rangle = 0 $ present. The right inset shows the overlap with the $\ket{\mathbb{Z}_2}$ product state among the energy eigenstates. The isolated states with extremely negative overlap are the zero-energy states. The left inset is the same plot but with a logarithmic $y$-axis. (b) Eigenvector component statistics for the PXP Hamiltonian, excluding the zero-energy states, again for $L=26$. The right inset shows the states that were excluded with red. These are the states that have an extremely low overlap with $\ket{\mathbb{Z}_2}$ and have energy $E=0$. The spike around $ \langle i | \psi \rangle = 0 $ has disappeared, implying that the spike is due to the zero modes in the spectrum of the PXP Hamiltonian. The left inset is the same plot but with a logarithmic $y$-axis.}
\label{bulk+scars}
\end{figure}

In Fig. \ref{bulk+scars} we demonstrate the above statement, that the peak is due to the zero-modes. We again calculate the distribution of eigenvector components for the PXP, but now excluding the zero-energy states. We identify these states using the $\ket{\mathbb{Z}_2}$ overlap we saw in Section \ref{sec:z2}.

\section{Scaling of the standard deviation}
\begin{figure} 
\includegraphics[width=\textwidth]{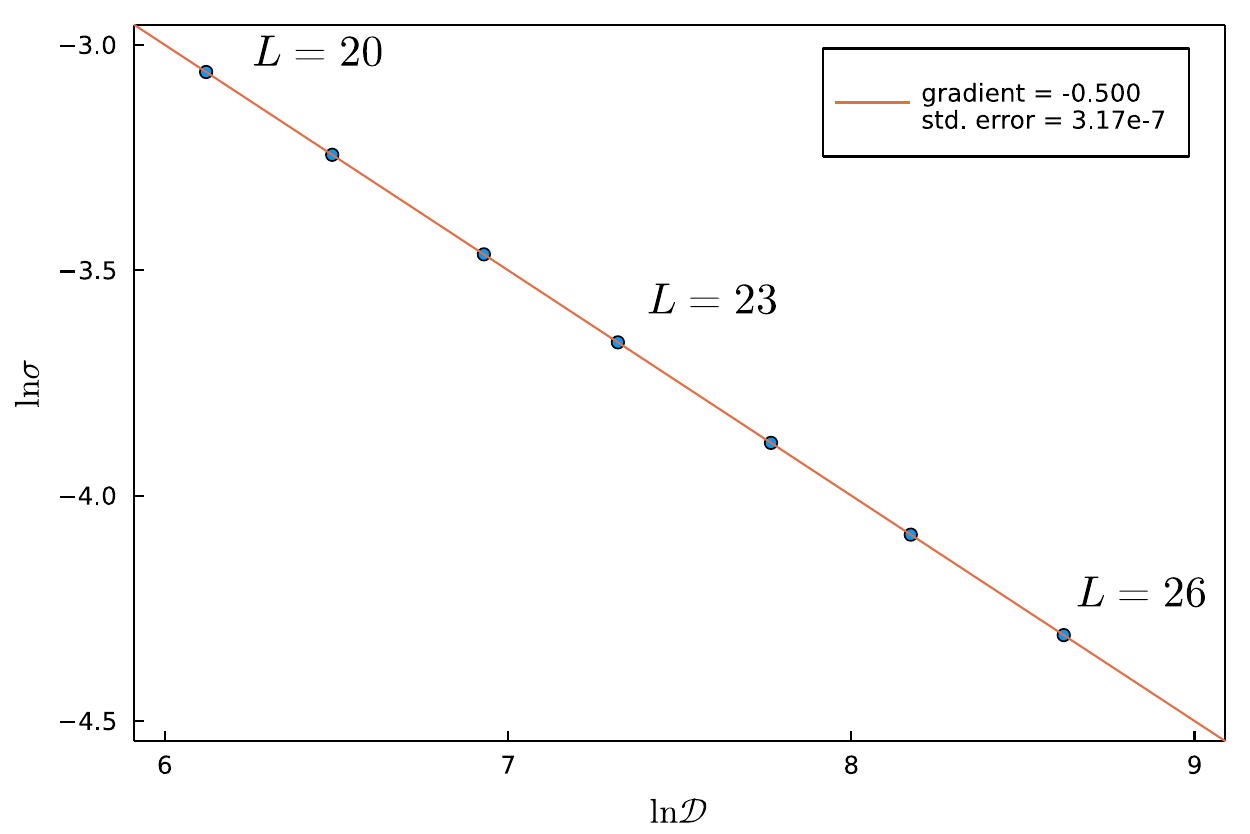}
\caption[Scaling of the standard deviation $\sigma$ of the eigenvector components]{Scaling of the standard deviation $\sigma$ of the eigenvector components of the PXP, with the Hilbert space dimension $\mathcal{D}$. System sizes between $L=20$ and $L=26$, in the zero-momentum, inversion-symmetric, no adjacent excitations sector, were used. The regression clearly verifies the ETH expectation for the scaling, Eq. (\ref{scaling}).}
\label{fig:sigma-regression}

\end{figure}

We then looked at the scaling of the standard deviation $\sigma$ of the eigenvector components with increasing the system size $L$ of the PXP model. The second term in the ETH ansatz, Eq. (\ref{eth}), is $ e^{-S(\bar{E})/2} f_{O}(\bar{E},\omega) R_{mn} $. Since the entropy scales with the logarithm of the system size $$ S(\bar{E}) \sim ln \mathcal{D} ,$$ we have that $$ e^{-S(\bar{E})/2} \sim \frac{1}{\sqrt{\mathcal{D}}}. $$ Therefore, for $\sigma $ we expect a scaling
\begin{equation} \label{scaling}
\sigma \sim \mathcal{D}^{-1/2},
\end{equation}
where $\mathcal{D}$ is the dimension of the Hilbert space. In Fig. \ref{fig:sigma-regression} we verify Eq. (\ref{scaling}). We exactly diagonalize systems with sizes between $L=20$ and $L=26$, again in the zero-momentum, inversion-symmetric, no adjacent excitations sector. For each system size $L$,  we pool the eigenvector components from all $\mathcal{D}$ eigenvectors and calculate the standard deviation $\sigma$. We plot the $\mathcal{D}$ vs $\sigma$ on a log-log plot and finally perform a linear regression. The results verify the expectation from the ETH ansatz.

\chapter{Transport of energy density fronts} \label{ch:transport}

Another new idea we had was to study how energy spreads in the PXP model, Eq. (\ref{pxp}). Because energy is a conserved quantity in our model, how this spreading occurs can indicate whether the system is diffusive or non-diffusive, and hence integrable or not. For an integrable system we expect ballistic (linear) fronts, while for a non-integrable system we expect diffusive (nonlinear) fronts. 

\section{Thermal states} \label{sec:thermal}

\par In order to see these fronts, we are going to prepare the system in a state that will be split in half, with each half having a different temperature, then evolve the state using the full Hamiltonian. This procedure is known as a quench. We can do this in the following way. Suppose that the system is of size $L$, an even number, with Hamiltonian $H$, and let us split the system in two smaller systems, each of size $L_{1/2} = \frac{L}{2}$ and Hamiltonian $H_{1/2}$. We can then generate a random initial state in the Hilbert space $\mathcal{H}_{1/2}$ of the $L_{1/2}$ system, $\ket{\psi_0}$, and then make it thermal with 
\begin{equation}\label{th:left}
\ket{\psi_{\text{left}}} = \frac{1}{\sqrt{A}}\ e^{-\beta_{\text{left}} H_{1/2} } \ket{\psi_0} ,
\end{equation}
where $A$ is a normalization constant, and similarly for the right half
\begin{equation}\label{th:right}
\ket{\psi_{\text{right}}} = \frac{1}{\sqrt{A'}}\ e^{-\beta_{\text{right}} H_{1/2} } \ket{\psi_0'} ,
\end{equation}
for another random initial state $\ket{\psi_0'}$, and where $ \beta_{\text{left}}/ \beta_{\text{right}} $ are the thermodynamic betas $\beta = \frac{1}{k_B T}$ for the left and the right halves of the system. Taking the tensor product (Kronecker product) of these two states, we create the state that we wanted, one where the left half has temperature $1/\beta_{\text{left}}$ and the right half has temperature $1/\beta_{\text{right}}$. However, we have to take into account the fact that the Hilbert space $\mathcal{H} $ for the $L$ system is not exactly equal to $\mathcal{H}_{1/2} \otimes \mathcal{H}_{1/2}$, because with OBC, some states that are allowed in the $\mathcal{H}_{1/2}$ are not necessarily allowed when we take their tensor product. For example, a state $ \ket{ \cdots \circ \bullet } $ on the left system is allowed, as well as a state $ \ket{ \bullet \circ \cdots} $ on the right. However, $$ \ket{ \cdots \circ \bullet } \otimes \ket{ \bullet \circ \cdots} $$ is not allowed in $\mathcal{H}$ since we have adjacent excitations at the point of contact. 

\par Finally we have to evolve the system in time using the full system Hamiltonian $H$, using
$$ \ket{\psi(t)} = e^{-i H t} \ket{\psi(t=0)}, $$
where $H$ is sparse, since our model is local. This operation of taking the exponential of a sparse matrix, and operating with it on a dense vector can be done in one step, i.e. without explicitly computing the exponential, very efficiently using Krylov subspaces methods. This was done using the \texttt{ExpmV} package in \textsf{Julia}.

\subsection{Fronts}

\begin{figure} 
     \centering
     \begin{subfigure}[t]{.48\textwidth}
         \centering
         \captionsetup{font=large, labelfont=bf}
         
         \includegraphics[width=\textwidth]{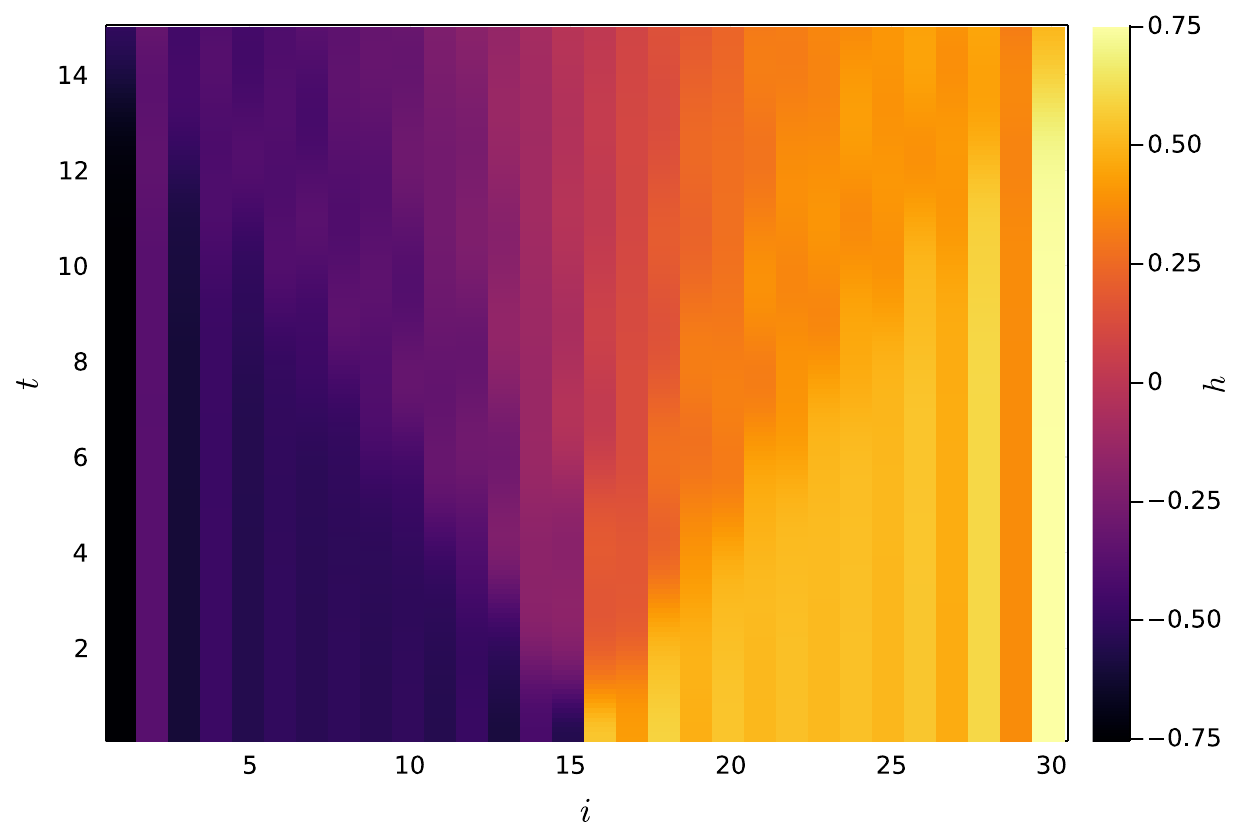}
         \caption{} \label{fig:fronts}
     \end{subfigure}     
     \hfill
     \begin{subfigure}[t]{.48\textwidth}
         \centering
         \captionsetup{font=large, labelfont=bf}
         
         \includegraphics[width=\textwidth]{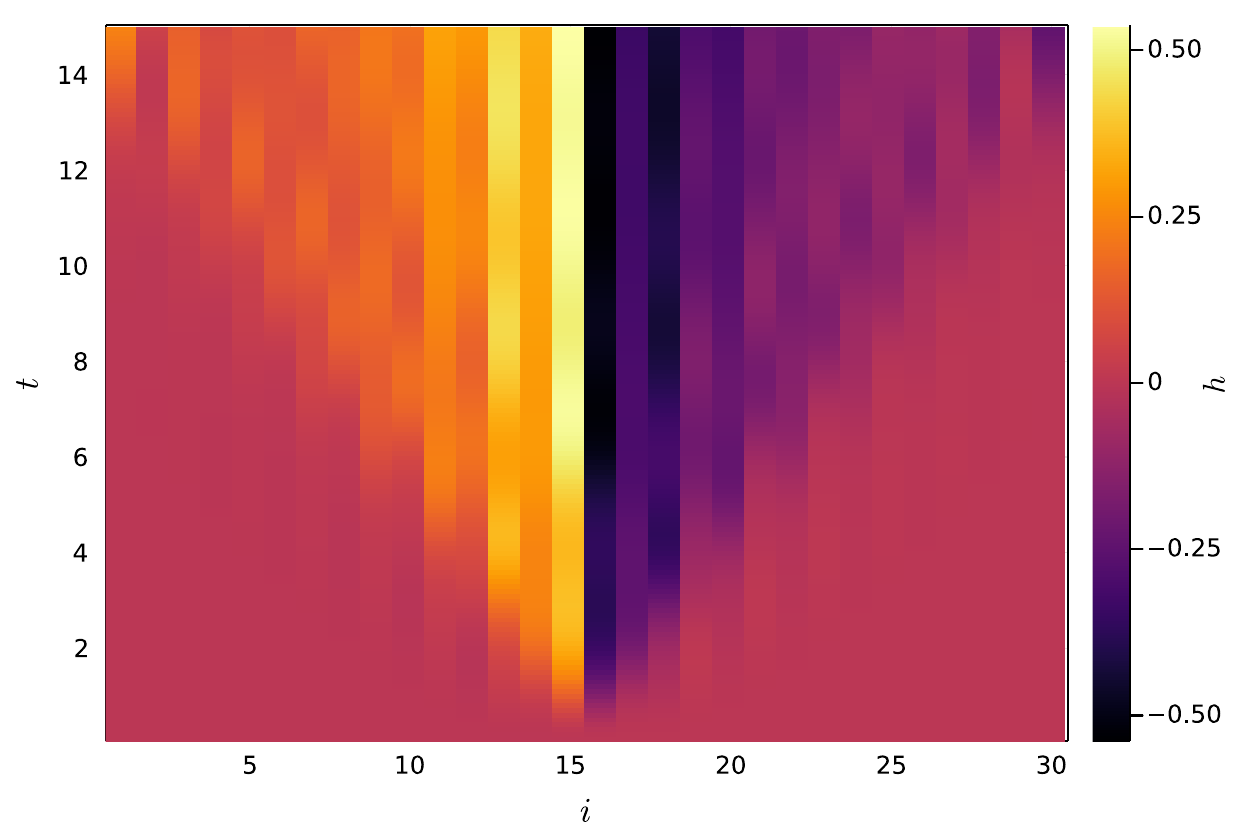}
         \caption{} \label{fig:fronts-subtract}
     \end{subfigure}
     \caption[Space-time diagram of energy density, thermal states]{(a) Space-time diagram of the evolution of the energy density of a PXP chain (with OBC) of size $L=30$, prepared in a state with the left half having an inverse temperature $\beta_{left} = -10$ and the right having $\beta_{right} = +10$. The $x$-axis is the index $i$ of the site on the chain, the $y$-axis is the time $t$ and the colorbar is the energy density $h$. It is evident that the fronts produced are ballistic. (b) The same space-time diagram, but in this case we also subtract the energy density at $t=0$ in order to highlight the linearity of the fronts.}
\label{fig:pxp-fronts}
\end{figure}

We study a system with size $L=30$. We follow the method described above, and the space-time diagram is presented in Fig. \ref{fig:pxp-fronts}, where the $x$-axis is the index $i$ of the site on the chain, the $y$-axis is the time $t$ and the colorbar is the energy density $h$. Therefore, we observe how the energy difference at $t=0$, equalizes and diffuses across the system. We use $\beta_{left} = -10$ and $\beta_{right} = +10$ in Eq. (\ref{th:left}) and (\ref{th:right}). It is evident that the fronts produced are ballistic. In Fig. \ref{fig:fronts-subtract}, we subtract the energy density at $t=0$ in order to highlight the unexpected linearity of the fronts observed.

\subsection{Regression}

\begin{figure}
\centering
\includegraphics[width=\textwidth]{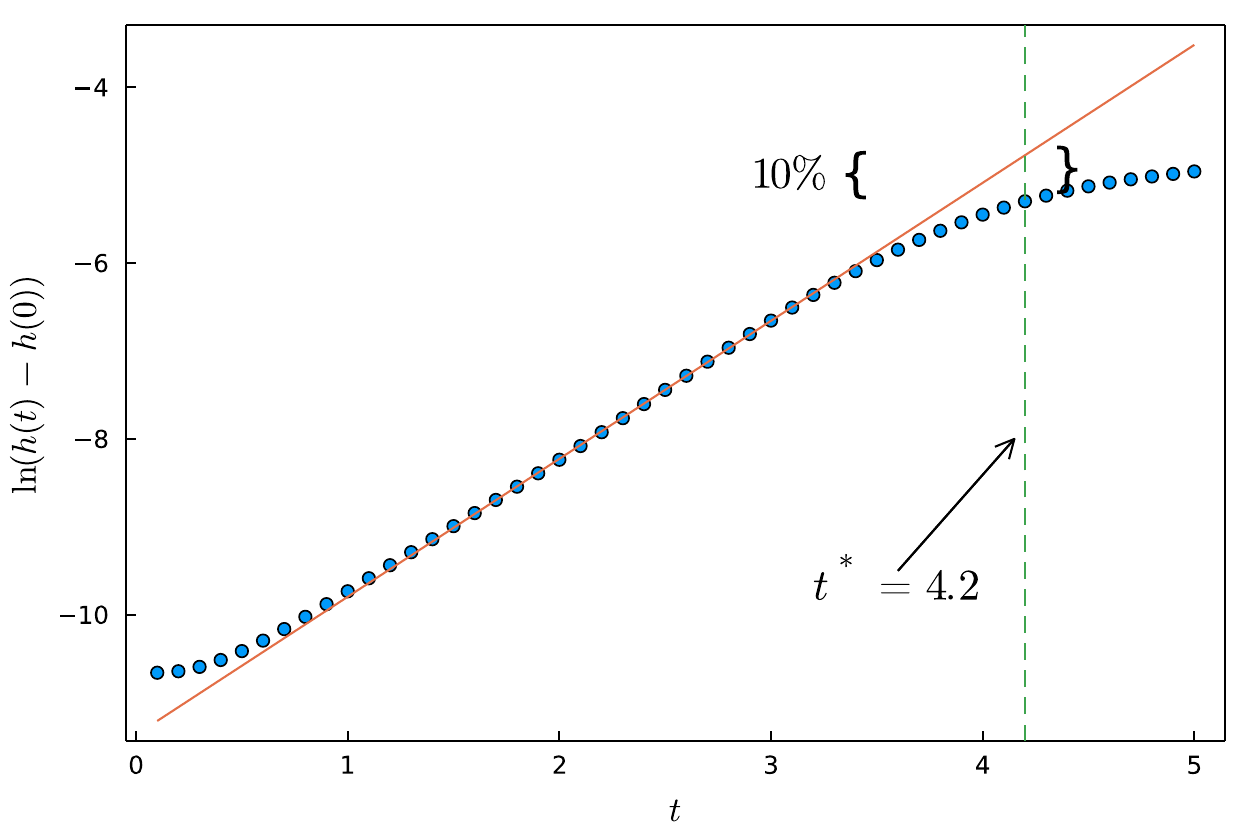}
\caption[Determination of the time of arrival of the front at a site $i$]{Determining the time of ``arrival'' of the front at a site $i$, for a system with $L=30$, and the other parameters as in Fig. \ref{fig:pxp-fronts}. This is a slice for the $i=7$ site of the space-time diagram, until time $t=5$. This is a curve that is very close to an exponential when the front is sufficiently far away. After subtracting the initial value such that all the points are above zero, and plotting the curve in a semi-log plot, we get a line. However, with the arrival of the front, the points deviate from this line, and we define the time of arrival $t^*$ as the point in time where this deviation is $>10\%$. In this case, for $i=7$, we have $t^* = 4.2$.} 
\label{regre}
\end{figure}

Also, we attempted to find a systematic way to determine when the front has ``arrived'' at some site. We tried looking at the space-time diagram for a specific site. This is a curve that is very close to an exponential when the front is sufficiently far away. After subtracting the initial value such that all the points are above zero, we plotted the curve in a semi-log plot. These points form a line, but after some time $t^*$ deviate from the line. Then we fitted a curve to the middle points where the points seem to follow a straight line. An example of this is shown in Fig. (\ref{regre}). 

\begin{figure}
\centering
\includegraphics[width=\textwidth]{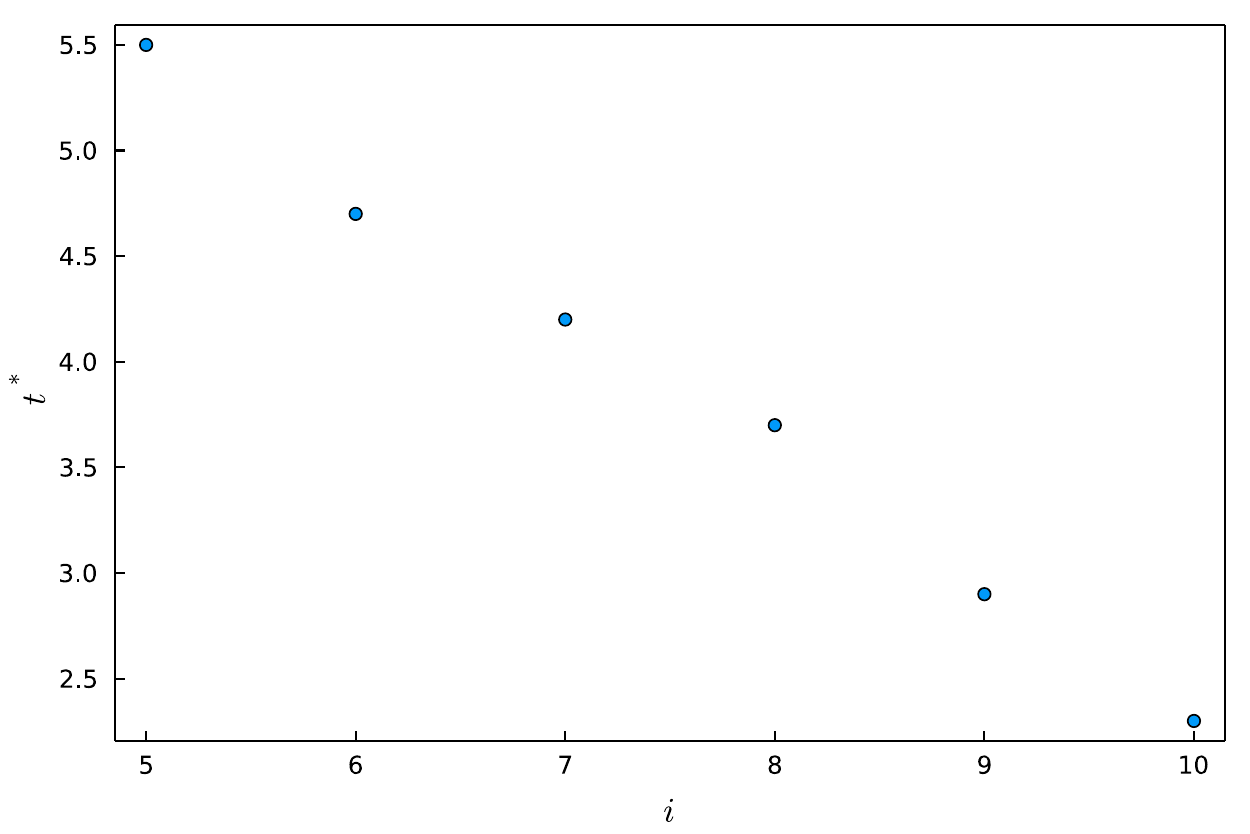}
\caption[Time of arrival of the front at a site $i$, $L=30$]{Time of arrival of the front at a site $i$, for a system with $L=30$, and the other parameters as in Fig. \ref{fig:pxp-fronts}, for $5 \leq i \leq 10$. We can clearly see linear behavior.} 
\label{regre-many}
\end{figure}

Then, we defined $t^*$ the point in time where the deviation is $>10\%$ from the regression line. For the even sites, we also had to flip the curve. We did that for the $L=30$ system and for sites $L = 5$ through $10$. The resulting plot is shown in Fig. (\ref{regre-many}); we clearly see linear behavior. We also tried using a least squares package for exponentials to do the same thing, and the results were similar.

\subsection{Multi-threading}

The algorithm was sped up further using \textsf{Julia}’s built-in multi-threading features. The decorator (macro) \texttt{Threads.@threads} was used to access for loops with multiple threads. The only problem is that appending to a list is not \textit{thread-safe}. Therefore, instead of appending an element to an empty list in each step, we have to initialize a list of zeroes with predetermined length, and then ``add'' elements by editing the list of zeroes. In this way, we make sure that no two threads can edit the same element at the same time, i.e. it is thread-safe.

\section{Microcanonical states}

To make our investigation more conclusive, we have to study the weights the initial state has with respect to the eigenstates of the Hamiltonian, and do the quench again for other initial states that do not have a lot of weight in the ground state, in order to exclude the possibility that this effect (linear spreading) has something to do with the ground state and is not a feature of the system.

In this case, similarly to Eq. (\ref{th:left}) and (\ref{th:right}), we prepare the system in
\begin{equation}\label{mi:left}
\ket{\psi_{\text{left}}} = \frac{1}{\sqrt{A}}\ e^{-\delta (H_{1/2}-E_0)^2}   \ket{\psi_0} ,
\end{equation}
\begin{equation}\label{mi:right}
\ket{\psi_{\text{left}}} = \frac{1}{\sqrt{A'}}\ e^{-\delta (H_{1/2}+E_0)^2}  \ket{\psi_0'} ,
\end{equation}
where $A$, $A'$ are normalization constants, and $\ket{\psi_0}$, $\ket{\psi_0'}$ are random initial states in the Hilbert space $\mathcal{H}_{1/2}$ of the $L_{1/2}$ system. In this way we can choose states in a very narrow window, of size $1/ \sqrt{\delta}$ and around energy $\pm E_0$, in the middle of the spectrum or even the ground state.  

\begin{figure}
\begin{subfigure}{.32\textwidth}
  \centering
  % include first image
  \includegraphics[width=1\linewidth]{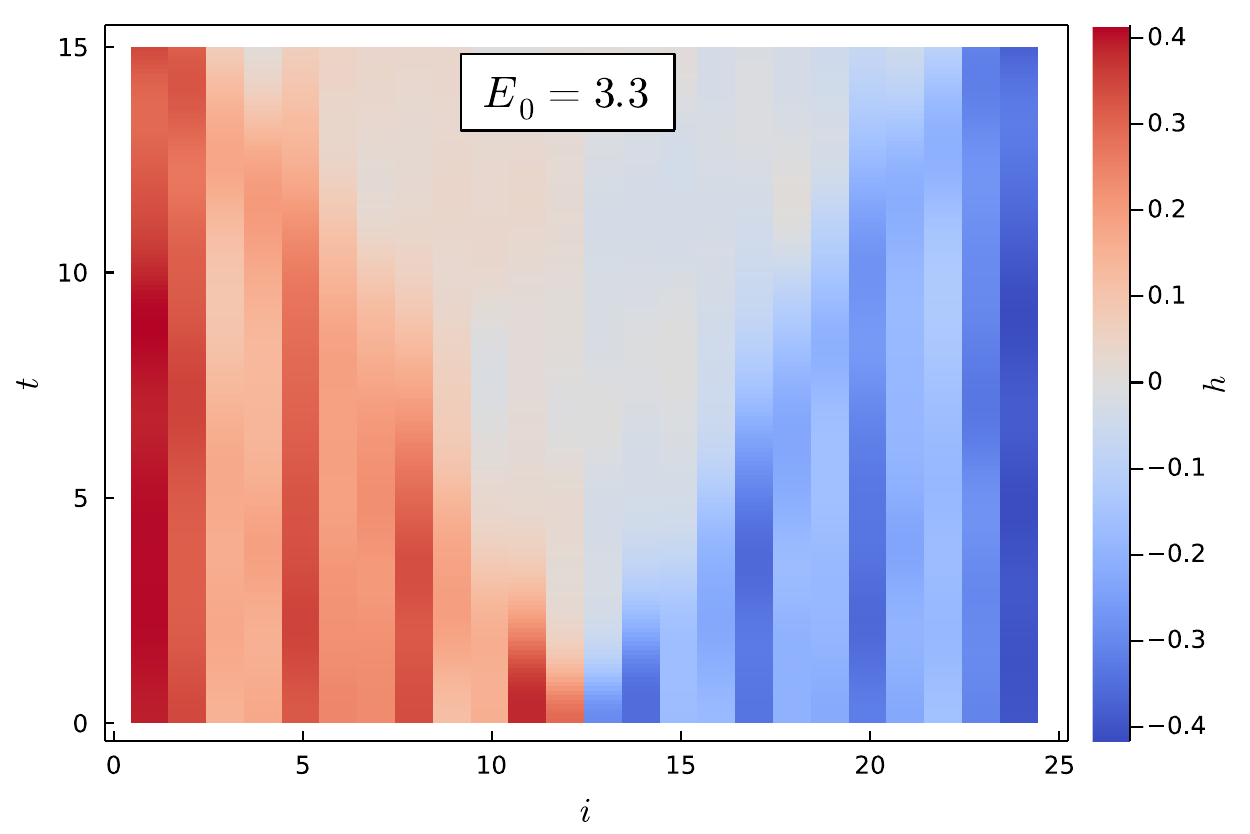}  
\end{subfigure}
\hfill
\begin{subfigure}{.32\textwidth}
  \centering
  % include second image
  \includegraphics[width=1\linewidth]{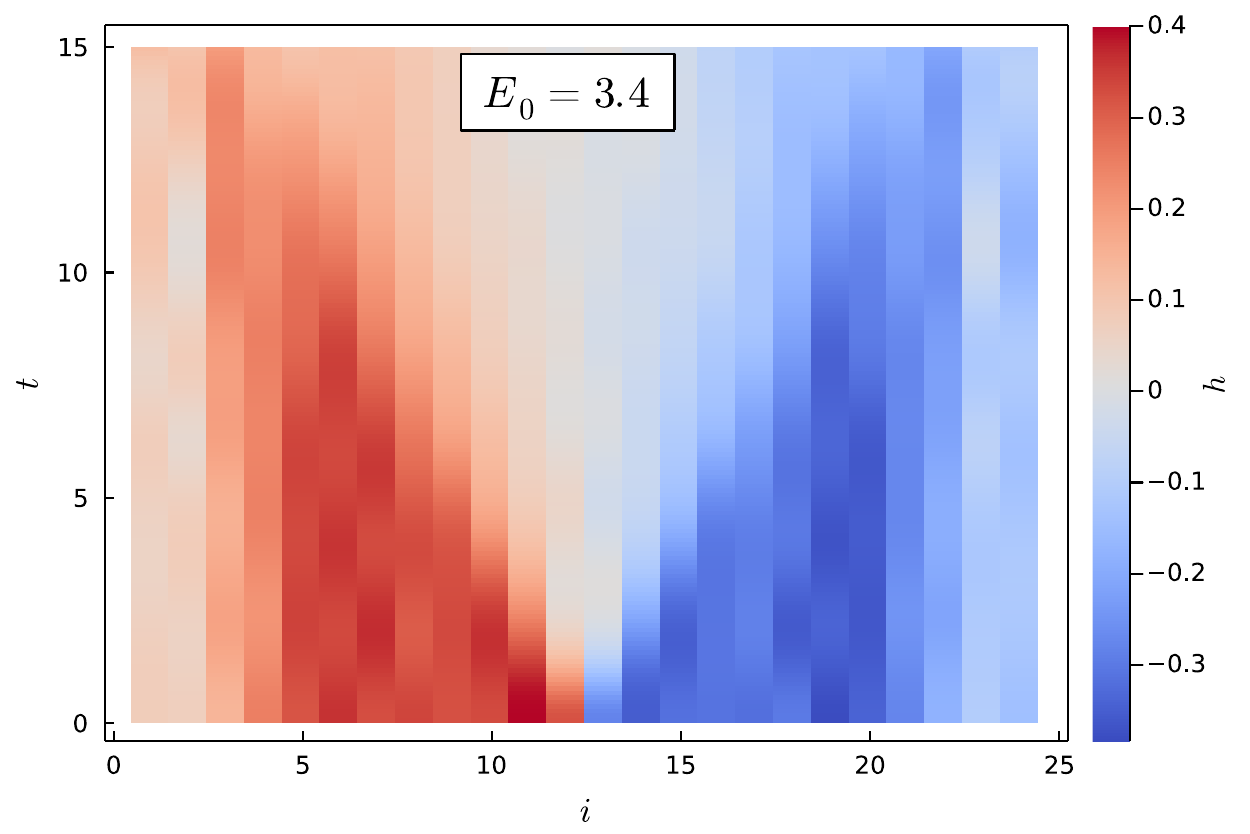}  
\end{subfigure}
\hfill
\begin{subfigure}{.32\textwidth}
  \centering
  % include first image
  \includegraphics[width=1\linewidth]{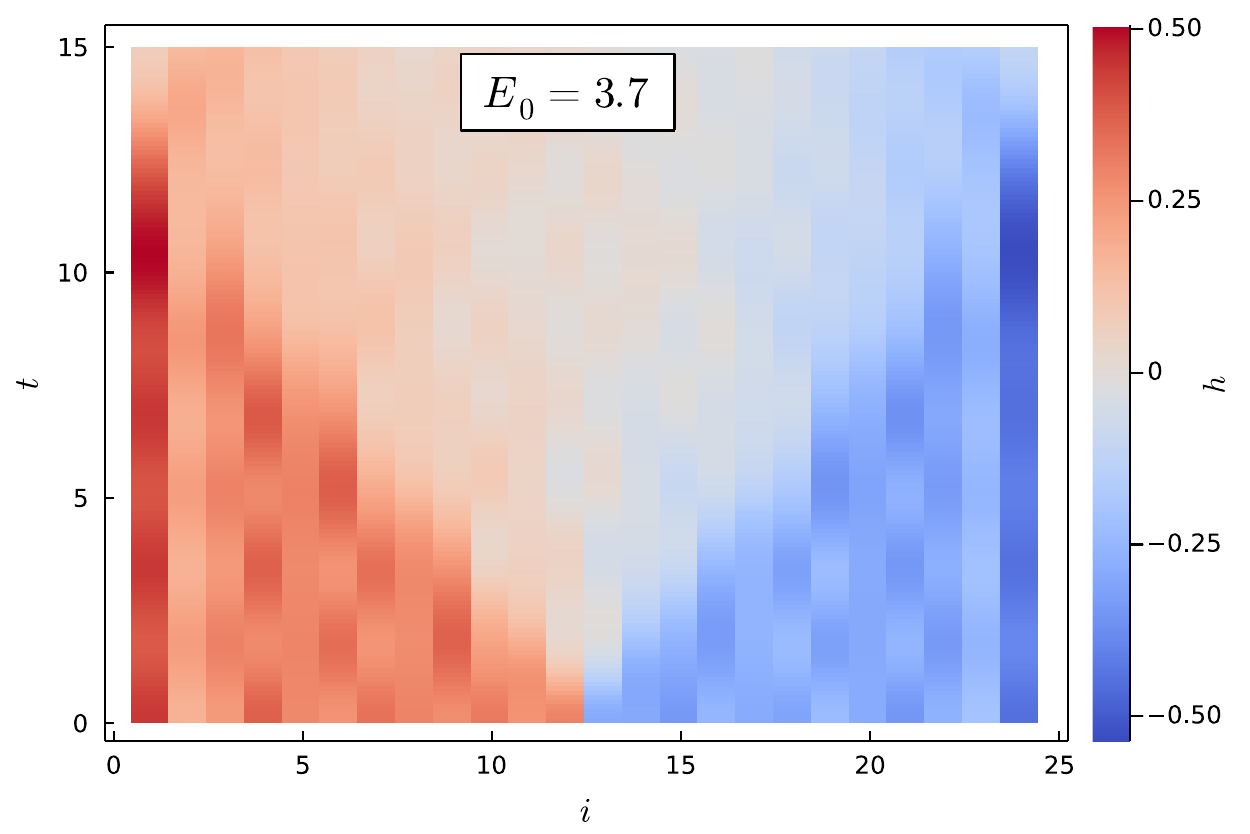}  
\end{subfigure}
\begin{subfigure}{.32\textwidth}
  \centering
  % include first image
  \includegraphics[width=1\linewidth]{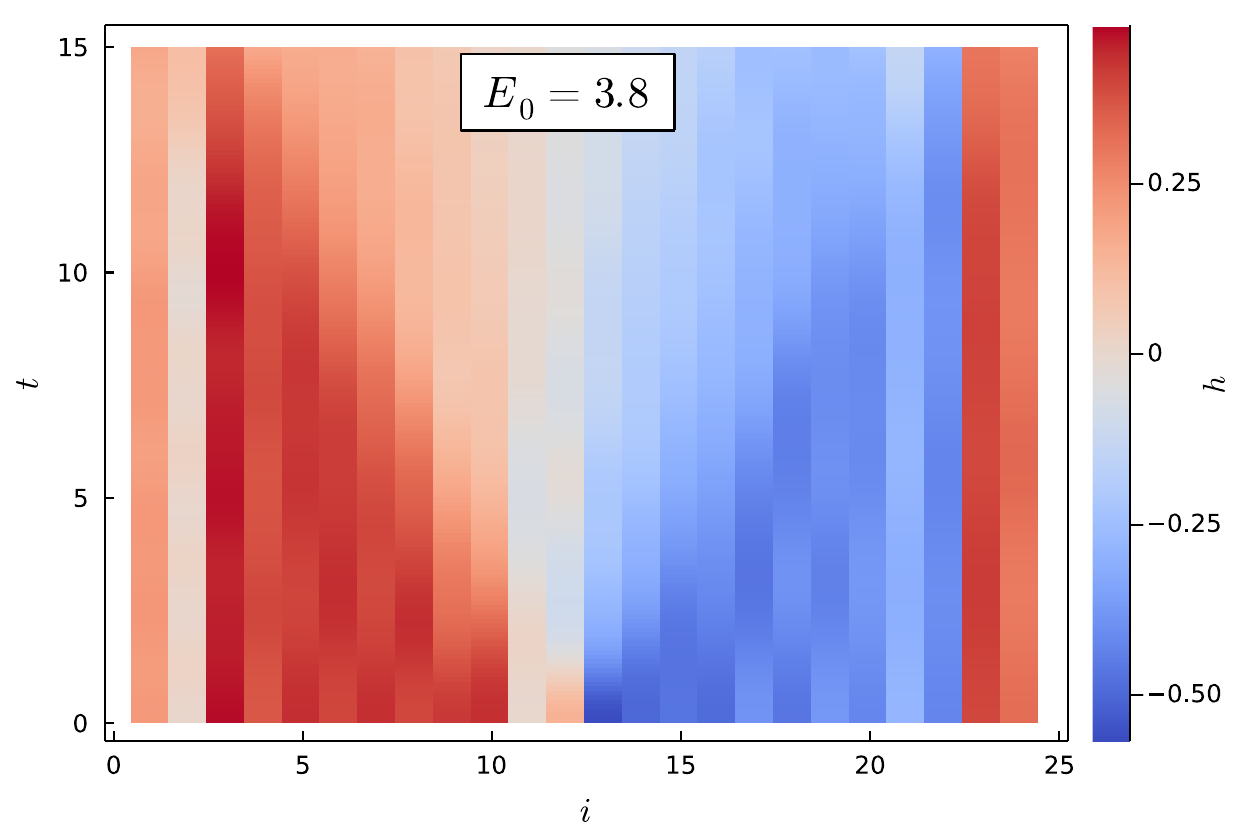}  
\end{subfigure}
\hfill
\begin{subfigure}{.32\textwidth}
  \centering
  % include first image
  \includegraphics[width=1\linewidth]{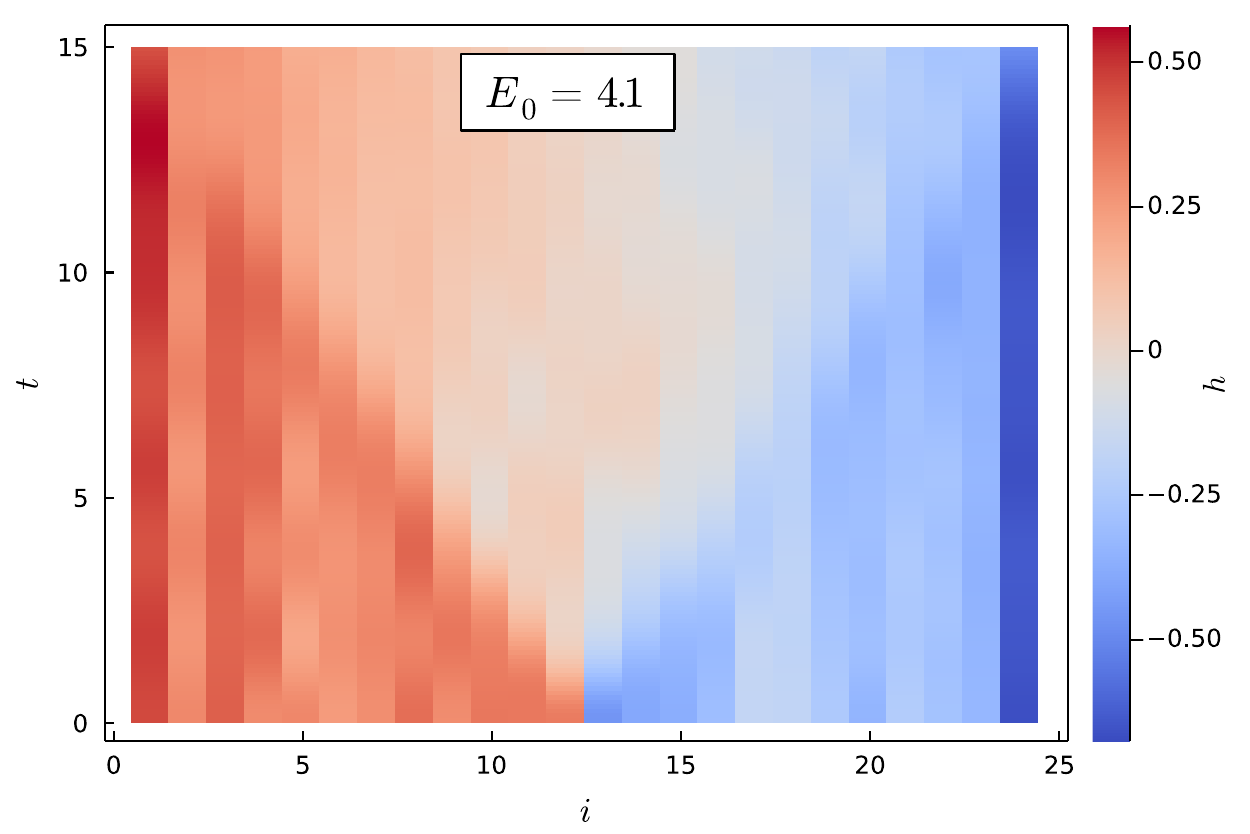}  
\end{subfigure}
\hfill
\begin{subfigure}{.32\textwidth}
  \centering
  % include first image
  \includegraphics[width=1\linewidth]{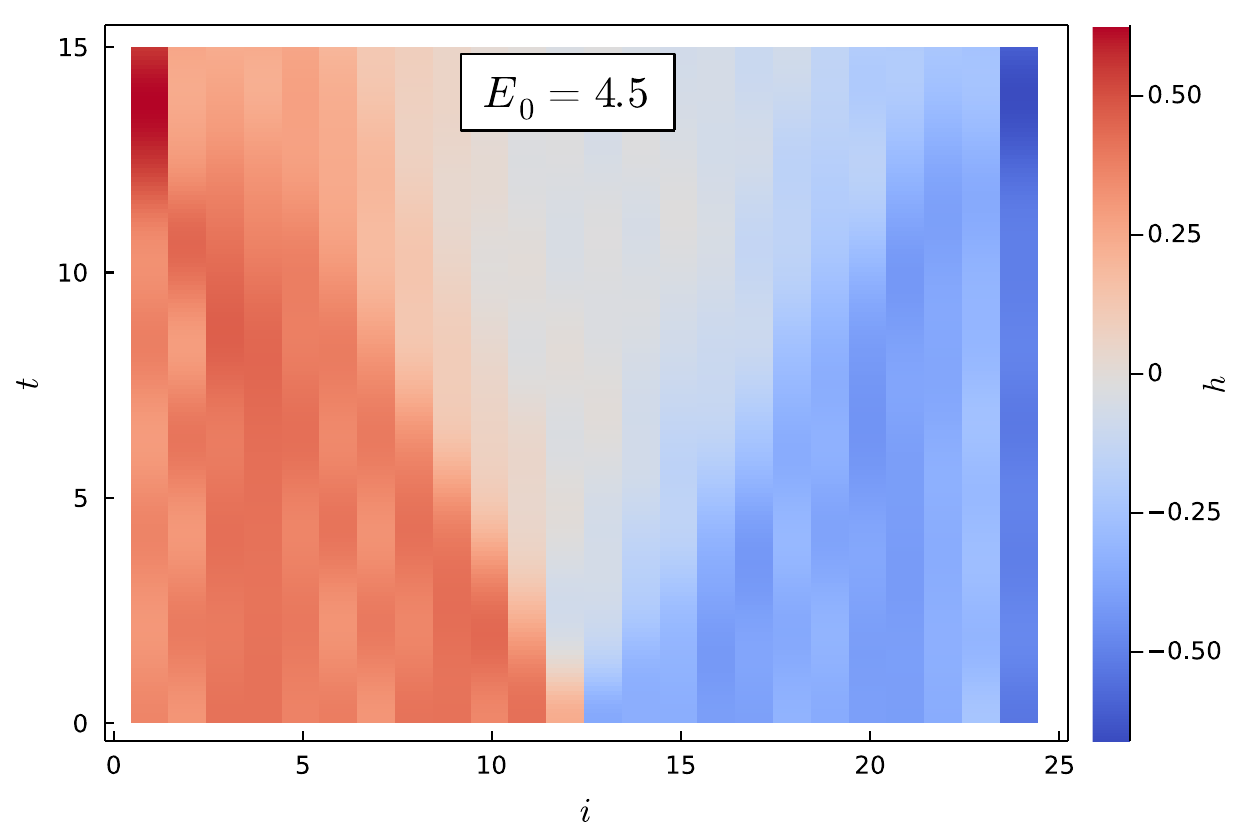}  
\end{subfigure}

\begin{subfigure}{.32\textwidth}
  \centering
  % include first image
  \includegraphics[width=1\linewidth]{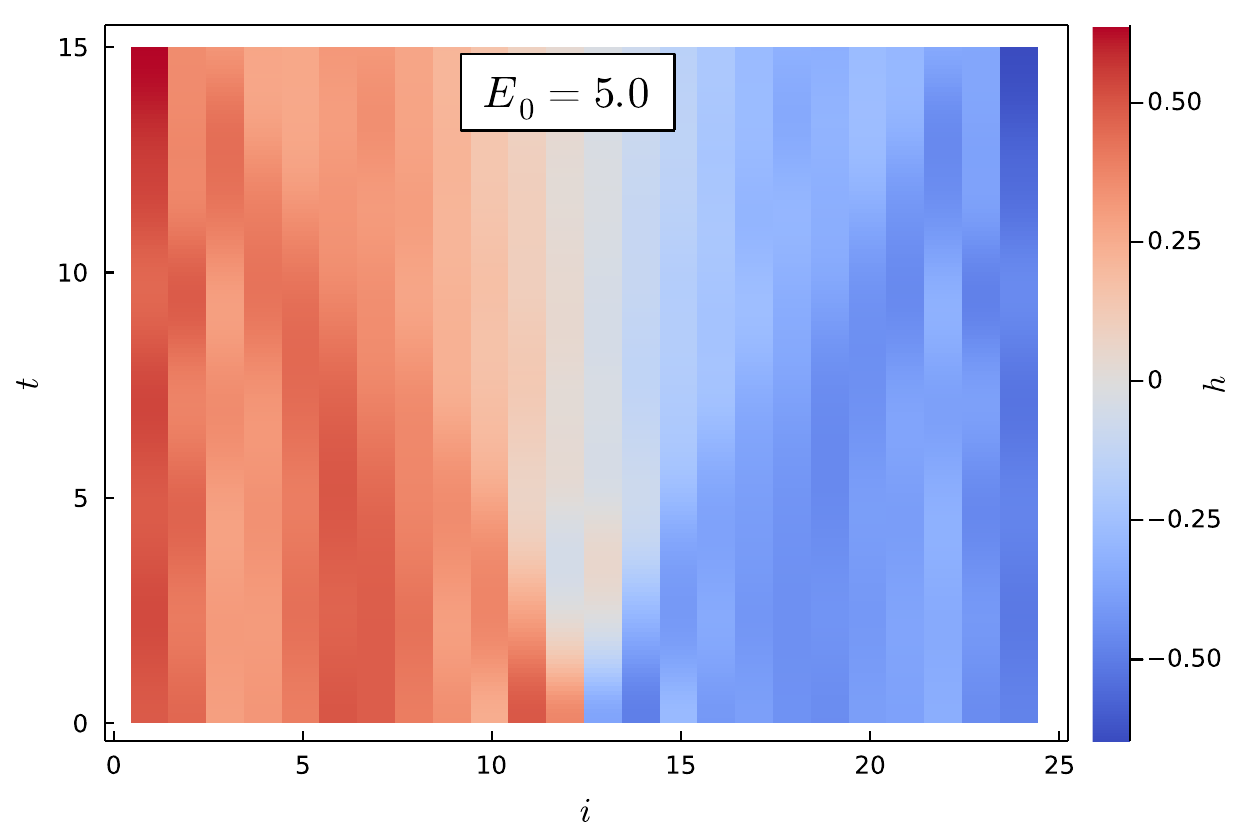}  
\end{subfigure}
\hfill
\begin{subfigure}{.32\textwidth}
  \centering
  % include first image
  \includegraphics[width=1\linewidth]{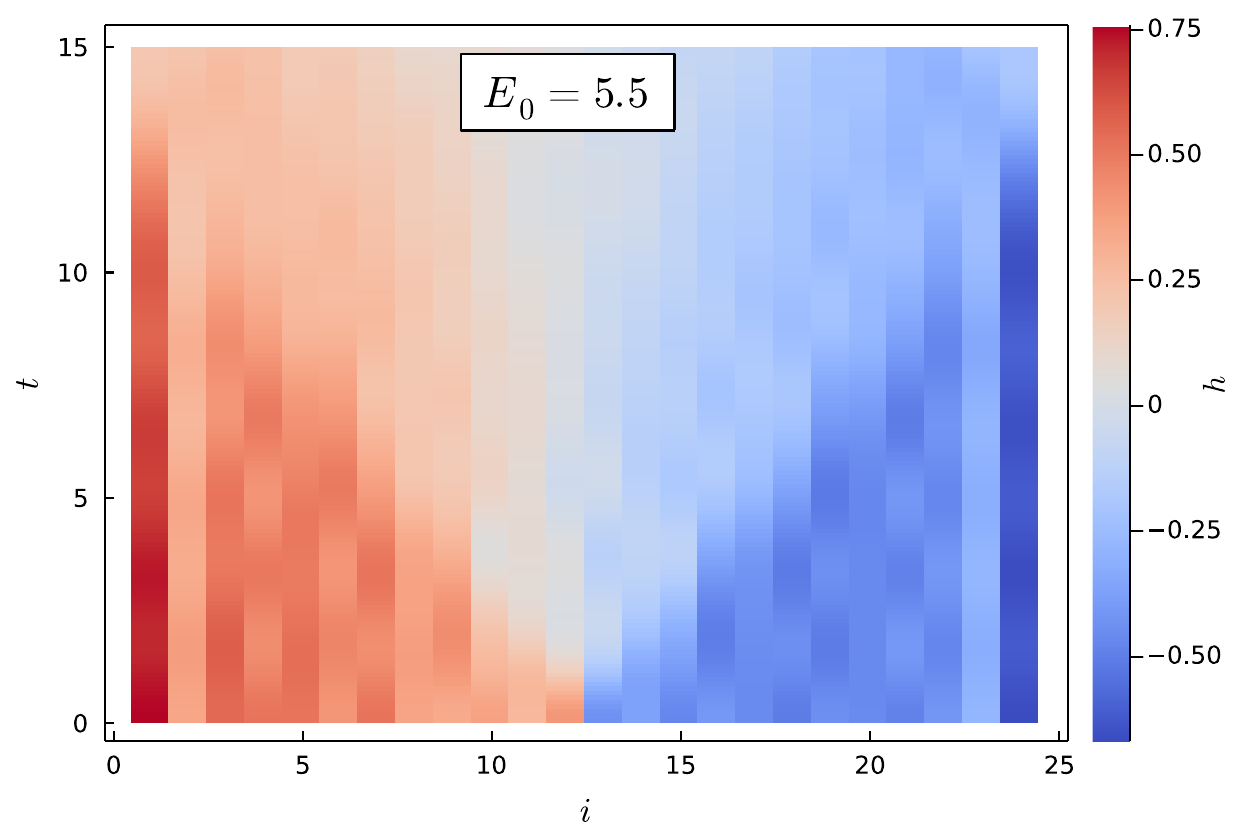}  
\end{subfigure}
\hfill
\begin{subfigure}{.32\textwidth}
  \centering
  % include first image
  \includegraphics[width=1\linewidth]{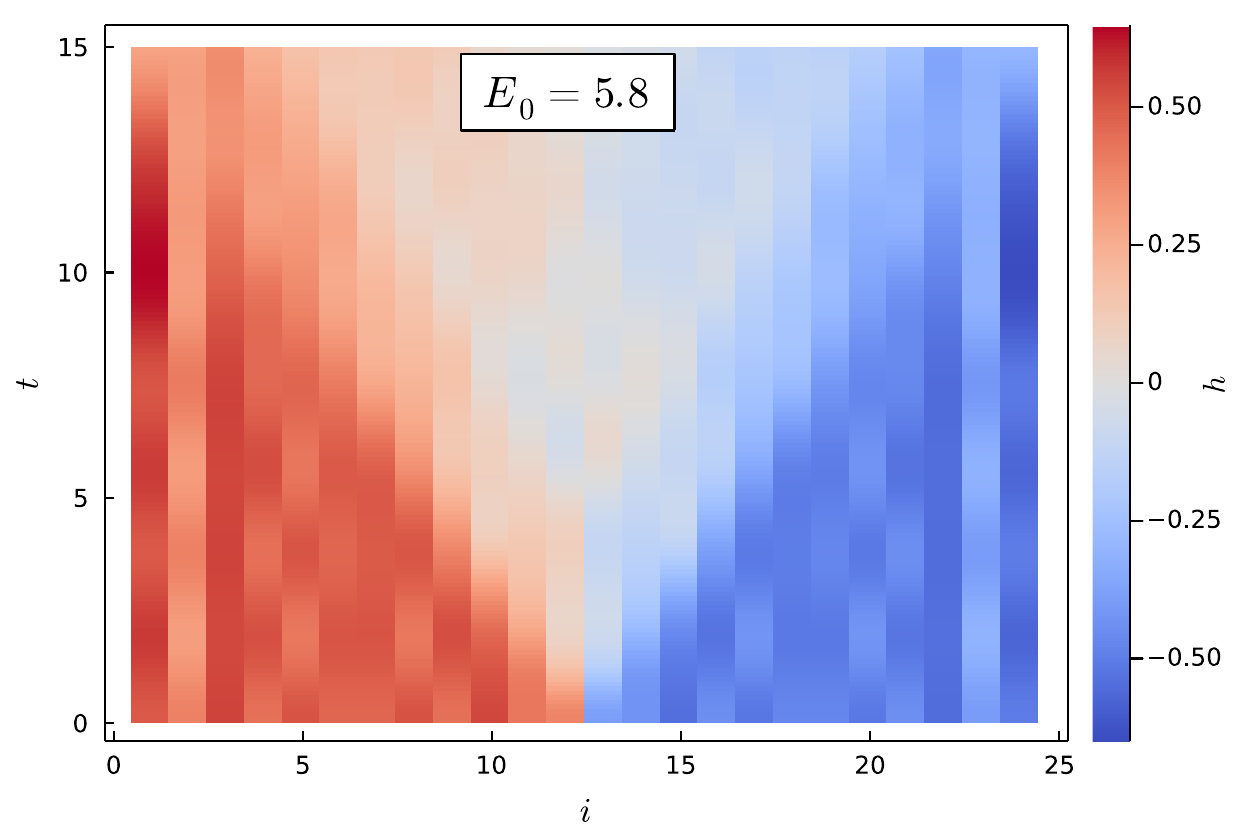}  
\end{subfigure}

\begin{subfigure}{.32\textwidth}
  \centering
  % include first image
  \includegraphics[width=1\linewidth]{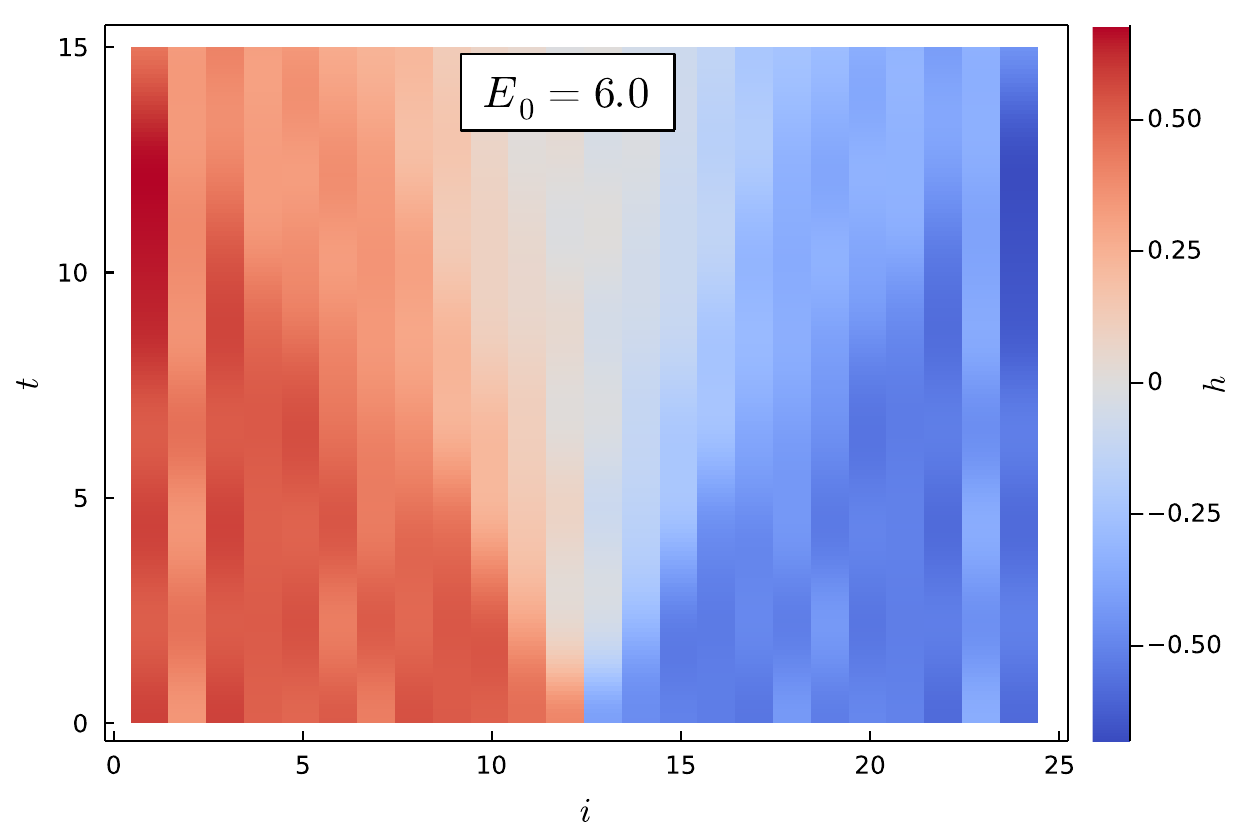}  
\end{subfigure}
\hfill
\begin{subfigure}{.32\textwidth}
  \centering
  % include first image
  \includegraphics[width=1\linewidth]{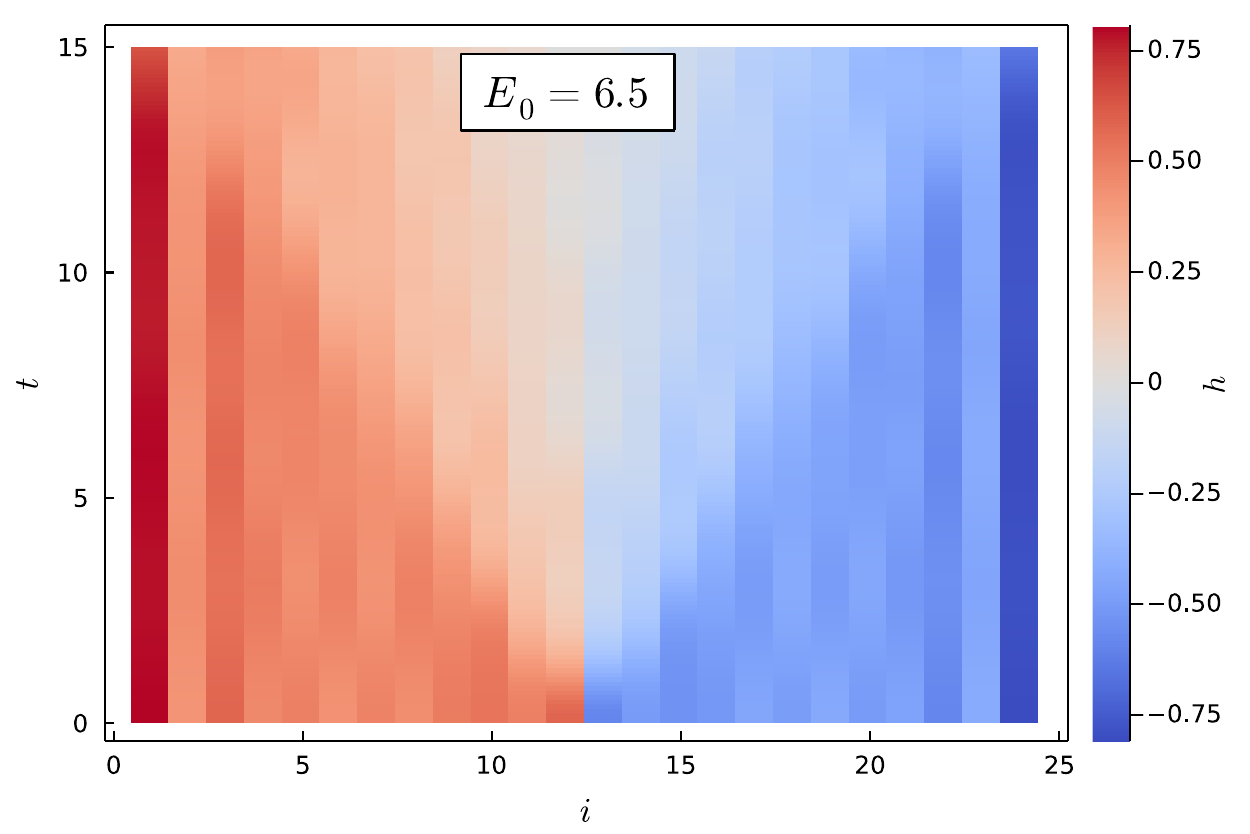}  
\end{subfigure}
\hfill
\begin{subfigure}{.32\textwidth}
  \centering
  % include first image
  \includegraphics[width=1\linewidth]{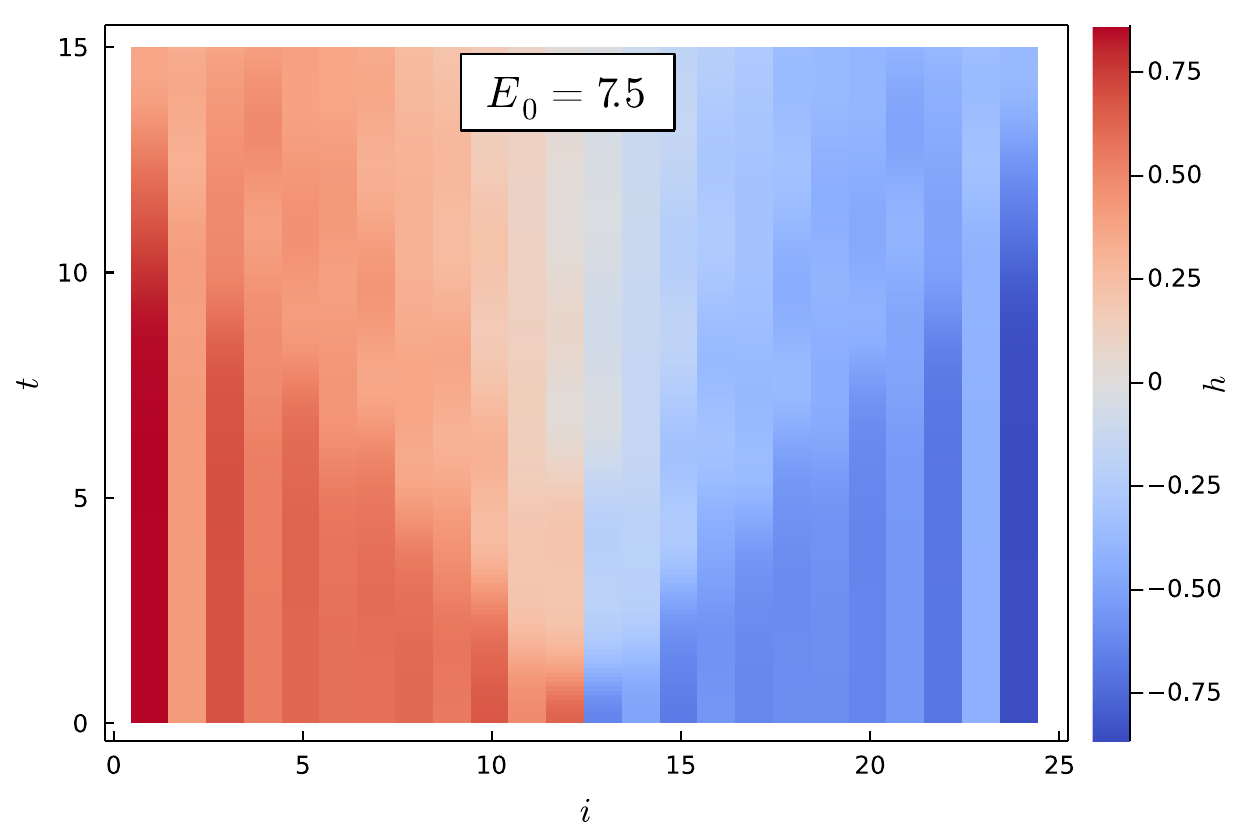}  
\end{subfigure}

\caption[Space-time diagram of energy density, microcanonical states]{Space-time diagrams of the evolution of the energy density of a PXP chain (with OBC) of size $L=24$ (ground state $E_{g.s.} \approx -7.5 $), where the left and right parts of the chain are prepared in a state according to Eq. (\ref{mi:left}) and (\ref{mi:right}) respectively, $\delta = 1,000$, and for various energies $E_0$. The $x$-axis is the index $i$ of the site on the chain, the $y$-axis is the time $t$ and the colorbar is the energy density $h$. It is evident that the fronts produced are ballistic. We see that for different $E_0$ the width of the cone is different, i.e. the velocities of the fronts at different energies $E_0$ are different. For example, for $E_0 = 3.7 $ the cone is particularly wide.}
\label{various-E0}
\end{figure}

For example, in Fig. \ref{various-E0}, for $L=24$, $\delta = 1,000$, which has a ground state of around $E_{g.s.} \approx -7.5 $, we choose various energies $E_0$. We see that for different $E_0$ the width of the cone is different, i.e. the velocities of the fronts at different energies $E_0$ are different. In particular for $E_0 = 3.7 $ the cone is particularly wide. 

\subsection{Pre-quench and post-quench weights}

We now also study the weights that the initial state has with the pre-quench, $H_{1/2}$, and post-quench Hamiltonian $H$. 

\begin{figure} 
     \centering
     \begin{subfigure}[t]{.48\textwidth}
         \centering
         \captionsetup{font=large, labelfont=bf}
         
         \includegraphics[width=\textwidth]{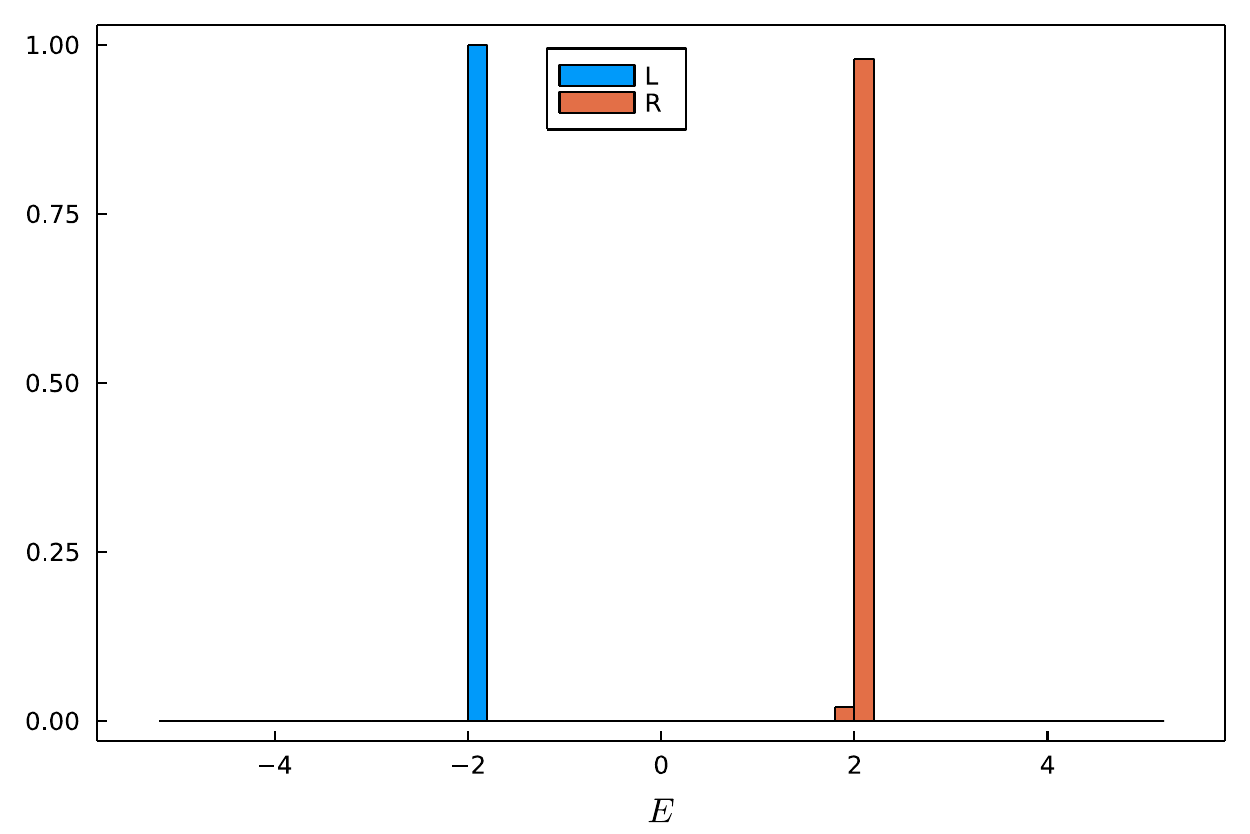}
         \caption{} \label{pre-quench}
     \end{subfigure}     
     \hfill
     \begin{subfigure}[t]{.48\textwidth}
         \centering
         \captionsetup{font=large, labelfont=bf}
         
         \includegraphics[width=\textwidth]{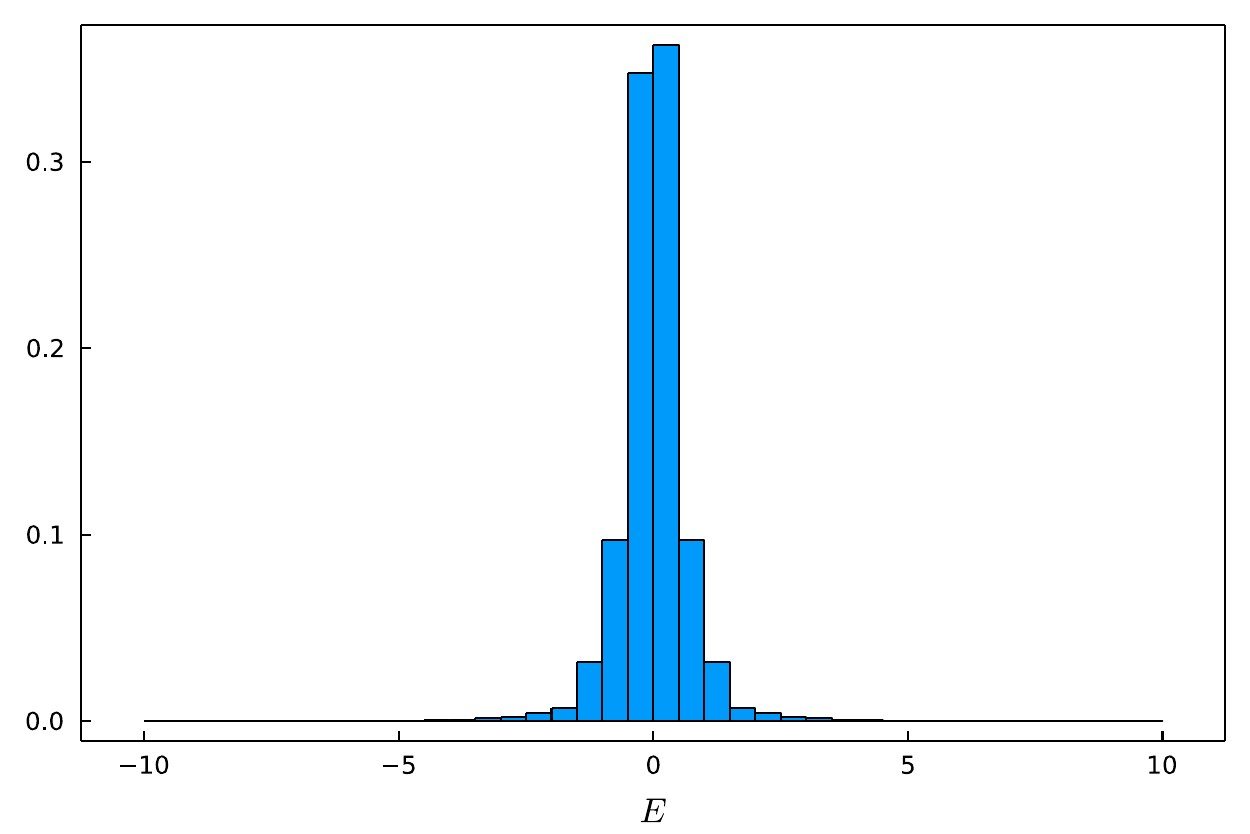}
         \caption{} \label{post-quench}
     \end{subfigure}
     \vfill
     \begin{subfigure}[t]{.48\textwidth}
         \centering
         \captionsetup{font=large, labelfont=bf}
         
         \includegraphics[width=\textwidth]{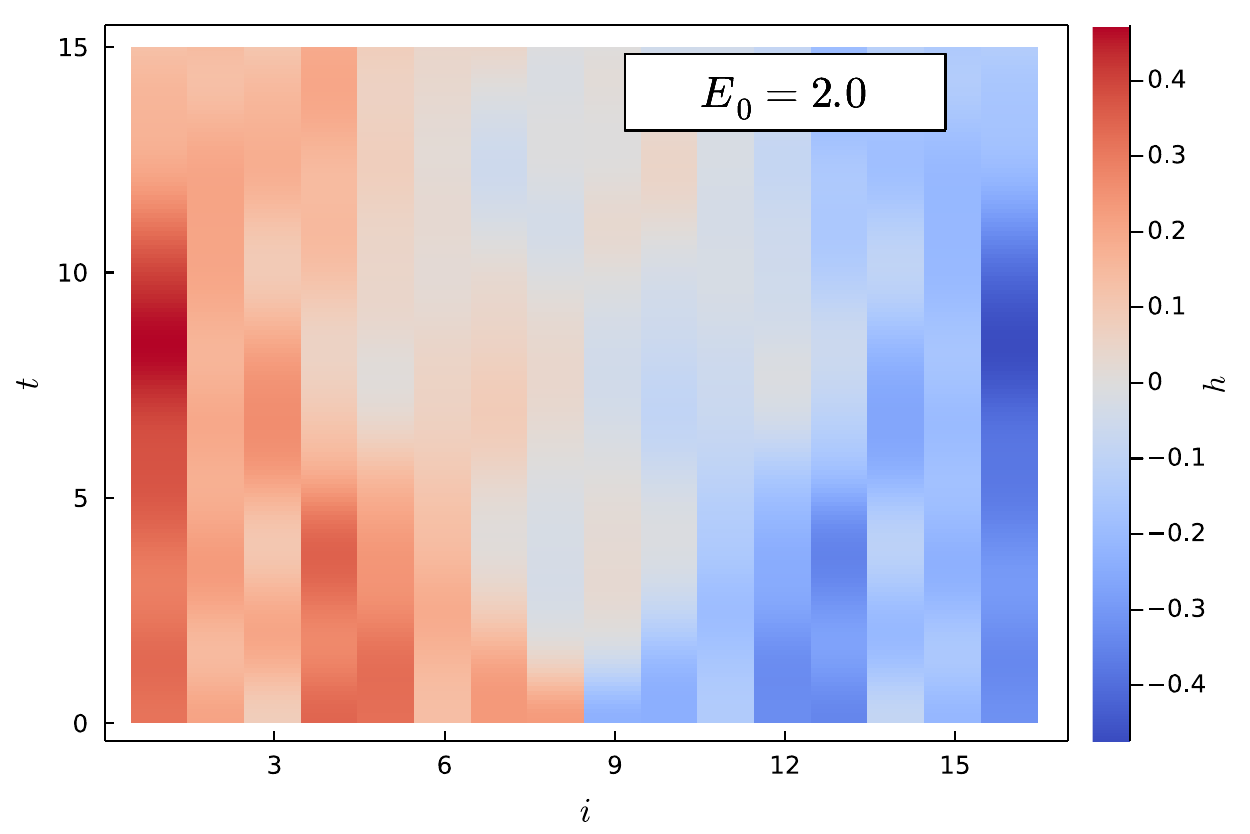}
         \caption{} \label{space-time}
     \end{subfigure}     
     \hfill
     \begin{subfigure}[t]{.48\textwidth}
         \centering
         \captionsetup{font=large, labelfont=bf}
         
         \includegraphics[width=\textwidth]{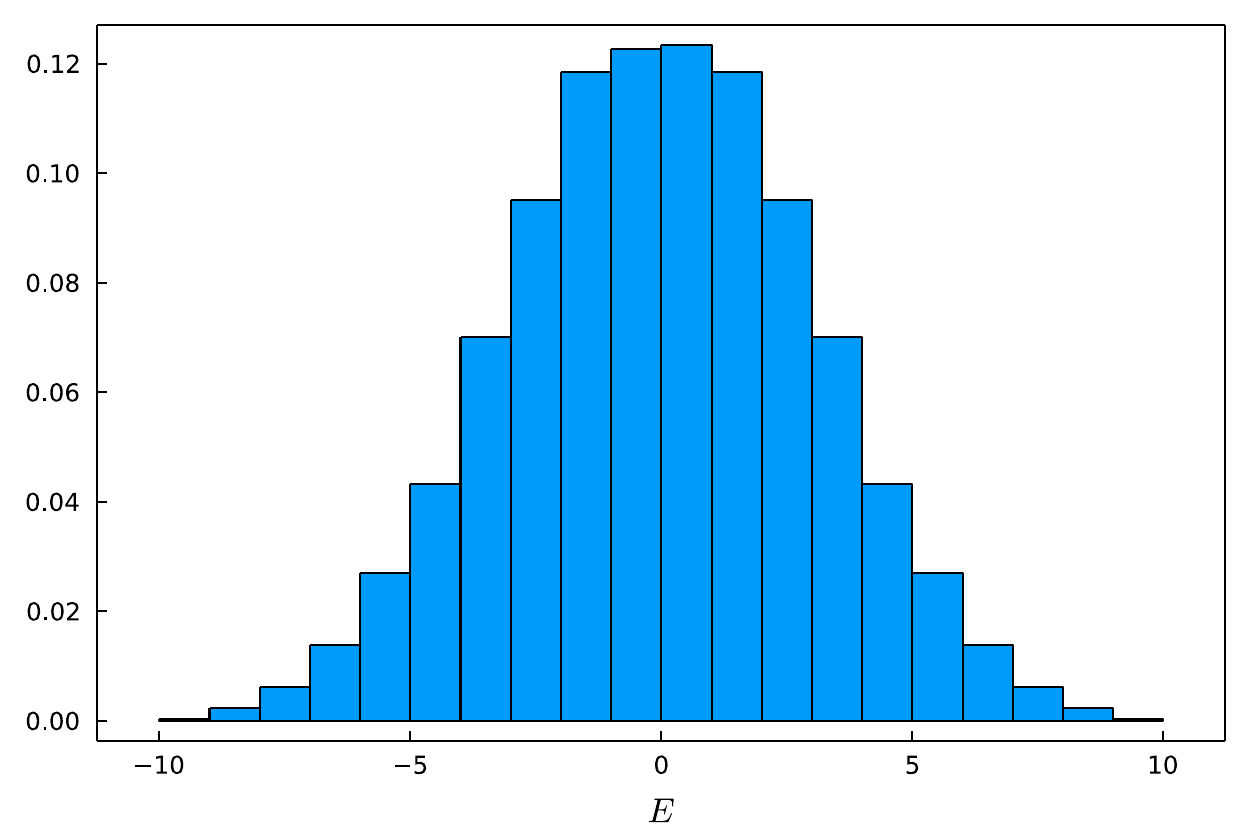}
         \caption{} \label{dos}
     \end{subfigure}
     \caption[Pre-quench and post-quench weights, $E_0 = 2.0$]{(a) Weights of the $\ket{\psi_{\text{left}}}$ and $\ket{\psi_{\text{right}}}$ in Eq. (\ref{mi:left}) and (\ref{mi:right}) with the pre-quench Hamiltonian $H_{1/2}$ for one half of the chain. We see that the state is prepared away from the ground state, in a narrow window somewhere in the middle of the spectrum. The system studied is of size $L=16$ ($L_{1/2} = 8$), with $\delta = 1,000$ and $E_0 = 2.0$ in Eq. (\ref{mi:left}) and (\ref{mi:right}), and has a ground state of $E_{g.s.} \approx -5.1 $. (b) Weights of the $\ket{\psi} = \ket{\psi_{\text{left}}} \otimes \ket{\psi_{\text{right}}} $ state with the post-quench Hamiltonian $H$ for the whole chain. It is clear that $\ket{\psi}$ has large weights around eigenenergies with $E=0$. (c) Space-time plot, as in Fig. \ref{various-E0}, where the $x$-axis is the index $i$ of the site on the chain, the $y$-axis is the time $t$ and the colorbar is the energy density $h$. (d) Density of states for the post-quench Hamiltonian $H$.}
\label{fig:weights}
\end{figure}

In Fig. \ref{fig:weights} we calculate the weights for a system of size $L=16$ ($L_{1/2} = 8$), and with $\delta = 1,000$ and $E_0 = 2.0$ in Eq. (\ref{mi:left}) and (\ref{mi:right}). The system under study has a ground state of $E_{g.s.} \approx -5.1 $. In Fig. \ref{pre-quench}, we see the weights of the $\ket{\psi_{\text{left}}}$ and $\ket{\psi_{\text{right}}}$ in Eq. (\ref{mi:left}) and (\ref{mi:right}) with the pre-quench Hamiltonian $H_{1/2}$ for one half of the chain. We see that the state is prepared away from the ground state, in a narrow window somewhere in the middle of the spectrum. In Fig. \ref{post-quench}, the weights of the $\ket{\psi} = \ket{\psi_{\text{left}}} \otimes \ket{\psi_{\text{right}}} $ state with the post-quench Hamiltonian $H$ for the whole chain are shown. There, it is clear that $\ket{\psi}$ has large weights around eigenenergies with $E=0$. In Fig. \ref{space-time}, we see the space-time plot, as in Fig. \ref{various-E0}, where the $x$-axis is the index $i$ of the site on the chain, the $y$-axis is the time $t$ and the colorbar is the energy density $h$. Finally, in Fig. \ref{dos}, we see the density of states for the post-quench Hamiltonian $H$. 

\begin{figure} 
     \centering
     \begin{subfigure}[t]{.48\textwidth}
         \centering
         \captionsetup{font=large, labelfont=bf}
         
         \includegraphics[width=\textwidth]{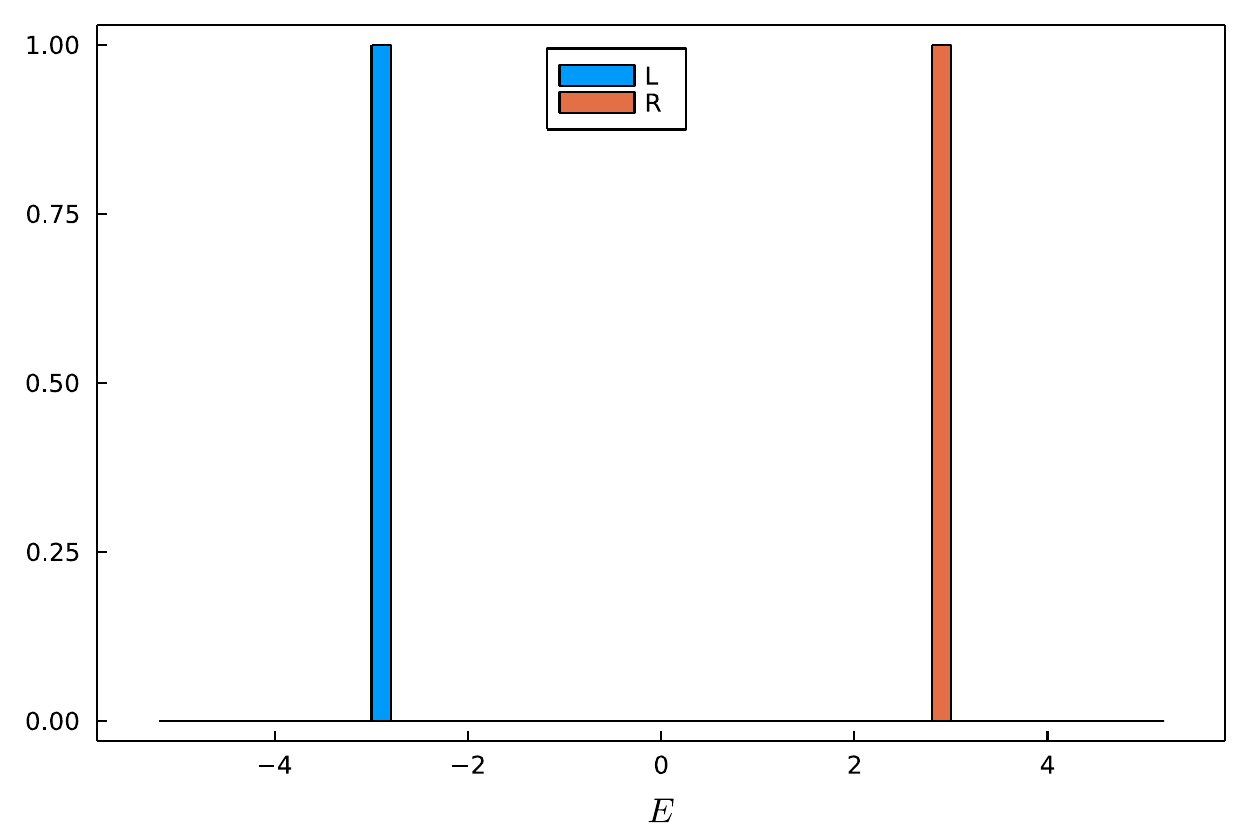}
         \caption{} \label{pre-quench-2}
     \end{subfigure}     
     \hfill
     \begin{subfigure}[t]{.48\textwidth}
         \centering
         \captionsetup{font=large, labelfont=bf}
         
         \includegraphics[width=\textwidth]{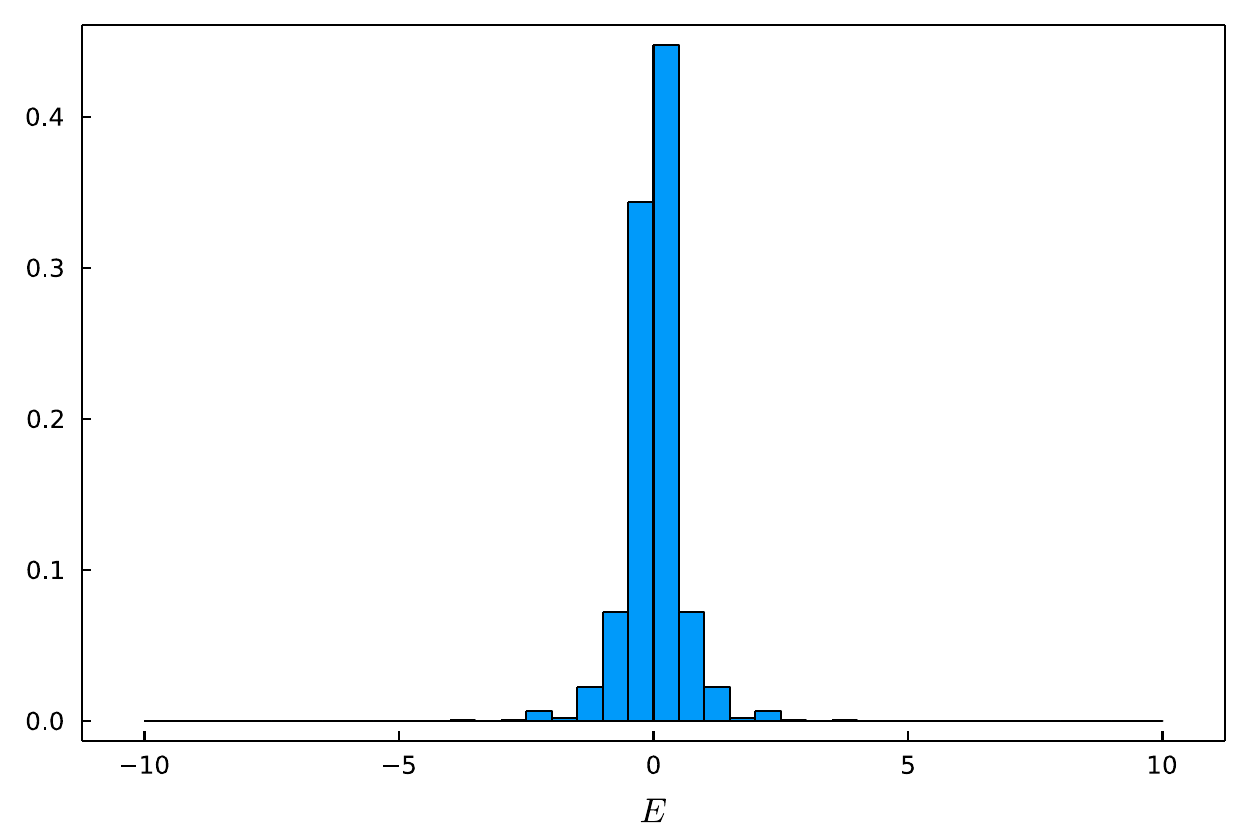}
         \caption{} \label{post-quench-2}
     \end{subfigure}
     \vfill
     \begin{subfigure}[t]{.48\textwidth}
         \centering
         \captionsetup{font=large, labelfont=bf}
         
         \includegraphics[width=\textwidth]{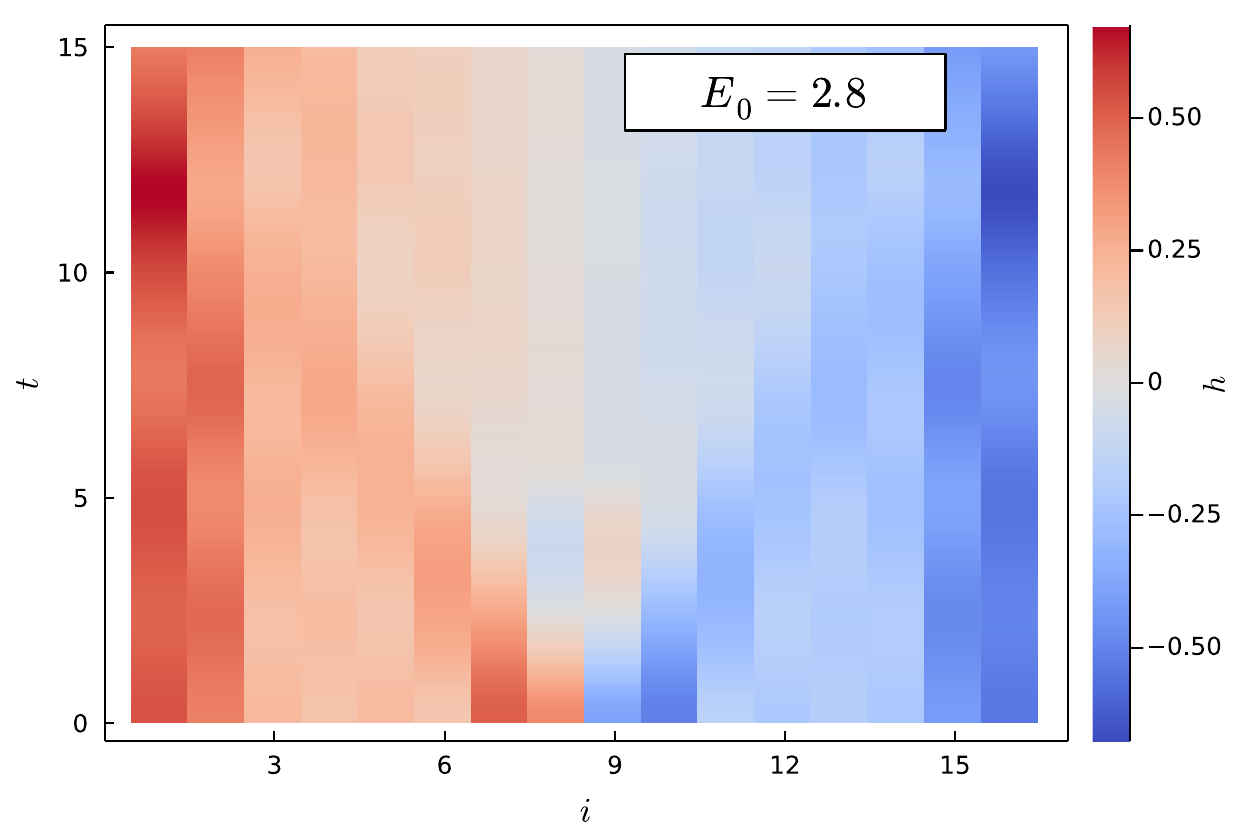}
         \caption{} \label{space-time-2}
     \end{subfigure}     
     \hfill
     \begin{subfigure}[t]{.48\textwidth}
         \centering
         \captionsetup{font=large, labelfont=bf}
         
         \includegraphics[width=\textwidth]{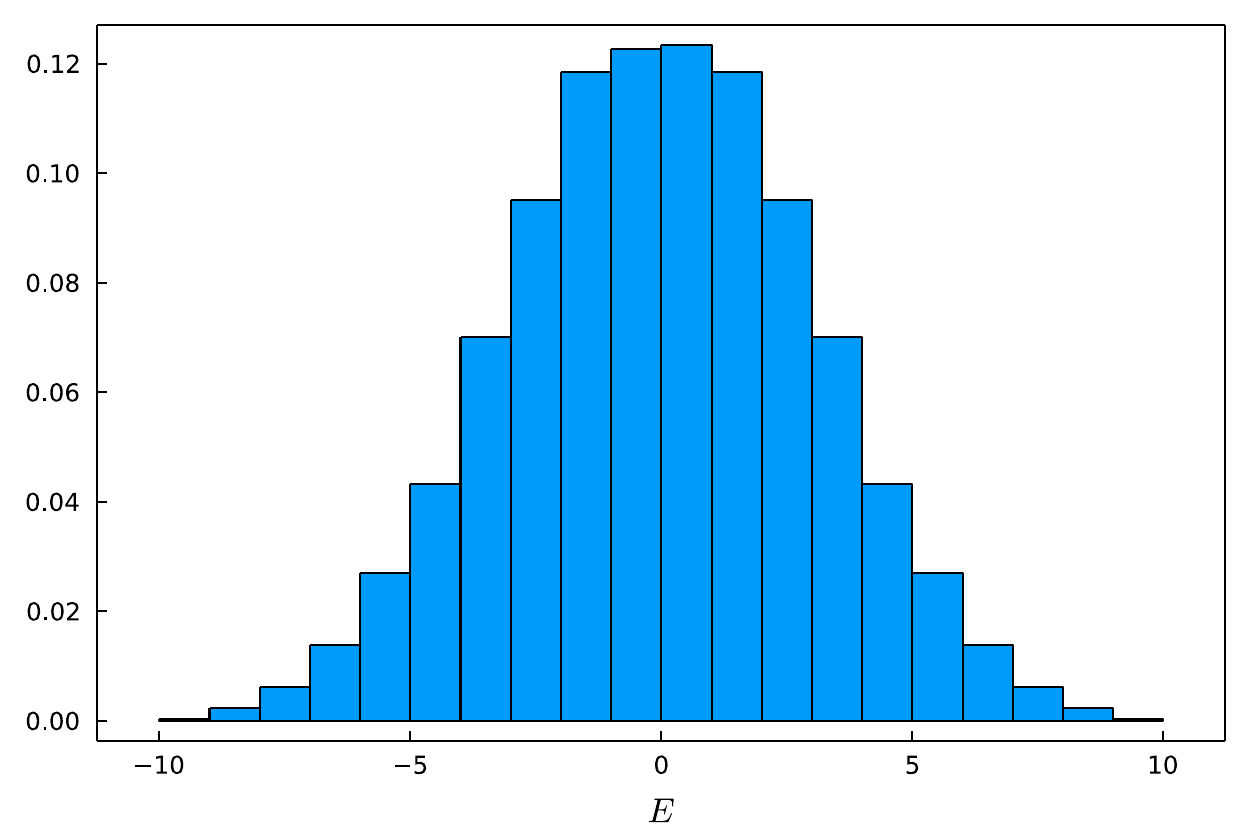}
         \caption{} \label{dos-2}
     \end{subfigure}
     \caption[Pre-quench and post-quench weights, $E_0 = 2.8$]{(a) Weights of the $\ket{\psi_{\text{left}}}$ and $\ket{\psi_{\text{right}}}$ in Eq. (\ref{mi:left}) and (\ref{mi:right}) with the pre-quench Hamiltonian $H_{1/2}$ for one half of the chain. We see that the state is prepared away from the ground state, in a narrow window somewhere in the middle of the spectrum. The system studied is of size $L=16$ ($L_{1/2} = 8$), with $\delta = 1,000$ and $E_0 = 2.8$ in Eq. (\ref{mi:left}) and (\ref{mi:right}), and has a ground state of $E_{g.s.} \approx -5.1 $. (b) Weights of the $\ket{\psi} = \ket{\psi_{\text{left}}} \otimes \ket{\psi_{\text{right}}} $ state with the post-quench Hamiltonian $H$ for the whole chain. It is clear that $\ket{\psi}$ has large weights around eigenenergies with $E=0$. (c) Space-time plot, as in Fig. \ref{various-E0}, where the $x$-axis is the index $i$ of the site on the chain, the $y$-axis is the time $t$ and the colorbar is the energy density $h$. (d) Density of states for the post-quench Hamiltonian $H$.}
\label{fig:weights-2}
\end{figure}

\begin{figure} 
     \centering
     \begin{subfigure}[t]{.48\textwidth}
         \centering
         \captionsetup{font=large, labelfont=bf}
         
         \includegraphics[width=\textwidth]{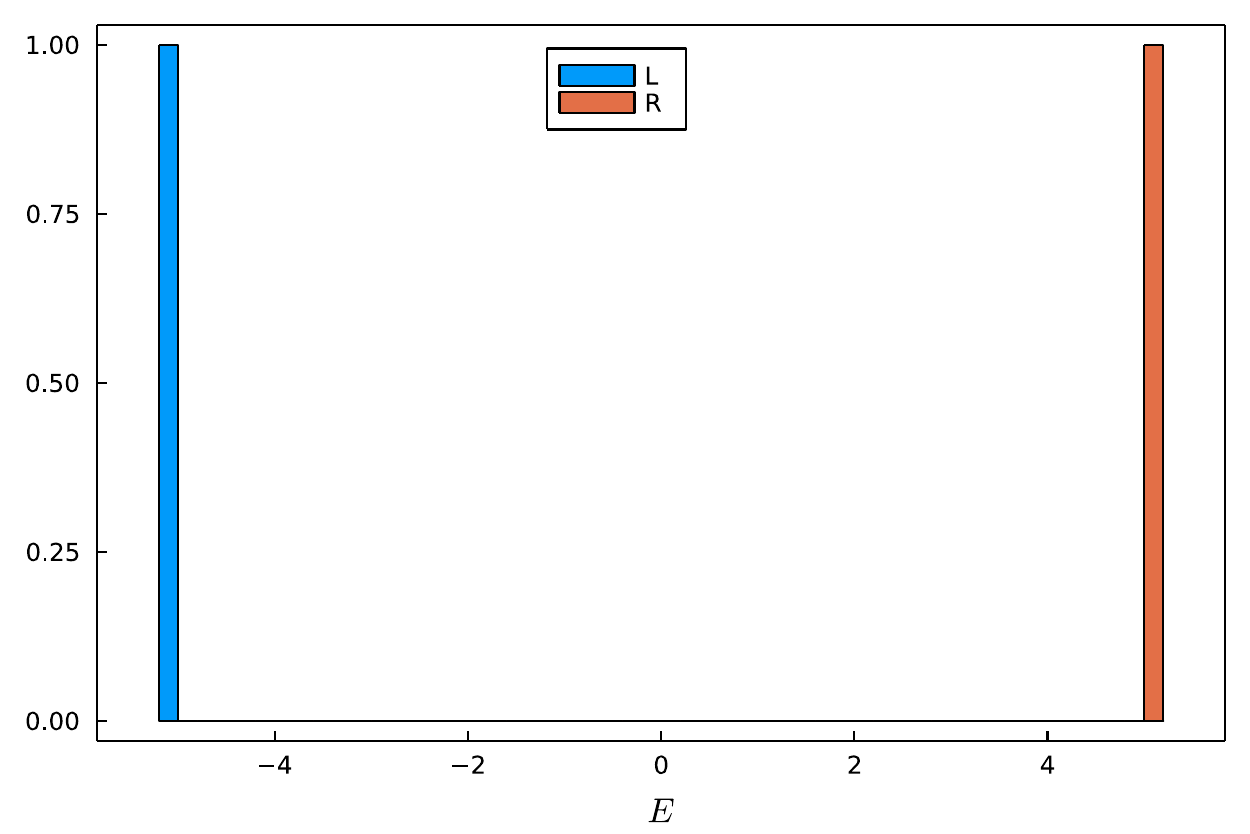}
         \caption{} \label{pre-quench-3}
     \end{subfigure}     
     \hfill
     \begin{subfigure}[t]{.48\textwidth}
         \centering
         \captionsetup{font=large, labelfont=bf}
         
         \includegraphics[width=\textwidth]{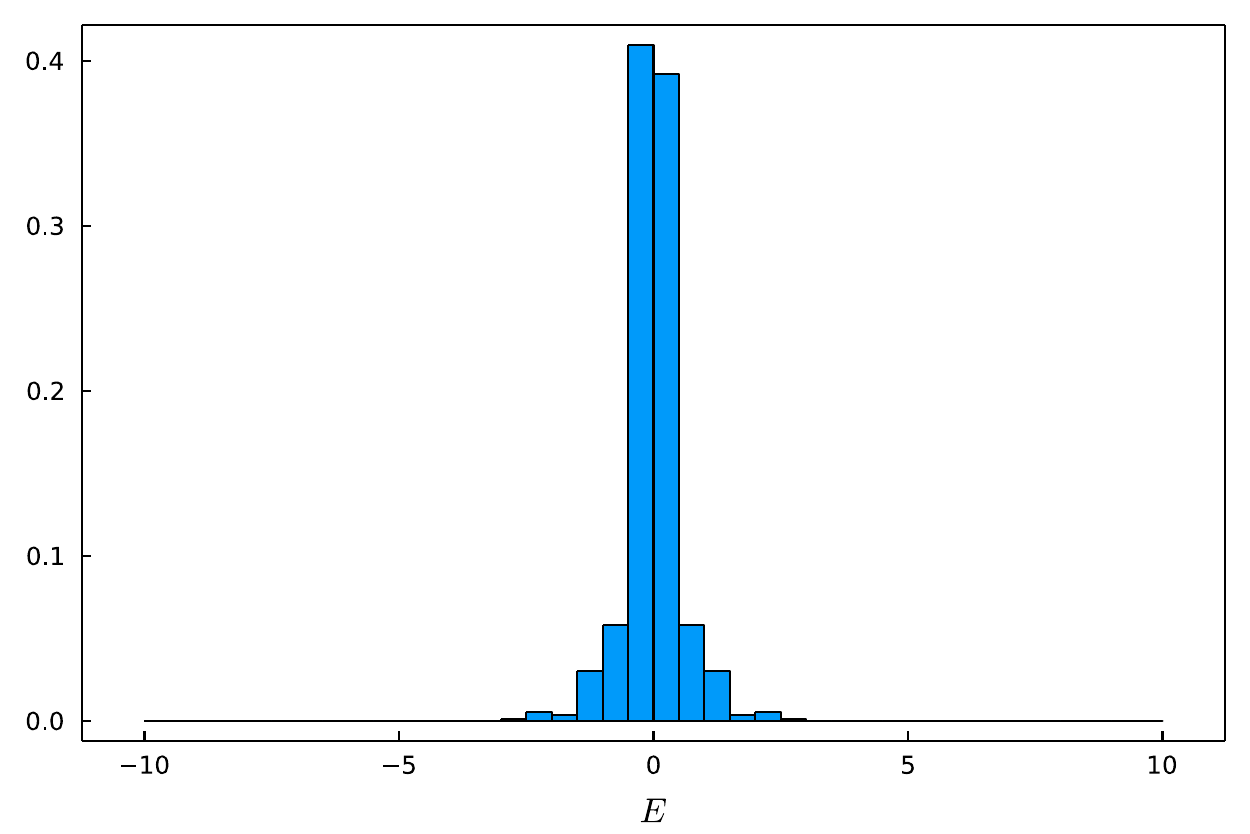}
         \caption{} \label{post-quench-3}
     \end{subfigure}
     \vfill
     \begin{subfigure}[t]{.48\textwidth}
         \centering
         \captionsetup{font=large, labelfont=bf}
         
         \includegraphics[width=\textwidth]{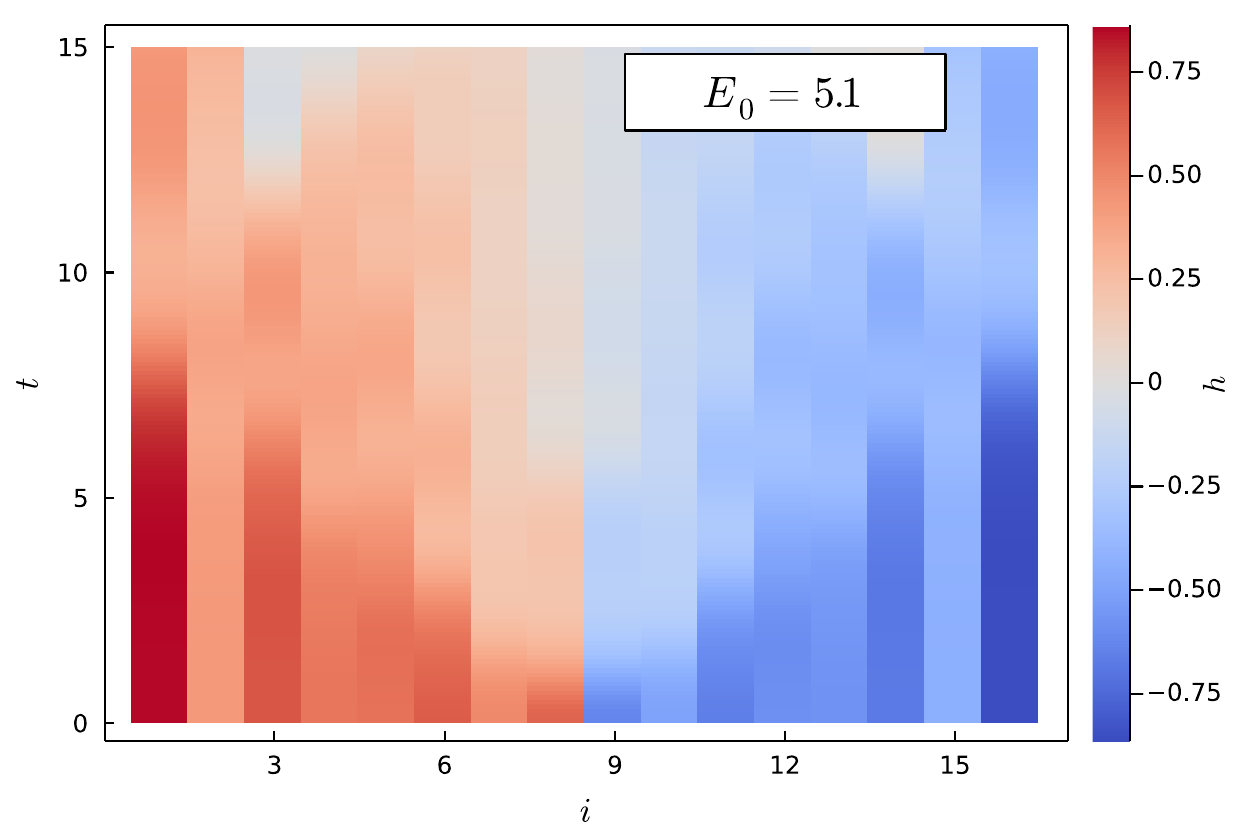}
         \caption{} \label{space-time-3}
     \end{subfigure}     
     \hfill
     \begin{subfigure}[t]{.48\textwidth}
         \centering
         \captionsetup{font=large, labelfont=bf}
         
         \includegraphics[width=\textwidth]{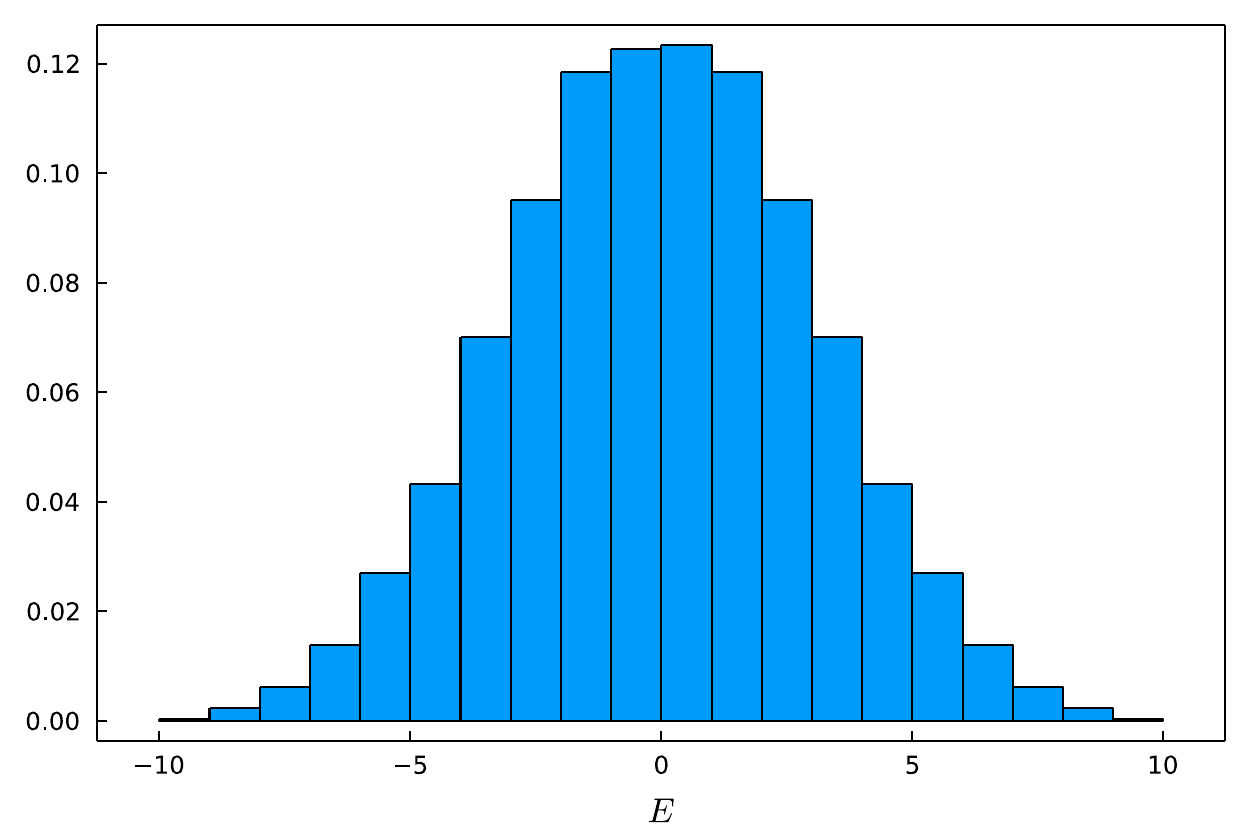}
         \caption{} \label{dos-3}
     \end{subfigure}
     \caption[Pre-quench and post-quench weights, $E_0 = 5.1$]{(a) Weights of the $\ket{\psi_{\text{left}}}$ and $\ket{\psi_{\text{right}}}$ in Eq. (\ref{mi:left}) and (\ref{mi:right}) with the pre-quench Hamiltonian $H_{1/2}$ for one half of the chain. We see that the state is prepared in the ground state $E_{g.s.} \approx -5.1 $. The system studied is of size $L=16$ ($L_{1/2} = 8$), with $\delta = 1,000$ and $E_0 = 5.1$ in Eq. (\ref{mi:left}) and (\ref{mi:right}). (b) Weights of the $\ket{\psi} = \ket{\psi_{\text{left}}} \otimes \ket{\psi_{\text{right}}} $ state with the post-quench Hamiltonian $H$ for the whole chain. It is clear that $\ket{\psi}$ has large weights around eigenenergies with $E=0$. (c) Space-time plot, as in Fig. \ref{various-E0}, where the $x$-axis is the index $i$ of the site on the chain, the $y$-axis is the time $t$ and the colorbar is the energy density $h$. (d) Density of states for the post-quench Hamiltonian $H$.}
\label{fig:weights-3}
\end{figure}

In Fig. \ref{fig:weights-2} and \ref{fig:weights-3}, we do the same, but with $E_0 = 2.8$ and $E_0 = 5.1$.

\section{Further splitting of the chain}
Another way to see the fronts, instead of splitting the $L$ chain in two halves of size $L/2$ each, is to split the chain in three sections, where the middle part is comprised of only three sites. In other words, the left part and right parts are going to be of size $\frac{L-3}{2}$ and Hamiltonian $H_{\frac{L-3}{2}}$ each, while the middle is going to have size $L=3$ and Hamiltonian $H_3$, provided that we use odd $L$. Then, we can prepare the middle part in a high energy state and the rest of the chain, the other two parts, in the ground state, or in other interesting configurations. Exactly as before, the Hilbert space of the chain $\mathcal{H}_L$ will be a restriction of the larger Hilbert space $$ \mathcal{H}^{\text{left}}_{\frac{L-3}{2}} \otimes \mathcal{H}^{\text{middle}}_{3} \otimes \mathcal{H}^{\text{right}}_{\frac{L-3}{2}} ,$$ where the three Hilbert spaces are for the left, the middle, and the right parts respectively, with OBC. The restriction is again due to the fact that we have to exclude the possibility of adjacent excitations at the point of contact between the sections of the chain that we are joining, see Section \ref{sec:thermal}.

For example, for a system with $L=23$, we prepare the whole chain in the ground state, using the method of Section \ref{sec:thermal} and $\beta = 10$, apart from the middle 3 sites, which we prepare in a high energy state using $\beta = -10$. In other words, the three parts of the chain are prepared in
\begin{equation}\label{th2:left}
\ket{\psi_{\text{left}}} = \frac{1}{\sqrt{A}}\ e^{-\beta_{\text{left}} H_{\frac{L-3}{2}} } \ket{\psi_0} ,
\end{equation}
\begin{equation}\label{th2:middle}
\ket{\psi_{\text{middle}}} = \frac{1}{\sqrt{A'}}\ e^{-\beta_{\text{middle}} H_{3} } \ket{\psi_0'} ,
\end{equation}
\begin{equation}\label{th2:right}
\ket{\psi_{\text{right}}} = \frac{1}{\sqrt{A''}}\ e^{-\beta_{\text{right}} H_{\frac{L-3}{2}} } \ket{\psi_0''},
\end{equation}
where the $A$'s are normalization constants, the $\ket{\psi_0}$'s are random initial states, $\beta_{\text{left}}=10$, $\beta_{\text{middle}}=-10$ and $\beta_{\text{right}}=10$.

\begin{figure} 
     \centering
     \begin{subfigure}[t]{.75\textwidth}
         \centering
         \captionsetup{justification=raggedright, singlelinecheck=false, font=large, labelfont=bf}
         \caption{} \label{fig:3-divs-short}
         \includegraphics[width=\textwidth]{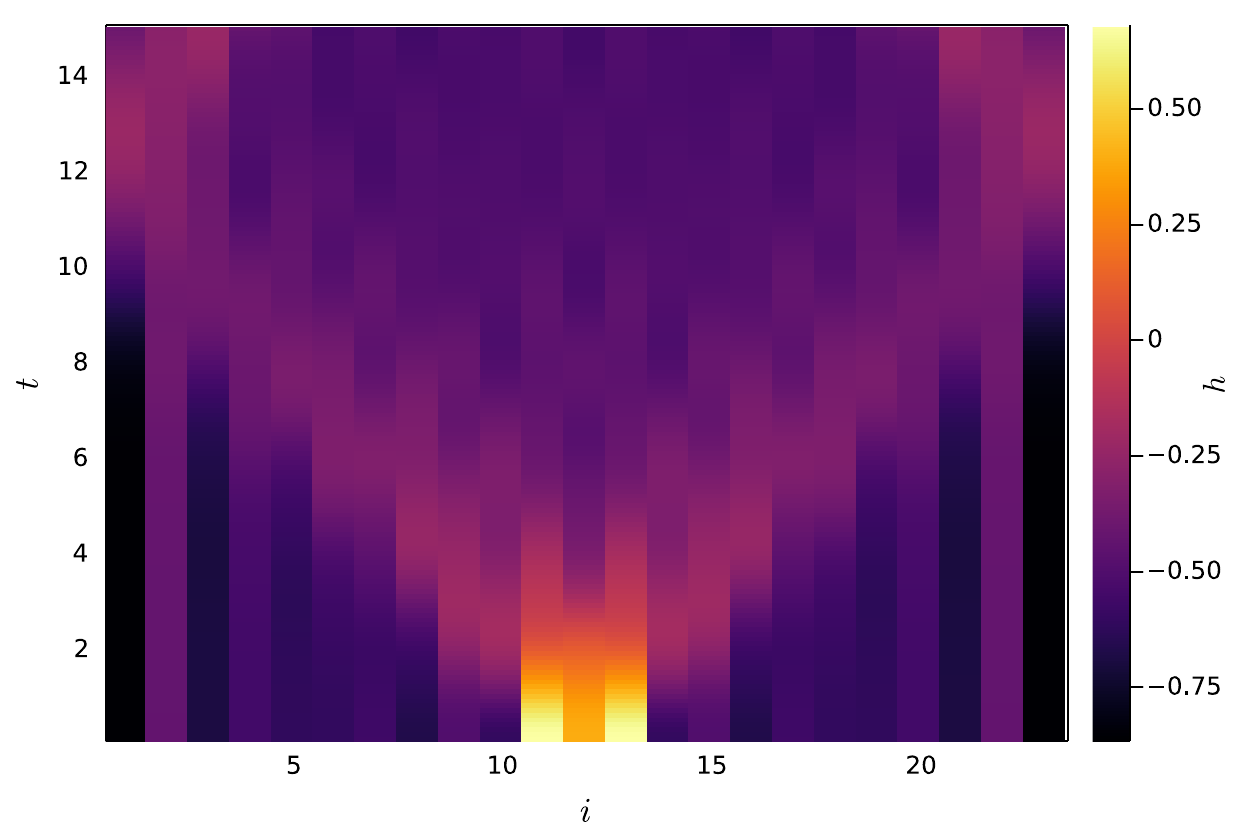}
     \end{subfigure}     
     \vfill
     \begin{subfigure}[t]{.75\textwidth}
         \centering
         \captionsetup{justification=raggedright, singlelinecheck=false, font=large, labelfont=bf}
         \caption{} \label{fig:3-divs-long}
         \includegraphics[width=\textwidth]{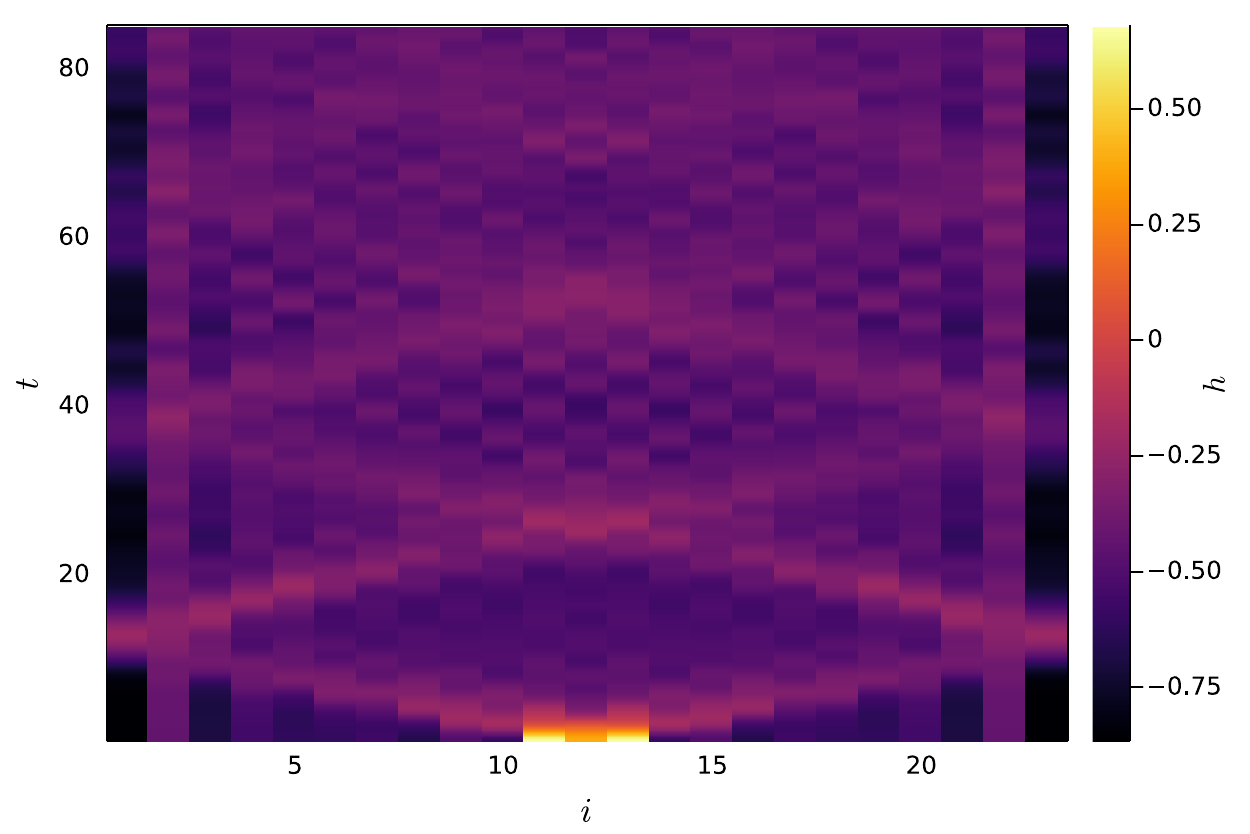}
     \end{subfigure}
     \caption[Space-time diagram of energy density, three sections, $L=23$]{(a) Space-time diagram of the evolution of the energy density of a PXP chain (with OBC) of size $L=23$, prepared in a state with the left and right 10 sites having an inverse temperature $\beta = 10$, and the middle 3 sites having $\beta = -10$. This results in the whole chain being prepared in the ground state apart from the middle three sites which are prepared in the highest excited state. The $x$-axis is the index $i$ of the site on the chain, the $y$-axis is the time $t$ and the colorbar is the energy density $h$. It is evident that the fronts produced are ballistic. (b) The same space-time diagram, but for longer time $t$.}
\label{fig:3-divs}
\end{figure}

The resulting space-time diagram is presented in Fig. \ref{fig:3-divs-short}, where the $x$-axis is the index $i$ of the site on the chain, the $y$-axis is the time $t$ and the colorbar is the energy density $h$. It is evident that the fronts produced are ballistic. In Fig. \ref{fig:3-divs-long}, we see the behavior of the fronts for larger $t$. The fronts start diffusing slightly after at least 2 reflections.

\begin{figure} 
\centering
\includegraphics[width=\textwidth]{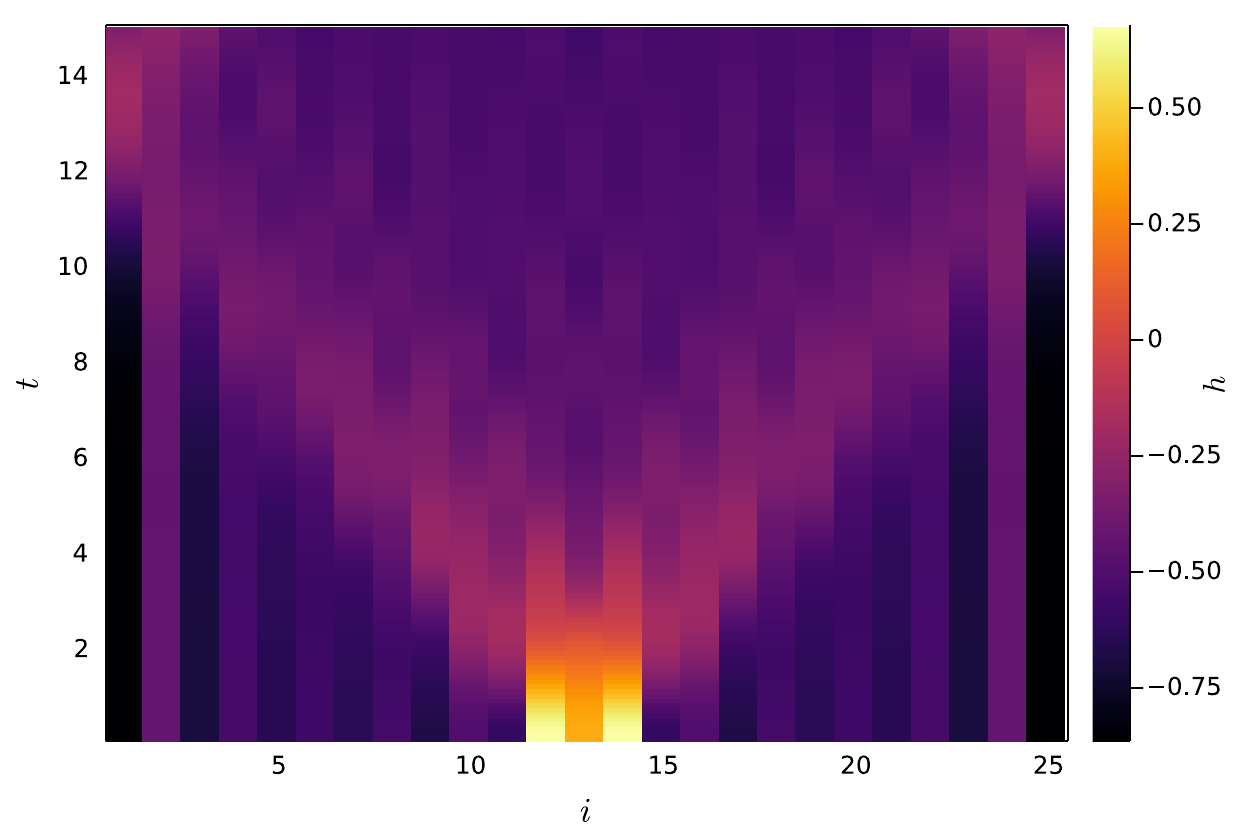}
\caption[Space-time diagram of energy density, three sections, $L=25$]{Space-time diagram of the evolution of the energy density of a PXP chain (with OBC) of size $L=25$, prepared in a state with the left and right 11 sites having an inverse temperature $\beta = 10$, and the middle 3 sites having $\beta = -10$. This results in the whole chain being prepared in the ground state apart from the middle three sites which are prepared in the highest excited state. The $x$-axis is the index $i$ of the site on the chain, the $y$-axis is the time $t$ and the colorbar is the energy density $h$. It is evident that the fronts produced are ballistic.}
\label{fig:3-divs-25}
\end{figure}

In Fig. \ref{fig:3-divs-25}, we do the same calculation but for $L = 25$.

\chapter{Discussion and Outlook}
In this work, we studied the eigenstate and dynamical properties of the PXP model, which describes a chain of Rydberg atoms realized in recent experiments \cite{Bernien:2017to}. We started by exactly diagonalizing the Hamiltonian of the system, utilizing the system's symmetries according to \cite{Sandvik}. Then, we used the eigenvectors and eigenvalues of this system to examine various aspects of the model. We tested the ETH for this model, and demonstrated the existence of a small number of states, throughout the PXP spectrum, that violate the ETH. Then we studied various quantum chaos diagnostics, like level-spacing statistics and eigenvector component statistics, for the model. The model has intermediate spectral statistics, so-called semi-Poisson, which seem to tend to Wigner--Dyson statistics for systems with size $L>28$. At the same time, the eigenvector component statistics of the model are non-Guassian, which comes as a surprise since chaotic systems have eigenvectors akin to Gaussian random vectors and hence Gaussian eigenvector component statistics. This non-Gaussianity might be related to the exotic dynamics observed and the breaking of the ETH. Finally, we performed a quench on the system and studied its response. We see linear fronts, which is also unexpected since the system, prepared in a generic state, is diffusive.

While our work sheds light on various properties of the special PXP states that strongly violate the ETH, many interesting questions remain open. Firstly, is the non-Gaussianity of the eigenvector component statistics related to the revivals observed in the PXP model when starting from the $\ket{\mathbb{Z}_2}$ state, or the violation of the ETH? Secondly, on the quenched system, why are the fronts linear? Could the scars be responsible for them? Finally, the issues discussed above naturally connect to questions about practical uses of quantum many-body scars and their dynamical signatures.

\chapter{Appendix -- Julia scripts}

The project can be found in the Github repository \href{https://github.com/fgias/quantum-chaos}{github.com/fgias/quantum-chaos}.

\section{Exact diagonalization of PXP Hamiltonian}
File name: \texttt{pxp-0+-no\_adjacent.jl}.
\begin{verbatim}
##==============================================================================
# zero-momentum, parity symmetric, no adjacent Rydberg PXP
##==============================================================================

using Plots, LinearAlgebra

## codecell ====================================================================
# functions definition

function fock2label(d, L)::Int64
    base = 2                        
    s = zero(eltype(d))
    mult = one(eltype(d))
    for val in d
        s += val * mult
        mult *= base
    end
    return s+1
end

function label2fock(α, L)
    return(digits(α-1, base=2, pad=L))
end

function flip_spin(α, j, L)
    state = label2fock(α, L)
    if state[j]==1
        state[j]=0
    else
        state[j]=1
    end
    return(fock2label(state, L))
end

function is_flip_allowed(α, j, L)
    jl, jr = mod1(j-1, L), mod1(j+1, L)
    state = label2fock(α, L)
    if state[jl]==0 && state[jr]==0
        return(true)
    else
        return(false)
    end   
end

# flip spin in bit representation
function flip_spin_state(d, j, L) 
    state = copy(d)     # assign with state = d * 1. or state = copy(d)
    if state[j]==1
        state[j]=0
    else
        state[j]=1
    end
    return(state)
end

# check if flip is allowed in bit representation
function is_flip_allowed_state(state, j, L)
    jl, jr = mod1(j-1, L), mod1(j+1, L)
    if state[jl]==0 && state[jr]==0
        return(true)
    else
        return(false)
    end   
end

# inversion
function reflect_bits(s, L)
    new_state = zeros(L)
    for i in 1:L
        new_state[i] = s[L-i+1]
    end
    new_state = Vector{Int64}(new_state)
    return(new_state)
end

# zero-momentum and parity
function check_state(s, L, k)
    R = -1;
    m = -1; # reflection-translation number
    for i = 1:L
        t = circshift(s, i)
        if fock2label(t, L) < fock2label(s, L)
            return (R, m)
        elseif fock2label(t, L) == fock2label(s, L)
            if mod(k, L/i) != 0
                return (R, m)
            end
            R = copy(i)
            break
        end
    end

    r = reflect_bits(s, L)
    for i = 0:(R-1)
        t = circshift(r, i)
        if fock2label(t, L) < fock2label(s, L)
            R = -1
            return (R, m)
        elseif fock2label(t, L) == fock2label(s, L)
            m = copy(i)
            return (R, m)
        end
    end
    return (R, m) # was getting an error when it was not returning anything
end

function construct_basis(k, p, basis, L) # for k = 0 sector
    reduced_basis = []
    Rs = []
    ms = []
    D = 2^L
    σ = 1 # k = 0 sector
    M = length(basis)
    a = 0
    for s = 1:M
        state = basis[s]
        (R, m) = check_state(state, L, k)
        
        if R > 0
            a = a + 1
            append!(reduced_basis, [state])
            append!(Rs, σ*R)
            append!(ms, m)
        end
    end
    return (reduced_basis, Rs, ms)
end

# find representative
function find_representative(s, L)
    r = copy(s)
    l = 0 # number of translations
    for i = 1:(L-1)
        t = circshift(s, i)
        if fock2label(t, L) < fock2label(r, L)
            r = copy(t);
            l = copy(i);
        end
    end

    m = 0
    s = reflect_bits(s, L)
    for i = 1:(L-1)
        t = circshift(s, i)
        if fock2label(t, L) < fock2label(r, L)
            r = copy(t)
            l = 0
            m = copy(i)
            #return (r, l, m)
        end
    end
    return (r, l, m)
end

# find position
function find_state(s, L, basis)
    b = -1
    position = findall(x->x==s, basis)
    if length(position) > 0
        b = position[1]
    end
    return(b)
end

# keep basis states with no adjacent Rydberg atoms
function no_adjacent_basis(L)
    state_list = []
    D = 2^L
    for α = 1:D
        add = true
        state = label2fock(α, L)
        for i = 1:L
            j = mod1(i+1, L)
            if state[i] == 1 && state[j] == 1
                add = false
                break
            end
        end
        if add == true
            append!(state_list, [state])
        end
    end
    return(state_list)
end

function construct_H_PXP_reduced(L)
    k = 0 # zero momentum
    p = 1 # inversion symmetry
    
    basis = no_adjacent_basis(L)

    (basis, Rs, ms) = construct_basis(k, p, basis, L);


    M = length(basis)

    H = zeros(Float64, (M,M))

    for α = 1:M
        state = copy(basis[α])

        Na = 2 * L^2 / Rs[α]
        if ms[α] >= 0
            Na = Na * 2
        end

        for i = 1:L
            if is_flip_allowed_state(state, i, L)
                new_state = flip_spin_state(state, i, L)
                (repr, l, m) = find_representative(new_state, L)
                β = find_state(repr, L, basis)

                if β >= 0
                    Nb = 2 * L^2 / Rs[β]
                    if ms[β] >= 0
                        Nb = Nb * 2
                    end

                    H[β,α] += sqrt(Nb/Na)
                end
            end
        end
    end
    return H
end

## codecell ====================================================================
# test calculation

L = 10;
H = construct_H_PXP_reduced(L)
pl = heatmap(H, yflip=true)
display(pl)

basis = no_adjacent_basis(L)
(basis, Rs, ms) = construct_basis(0, 1, basis, L);

## codecell ====================================================================
# check symmetry

isapprox(transpose(H), H) # has to be true

esys = eigen(Hermitian(H)) # eigen for Hermitian matrices
evals, evecs = esys.values, esys.vectors;

size(H)

## codecell ====================================================================
\end{verbatim}
\pagebreak

\section{Breakdown of ETH in special eigenstates}
File name: \texttt{eev.jl}.
\begin{verbatim}
##==============================================================================
# zero-momentum, parity symmetric, no adjacent Rydberg PXP
# calculation of EEVs
##==============================================================================

using Plots, LinearAlgebra, LaTeXStrings

## codecell ====================================================================
# functions definition

include("pxp-0+-no_adjacent.jl");

function construct_O_Z_PXP_reduced(L)
    k = 0 # zero momentum
    p = 1 # inversion symmetry
    
    basis = no_adjacent_basis(L)

    (basis, Rs, ms) = construct_basis(k, p, basis, L);

    M = length(basis)

    O = zeros(Float64, (M,M))

    for α = 1:M
        state = copy(basis[α])
        for i = 1:L
            O[α,α] += (-1)^(state[i]+1)/L # normalization factors cancel out
        end
    end
    return O
end

## codecell ====================================================================
# test calculation

L = 20;
H = construct_H_PXP_reduced(L) # from fully reduced PXP
esys = eigen(Hermitian(H))
evals, evecs = esys.values, esys.vectors;
O = construct_O_Z_PXP_reduced(L)

basis = no_adjacent_basis(L);
(basis, Rs, ms) = construct_basis(0, 1, basis, L);

## codecell ====================================================================

eev = []
for i = 1:length(basis)
    e = transpose(evecs[:,i]) * O * evecs[:,i]
    append!(eev, e)
end

## codecell ====================================================================

plt = plot(evals, eev, seriestype=:scatter, legend=false, 
title=L"L = %$L", markersize=3, xlabel=L"E", ylabel=L"\langle O^Z \rangle", 
framestyle=:box, grid=false)

# savefig(plt, "O^Z, L = $L.pdf")

## codecell ====================================================================
\end{verbatim}
\pagebreak

\section{Calculation of the canonical curve from the Gibbs state}
File name: \texttt{thermal.jl}.
\begin{verbatim}
##==============================================================================
# zero-momentum, parity symmetric, no adjacent Rydberg PXP
# thermal states
##==============================================================================

using LinearAlgebra, Plots, LaTeXStrings, ProgressBars

## codecell ====================================================================
# includes

include("pxp-0+-no_adjacent.jl");
include("eev.jl");

## codecell ====================================================================

L = 20
O = construct_O_Z_PXP_reduced(L)
H = construct_H_PXP_reduced(L)

evals, evecs = eigen(Hermitian(H))

# thermal curve
E_list = []
o_list = []
for β = ProgressBar(-2:.1:2)
    E = 0; o = 0;
    Z = sum(exp.(-β * evals))
    ρ = Matrix(Diagonal(exp.(-β * evals)))/Z
    
    E = tr(ρ * transpose(evecs) * H * evecs)
    o = tr(ρ * transpose(evecs) * O * evecs)
    append!(E_list, E)
    append!(o_list, o)

end

# the canonical curve
plot(E_list, o_list, legend=false, ylim=[-0.6, -0.3], linewidth=4)

## codecell ====================================================================
# plot with EEVs

L = 20
plt = plot(evals, eev, seriestype = :scatter,
title = L" \langle O^Z \rangle,\ L = %$L",
labels=false, xaxis=L"E", yaxis=L"\langle O^Z \rangle")
plot!(E_list, o_list, ylim=[-0.6, -0.3], linewidth=4, labels="Canonical")

# savefig(plt, "O^Z and canonical, L = $L.pdf")

## codecell ====================================================================
\end{verbatim}
\pagebreak

\section{Scaling of the standard deviation of \texorpdfstring{$O^Z$}{O Z} EEVs}
File name: \texttt{rolling-argparse.jl}.
\begin{verbatim}
##==============================================================================
# rolling mean and variance for eevs
##==============================================================================

using Plots, LinearAlgebra, LaTeXStrings, Statistics, DataFrames, GLM, 
JLD2, ArgParse

## codecell ====================================================================
# command line parsing

function parse_commandline()
    s = ArgParseSettings()

    @add_arg_table s begin
        "L1"
            help = "lower limit on the number of spins"
            arg_type = Int
            required = true
        "L2"
            help = "upper limit on the number of spins"
            arg_type = Int
            required = true
    end

    return parse_args(s)
end

parsed_args = parse_commandline()
L1 = parsed_args["L1"]
L2 = parsed_args["L2"]

## codecell ====================================================================
# includes
# PXP 0(+)

include("pxp-0+-no_adjacent.jl");
include("eev.jl");

## codecell ====================================================================
# variance scaling with D
# PXP

function eevs_rolling(L, no_windows=18, window_threshold=20)
    t1 = time_ns()
    println("---------------------------------------")
	println("Starting calculation: L = $L")
	println("---------------------------------------")
    H = construct_H_PXP_reduced(L) # for reduced PXP
    println("constructed H")
    esys = eigen(Hermitian(H))
    evals, evecs = esys.values, esys.vectors
    mkpath("data_run_id=$id")
    save_object("data_run_id=$id//evals_L=$L.jld2", evals)
    O = construct_O_Z_PXP_reduced(L) # for reduced PXP
    println("constructed O")
    basis = no_adjacent_basis(L)
    (basis, Rs, ms) = construct_basis(0, 1, basis, L)
    M = length(basis)

    eev = []
    for i = 1:M
        e = transpose(evecs[:,i]) * O * evecs[:,i]
        append!(eev, e)
    end
    println("calculated EEVs")
    save_object("data_run_id=$id//eevs_L=$L.jld2", eev)

    e_min = minimum(evals)
    e_max = maximum(evals)
    width = (e_max - e_min)/no_windows

    evals_windows = []
    means = []
    variances = []
    adjusted = []
    for i_window = 1:no_windows
        s = eev[e_min + (i_window-1)*width .< evals .< e_min + i_window*width]
        if length(s) > window_threshold
            append!(evals_windows, e_min + (i_window - 1)*width + width/2)
            append!(means, mean(s))
            append!(variances, var(s))
            append!(adjusted, s .- mean(s))
        end
    end
    println("completed rolling calculations")
    println("---------------------------------------")
	println("End of calculation: L = $L")
	println("---------------------------------------")
    σ = sqrt(var(adjusted))

    t2 = time_ns()
    time_calc = (t2 - t1)/1.0e9

    ## saving data ==========
    logm = log(M)
    logsigma = log(σ)
    log_file = open("data_run_id=$id//log_L=$L.txt", "w")
    write(log_file, "Total calculation time: $time_calc s \n")
    write(log_file, "L = $L \n")
    write(log_file, "no_windows = $no_windows \n")
    write(log_file, "window_threshold = $window_threshold \n")
    write(log_file, "M = $M \n")
    write(log_file, "σ = $σ \n")
    write(log_file, "log M = $logm \n")
    write(log_file, "log σ = $logsigma \n")
    close(log_file)

    return (evals_windows, means, M, σ)
end

## codecell ====================================================================
# PXP
# linear regression

# generating a random integer to distinguish between runs
id = rand((100000:999999)) 

x = Float64[]
y = Float64[]
for L = L1:L2
    c = log.(eevs_rolling(L)[3:4])
    append!(x, c[1])
    save_object("data_run_id=$id//x.jld2", x)
    append!(y, c[2])
    save_object("data_run_id=$id//y.jld2", y)
end

## codecell ====================================================================
\end{verbatim}
\pagebreak

\section{Overlap of special eigenstates with product states}
File name: \texttt{z2\_overlap.jl}.
\begin{verbatim}
##==============================================================================
# overlap with Z2
##==============================================================================

using LinearAlgebra, Plots, LaTeXStrings

## codecell ====================================================================
# functions definition

include("pxp-0+-no_adjacent.jl");

## codecell ====================================================================

esys = eigen(Hermitian(H)) # H from fully reduced PXP
evals, evecs = esys.values, esys.vectors;
M = size(H)[1]

init_fock = []     # |Z2>
for i = 1:L
    if mod(i,2)==0
        append!(init_fock, 0)
    else
        append!(init_fock, 1)
    end
end
init_fock = Vector{Int64}(init_fock)
init_α = findall(x->x==init_fock, basis)[1]
init_state = zeros(M)
init_state[init_α] = 1

overlap = zeros(M)
for k = 1:M
    overlap[k] = log(abs(transpose(evecs[:,k])*init_state)^2)/log(10)
end


plt = plot(evals, overlap, seriestype = :scatter, 
    title = L"L = %$L", ylims=[-9,0], legend=false, markersize=3, 
    xaxis=L"E", yaxis=L"\log_{10} |\langle \mathbb{Z}_2 | \psi \rangle|^2",
    framestyle=:box, grid=false
)
display(plt)

# savefig(plt, "Z2 overlap, L = $L.pdf")

## codecell ====================================================================
\end{verbatim}
\pagebreak

\bibliography{bibliography/bibliographyQMBS}

\end{document}